\documentclass[preprint]{aastex}
\begin{document}

\title{Cosmic Infrared Background Fluctuations in Deep {\it Spitzer} IRAC
Images: Data Processing and Analysis}
\author{Richard G. Arendt\altaffilmark{1,2,3}, A. Kashlinsky\altaffilmark{1,2},
S. H. Moseley\altaffilmark{1,4}, J. Mather\altaffilmark{1,4}}
\altaffiltext{1}{Observational Cosmology Laboratory, Code 665,
Goddard Space Flight Center, 8800 Greenbelt Road, Greenbelt, MD
20771; Richard.G.Arendt@nasa.gov, Alexander.Kashlinsky@nasa.gov,
Harvey.Moseley@nasa.gov, John.C.Mather@nasa.gov}
\altaffiltext{2}{Science Systems \& Applications Inc.}
\altaffiltext{3}{University of Maryland -- Baltimore County}
\altaffiltext{4}{NASA}

\begin{abstract}
This paper provides a detailed description of the data reduction
and analysis procedures that have been employed in our previous
studies of spatial fluctuation of the cosmic infrared background
(CIB) using deep {\it Spitzer} IRAC observations. The
self-calibration we apply removes a strong instrumental signal
from the fluctuations which would 
otherwise corrupt our results. The
procedures and results for masking bright sources, and modeling
faint sources down to levels set by the instrumental noise are
presented. Various tests are performed to demonstrate that the
resulting power spectra of these fields are not dominated by
instrumental or procedural effects. These tests indicate that the
large scale ($\ga 30'$) fluctuations that remain in the deepest
fields are not directly related to the galaxies that are bright
enough to be individually detected. We provide the
parameterization of these power spectra in terms of separate
instrument noise, shot noise, and power law components. We discuss
the relationship between fluctuations measured at different
wavelengths and depths, and the relations between constraints on
the mean intensity of the CIB and its fluctuation spectrum. 
Consistent with growing evidence that the $~\sim 1 - 5\ \micron$ 
mean intensity of the CIB may not be as far above the integrated 
emission of resolved galaxies as has been reported in some analyses of
DIRBE and {\it IRTS} observations, our measurements of spatial fluctuations
of the CIB intensity indicate the mean emission from the objects producing
the fluctuations is quite low ($\gtrsim 1$ nW m$^{-2}$ sr$^{-1}$ at 
$3 - 5 \micron$), and thus consistent with current $\gamma$-ray 
absorption constraints. The source of the fluctuations may be high-$z$ 
Population III objects, or a more local component of very low 
luminosity objects with clustering properties that differ 
from the resolved galaxies. Finally, we discuss the prospects of the upcoming
space-based surveys to directly measure the epochs inhabited by the populations
producing these source-subtracted CIB fluctuations, and to isolate the
individual fluxes of these populations.
\end{abstract}


%%%%%%%%%%%%%%%%%%%%%%%%%%%%%%%%
\section{Introduction}
The cosmic infrared background (CIB) is comprised of radiation emitted 
throughout the entire history of the Universe (e.g. Bond, Carr \&
Hogan 1986). 
The CIB contains emission from objects which may be too 
faint to be individually detected or too numerous to be individually 
resolved with current (or even future) instruments. However, since the 
collective emission is detectable, the CIB provides unique 
information on the history of the Universe at very early times.
Analogous to studying the brightness and structure of individual 
galaxies in which the stars cannot be resolved, in
recent years we have witnessed new CIB measurements
identifying and constraining both its mean level (isotropic
component) and spatial fluctuations (see Kashlinsky 2005a for a
recent review). The near-IR CIB (hereafter taken to span
wavelengths from $1 - 10 \micron$) probes stellar emission, whereas
at longer wavelengths the CIB is generated by dust. Foregrounds,
such as Galactic stars, interstellar dust emission (cirrus),
zodiacal light, and atmospheric emission, represent formidable
obstacles to isolating the true CIB (see review by Leinert et al
1998). Significant progress in CIB research was made possible due
to dedicated space experiments conducted by COBE/DIRBE (Hauser et
al. 1998; see review by Hauser \& Dwek 2001) and {\it IRTS} (Matsumoto et
al. 2005).

Theoretically, the most plausible candidates for the bulk of the
near-IR CIB are evolving stellar populations in galaxies. These
nucleosynthetic energy sources would include the first generation
of stars, known as Population III. A fraction of the CIB must also
be generated by accretion onto black holes in active galactic
nuclei (AGN) rather than by stellar nucleosynthesis. 
It is now thought that the first stars were very massive (see review by
Larson \& Bromm 2004), in which case theoretical models indicate
they may produce a detectable contribution to the mean level 
of the near-IR CIB (Santos et al.
2002, Salvaterra \& Ferrarra 2003; Kashlinsky 2005b, Dwek, Arendt
\& Krennich 2005, Fernandez \& Komatsu 2005). 
They are also 
expected to have left a measurable imprint in CIB anisotropies
(Cooray et al. 2004, Kashlinsky et al. 2004). The intuitive reasons
why these fluctuations would be significant are: 1) if massive,
such stars would emit at light to mass ratios $\sim 10^4-10^5$
higher than the present-day stellar populations leading to
significant CIB flux levels; 2) assuming that the Pop III era
occupied a comparatively narrow epoch in time (say $\Delta t\sim $
a few hundred million years) there should be a higher amplitude of
relative CIB fluctuations ($\propto 1/\sqrt{\Delta t}$); and 3) it
is expected that within the framework of the concordance
$\Lambda$CDM model the first stars formed out of rare high peaks
of the underlying density field and, hence, their correlation
properties would be amplified. The CIB fluctuations from such
early populations are distinguishable from those produced by more
recent populations. Their spatial spectrum should reflect the
$\Lambda$CDM matter spectrum rising to a peak at $\sim
0.3^\circ-0.5^\circ$ and its spectral energy distribution should
be cutoff due to the Lyman break at wavelengths $\lesssim 1
[(1+z)/10] \micron$ (Cooray et al. 2004, Kashlinsky et al. 2004).

Present measurements of the mean CIB levels are based on the DIRBE
and {\it IRTS} data and suggest a substantial excess over the
contribution from known galaxy populations (Dwek \& Arendt 1998,
Gorjian et al. 2001, Arendt \& Dwek 2003, Matsumoto et al. 2005).
This excess of $\sim 30$ nW m$^{-2}$ sr$^{-1}$ at $\lambda \gtrsim
1 \micron$\ (Kashlinsky 2005a) is commonly known as the NIRBE
(Near-IR Background Excess). If produced by the first massive
stars, it is proportional to the fraction of baryons processed
through these stars; a fraction of $\sim 2-4\%$ is necessary to
explain the levels claimed in the above studies. On the other
hand, much of this excess may be due to inaccurate zodiacal light modeling
(Dwek et al. 2005) and the remaining NIRBE may be much smaller, in
agreement with the recent analysis of the deep {\it HST} NICMOS
data at 1.6 $\micron$ (Thompson et al. 2007a). Further limits on
the CIB come from the amount of photon absorption at $\gamma$-ray
energies in blazars at moderate $z\sim 0.2$ (Dwek et al. 2005,
Aharonian et al. 2005). However such limits are sensitive to the
assumptions on the intrinsic unabsorbed blazar spectrum and the
fine details of the CIB spectral distribution (Kashlinsky \& Band
2007); they are discussed later in the paper.

At certain wavelengths CIB fluctuations can be more readily
measurable than the mean levels. 
As differential rather than absolute 
measurements, the study of fluctuations places different requirements 
on instrument capabilities and calibration, and on the precision
of removal of foreground emission (zodiacal and Galactic).
Shectman (1974) applied
fluctuations analysis to constrain the diffuse light in the
optical bands. Kashlinsky et al. (1996a,b) have pioneered such
studies in the IR using DIRBE data with a further analysis by
Kashlinsky \& Odenwald (2000) isolating the degree-scale CIB
fluctuation at 1-5 $\micron$. The {\it IRTS} results on CIB fluctuations
at $\sim 2 \micron$ agree with the latter study and extend to
larger angular scales (Matsumoto et al. 2005). Because of their
wide beams the DIRBE- and {\it IRTS}-based data sets did not allow the
removal of many foreground galaxies and the isolation of the
contribution from fainter sources. Using ground-based deep 2MASS
measurements in the J,H,K photometric bands enabled removal of
galaxies to $m_{\rm Vega}\sim 18-19$ and led to detecting the CIB
fluctuations signal from galaxy populations below that magnitude
threshold on subarcminute scales (Kashlinsky et al. 2002, Odenwald
et al. 2003). However, all of these studies involved data sets with
either low angular resolution and/or relatively shallow
integrations so that the CIB fluctuations from the remaining
galaxies and instrument noise prevented isolating any signal
arising from the first stars epochs.

In the past several years we have used deep-integration {\it
Spitzer} data to measure the CIB fluctuations component 
(Kashlinsky, Arendt, Mather \& Moseley
2005, 2007a,b,c - hereafter KAMM1, KAMM2, KAMM3, KAMM4). 
They revealed
significant CIB fluctuations at the IRAC wavelengths (3.6 to 8
$\micron$) which remain after removing galaxies down to very faint
levels (KAMM1, KAMM2). These fluctuations must arise from
populations that have a significant clustering component, but only
low levels of shot noise (KAMM3). Furthermore, it was shown
that there are no correlations between source-subtracted IRAC
maps and the corresponding fields observed with the {\it HST} ACS
at optical wavelengths (KAMM4), which means that the sources
producing these CIB fluctuations are {\it not} in the ACS source
catalog extending to $m_{\rm AB} \lesssim 28$ at wavelengths $\lesssim
0.9 \micron$.

KAMM found statistically significant cross-correlations between
the different IRAC channels, indicating the presence of a common
component in all the channels, and determined the color of
unresolved fluctuations (KAMM1). Their analysis allowed the
separation of various noise and systematic effects individually in
each IRAC channel (KAMM1, KAMM2), thus characterizing the
statistical uncertainties and systematic errors in fluctuation
measurements. The results were further verified with simulated
patterns of first star galaxies (KAMM1). The simulations recovered
the input fluctuations and established the good accuracy of the
determined diffuse backgrounds in the assembled images. These
techniques directly showed that our existing procedure based on
Fourier transforms {\it and} correlation function analysis does
not lead to biased estimates of the CIB fluctuations \citep[see
also][]{Kash2007}.

In this paper we describe, illustrate, and further verify many
details of our analysis efforts. In \S 2, we describe the {\it
Spitzer} data sets we have analyzed and the self-calibration
procedures we have applied in order to generate maps with accurate
large scale structure. Section 3 details the steps used in the
analysis of the power spectra of the backgrounds. Section 4
illustrates the extent to which the derived CIB power spectra
depend on the details of the analysis steps, particularly the
masking and removal of resolved foreground sources. Section 5
discusses the results, with future prospects described in Section
6. The paper concludes with a summary (\S7).

%%%%%%%%%%%%%%%%%%%%%%%%%%%%%%%%
\section{Data Processing and Mosaic Construction}

\begin{figure}[ht] 
   \plotone{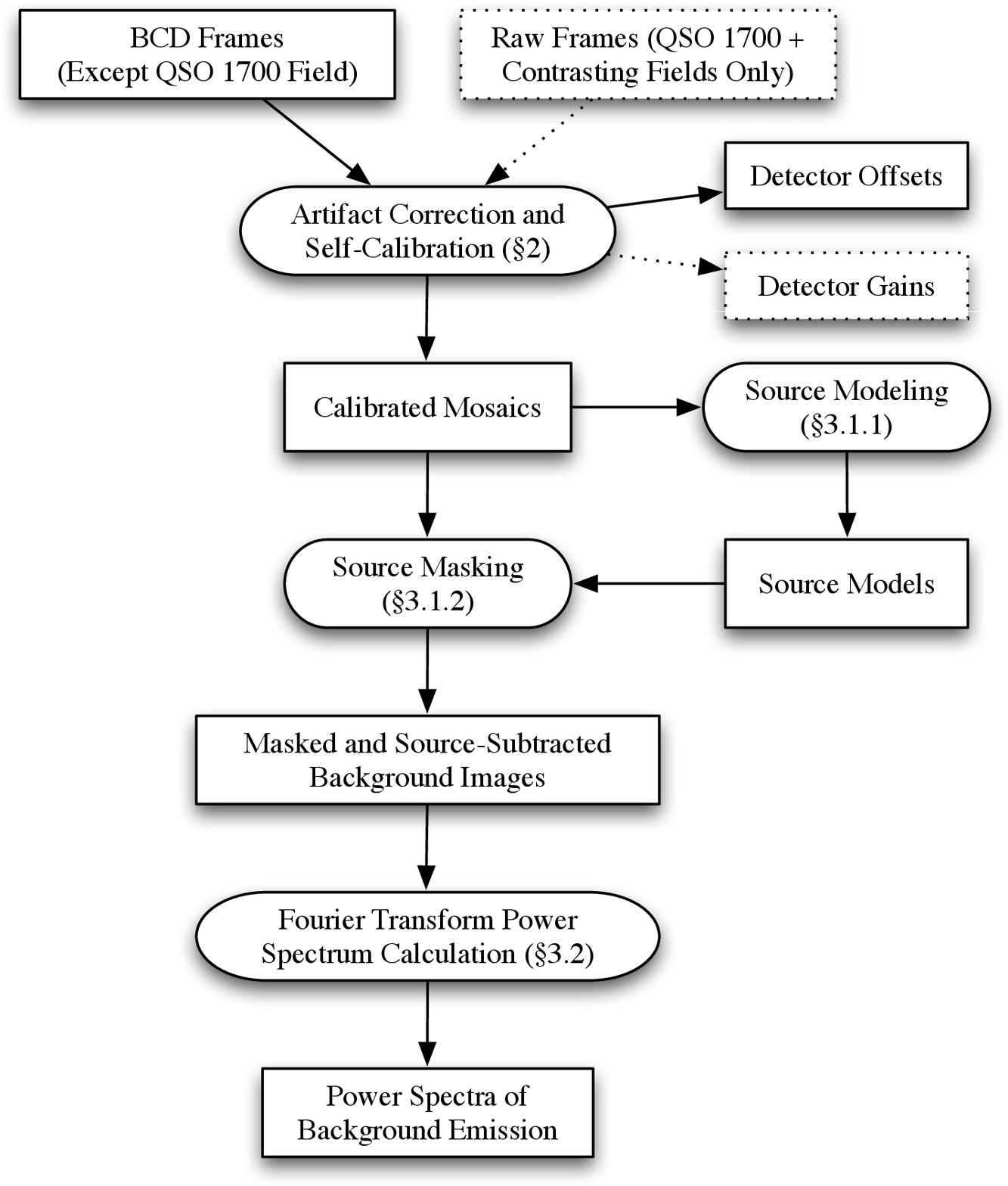}
   \caption{Flow chart of the data processing steps. Only for the QSO 1700 field
   was the process started with raw data rather than the BCD. This requires self-calibration
   for detector gains as well as offsets. }
   \label{fig:flow}
\end{figure}

In this section we present the steps taken to process the individual 
frames of IRAC data into integrated mosaicked images for each 
field, wavelength and epoch (see also the Supplementary Information of KAMM1). 
This is the first stage of the overall data processing and analysis
which is shown schematically in Figure \ref{fig:flow}.
Discussion of the effects that the data reduction may have on the results,
and of the comparison of results between different fields and different epochs
is deferred to \S4, after we present the analysis procedures in \S3.

\subsection{{\it Spitzer's} Infrared Array Camera (IRAC)}
Our research has used data from {\it Spitzer's} IRAC instrument.
This camera contains two parallel optical systems. Each system
images a separate $5'\times5'$ field of view, with the fields
separated by $\sim6'$. Using beam splitters, each optical system
collects images in two channels (at two wavelengths)
simultaneously. So while one field of view is being observed at
3.6 and 5.8 $\micron$, a nearly adjacent field is being observed
at 4.5 and 8 $\micron$. The detector
for each channel is a $256\times256$ pixel array, with a scale of
$\sim1.2''$/pixel. This pixel scale slightly undersamples the
instrument point spread function at the shortest wavelengths. The
paper by Fazio et al. (2004a) is the primary description of the
IRAC instrument. Many other details are contained within the {\it
Spitzer} Observer's Manual and the IRAC Data Handbook, which are
found on the Spitzer Science Center
website.\footnote{http://ssc.spitzer.caltech.edu}

Normal observing procedures entail dithering the telescope
pointing between successive exposures or frames. Altering the
pointing by various fractions of the array size prevents any
detector defects (e.g. bad pixels) from completely eliminating
data from a particular point on the sky. Dithering also serves to
alter the pattern of stray light artifacts (most prominently
occurring near bright sources that lie just outside the field of
view). Most importantly, dithered data can be used to derive the
relative detector offsets (and gains) as the sky itself can serve
as a stable relative calibration source. Mapping fields larger
than the instantaneous $5'\times5'$ field of view is usually done
by stepping though a rectangular grid of $N\times M$ positions
separated by $\lesssim5'$, with many dithered frames collected at
each raster position. Because of the offset between the two
instrument fields of view, the overall coverage at 3.6 and 5.8
$\micron$ is displaced from that at 4.5 and 8 $\micron$ by
$\sim6'$. In some programs, the field is re-observed after 6 months
have elapsed. At that time the relative locations of the two
fields of view are transposed, and thus if both epochs of data are
combined then the same area can be covered equally in all four
channels (wavelengths).

The properties of the IRAC data sets we have examined are listed in Table \ref{tab:data}.

\subsection{IOC Deep Image = QSO 1700 Field}
The IRAC IOC Deep Image observations were a test to verify that a
deep (close to confusion--limited) integration could detect moderately
high redshift sources. They were also intended to verify that the noise
in an image would scale inversely with the square root of the
integration time, even for very deep integrations.
A secondary goal of these observations was to investigate
the effects of different dither patterns and observing strategies on the results.
Analyses of the resolved sources in this field have been published by Fazio
et al. (2004b) and Barmby et al. (2004).

The nominal target field for these observations was a field including
the quasar HS 1700+6416 and several known Lyman break galaxies.
Each channel observed a $\sim11'\times5'$ field, or $2\times1$ IRAC fields
of view. The field observed at 3.6 and 5.8 $\micron$ only overlaps
with the field covered at 4.5 and 8 $\micron$ in a $5'\times 5'$ region.
The dithering used during the observations extended the coverage over
a wider region than the $\sim5'\times 5'$ IRAC field of view, but this
coverage is at a lower depth than the center of the field.

Scheduling constraints required that the observations be broken up into
several astronomical observation requests (AORs), so each AOR employed
a different dithering or coverage strategy.
The observations were carried out over an interval of less than 2 days.
Therefore any changes in the zodiacal light were small, e.g. the
8 $\micron$ zodiacal light intensity changes from 4.595 MJy sr$^{-1}$ to 4.600
MJy sr$^{-1}$ between the start of the first and last AORs according to
the Spitzer Science Center's zodiacal light model.

At the time we began this project (KAMM1), the basic calibrated data (BCD) pipeline
was not so well developed, and calibration observations had not been accumulated
over a long enough time to provide the best flat field and dark frame calibration.
Therefore we performed our own data reduction beginning with the raw data.
We applied the least--squares self--calibration procedure as described by Fixsen,
Moseley \& Arendt (2000). The approach formalizes the calibration procedure
by describing the data with a model whose parameters include both the
detector characteristics and the true sky intensity. The derivation
of the model parameters via a least--squares algorithm yields an optimal
solution for the calibration and the sky intensity. In this case our chosen
model is given by
\begin{equation}
D^i = S^{\alpha} G^p + F^p + F^q
\end{equation}
where $D^i$ represents the raw data from a single pixel of a single frame,
$S^{\alpha}$ is the sky intensity at location $\alpha$, $G^p$ and $F^p$ are
the gain and offset for detector pixel $p$, and $F^q$ is a variable offset
for each of the 4 readouts (alternate vertical columns of the detector) and
each frame. This model assumes that the sky intensity ($S^{\alpha}$) and
the detector gains and offsets ($G^p$ and $F^p$) are invariant during the
course of the observations. 
For a data set with fixed frame times (as our IRAC data), the detector dark 
current is included in the $F^p$ term as it is indistinguishable from an offset.
For data sets with multiple frame times, a relatively simple extension of
this data model could be applied.
The variable offset $F^q$ can absorb time--dependent
behavior of the detector, but only to the extent that it can be characterized
with a single value frame, or in some cases, a single value per readout per frame.

In order to be able to self--calibrate the raw data for both gain and offset
effects, the procedure requires a higher intensity contrast than is found
in the QSO 1700 data alone. Therefore additional AORs taken at low ecliptic
latitude (and hence high zodiacal brightness) were combined with the IOC Deep
Image AORs for the self--calibration. Ideally, these ``hizodi'' observations
would have been performed just before or after the QSO 1700 observations.
However in fact, the nearest suitable 200--sec frame time data for Channels
1 -- 3 (3.6 -- 5.8 $\micron$) were observed over a
month later, although suitable 100--sec (2$\times$50sec)
frame time data for Channel 4 (8 $\micron$) were observed shortly prior to the
QSO 1700 AORs.

The self--calibration was initially applied to each of the QSO
1700 AORs (combined with the hizodi AOR) separately. However, as
variations proved to be relatively small, our final results were
obtained by running the self-calibration on the complete set of
200--sec frame time data at each wavelength. In the case of the 8
$\micron$ data, which used four 50--sec frames for each 200 sec
frame in the other channels, we self-calibrated the data in 4
subsets, which were combined as a weighted average of the
resulting mosaics.

After the derived gain and offset calibrations are applied to the
individual frames and before they are mosaicked, each frame at 3.6
and 4.5 $\micron$ is corrected for the ``column pulldown'' effect
in the columns of bright point sources (see the IRAC Data
Handbook) using a version of the algorithm that is available from
the SSC as a user--contributed
tool\footnote{http://ssc.spitzer.caltech.edu/irac/pulldown/}. For
8 $\micron$ frames we applied a similar correction of our own
development to correct for the ``banding'' artifact which affects
detector rows containing bright sources. No artifact correction
was applied for 5.8 $\micron$ data, as its banding is less severe
than at 8 $\micron$, and it is not substantially improved by our
procedure. 
We have also not made any extra corrections for 3.6 and 4.5 $\micron$ 
muxbleed artifacts (see the IRAC Data
Handbook), which appear as a decaying excess intensity 
in consecutive detector pixels sampled by a readout after sampling a 
very bright (saturated) source (i.e. in every 4th pixel of a detector 
row following the saturated source). Because the muxbleed effect is 
strictly periodic, i.e. a shah function (Bracewell 1986) in the spatial 
domain, it transforms to another shah function in the frequency domain.
Thus excess power is found at the fundamental frequency 
(0.25 pixels$^{-1}$), and at the first harmonic which corresponds 
to the Nyquist frequency (0.5 pixels$^{-1}$). These spikes are evident 
at spatial scales of $\sim 4.8''$ and $\sim 2.4''$ in the power spectra 
which are shown below. The slow decay of the muxbleed effect transforms
to a slight broadening of the spikes in the power spectra. 
Power at these scales is not important to the present research, in which the 
primary signal of interest is found at scales $\gtrsim 40''$.

This first step in the overall data processing flow is indicated in Figure \ref{fig:flow}.
Subsequent processing steps of source modeling and source masking (see \S 3), will
additionally affect the photometry of the images to be analyzed.

Continued improvements to the BCD pipeline since the KAMM1 analysis now allow
good results when starting with the BCD (or corrected BCD, cBDC) rather than 
the raw data. Verification that similar power spectra are derived from either
data set is presented in \S 4.3.

\subsection{Extended Groth Strip}

The Extended Groth Strip observations are part of a large extragalactic GTO
project (Spitzer Program ID number 8). The full observations
cover a $10' \times 125'$ region with a
depth of 26 200-sec frames (1.4 hr), repeated after a 6--month interval. 
The data are well--dithered for our purposes, using the medium--scale cycling 
pattern\footnote{http://ssc.spitzer.caltech.edu/irac/dither.html},
which is based on a 2-dimensional gaussian distribution ($\sigma = 32$ pixels) 
of dither positions with a maximum offset of 119 pixels $\approx 145''$. 
Our data reduction did not include the full data set, but
only two $10' \times 5'$ portions of the strip, at both epochs.
Data from the two epochs were reduced separately.

The data reduction was similar to that described above for the QSO 1700 field,
except that for these data (and all subsequent data sets) our analysis began
with the individual BCD frames rather than the raw data. 
If an imperfect calibration, designated by $\{G'^p, F'^p, F'^q\}$
is applied to the raw data, $D^i$, then equation (1) becomes
\begin{eqnarray}
\frac{D^i - F'^p - F'^q}{G'^p} & = & \frac{S^{\alpha}G^p + (F^p-F'^p) + (F^q-F'^q)}{G'^p}\\
D^i_{\rm{BCD}}                 & = & S^{\alpha} + \frac{S^{\alpha}\delta G^p + \delta F^p}{G'^p} + \frac{\delta F^q}{G'^p}\\
D^i_{\rm{BCD}}                 & = & S^{\alpha} + \Delta F^p + \Delta F^q
\end{eqnarray}
where $D^i_{\rm{BCD}}$ is the BCD data, $\delta G^p \equiv G^p - G'^p$, 
$\delta F^p \equiv F^p - F'^p$ and $\delta F^p \equiv F^p - F'^p$.
In the last equation we make the approximations that $S^{\alpha}\ \sim$ constant
so that we can ignore any dependence of $\Delta F^p$ on $\alpha$, and that
$G'^p\ \sim$ constant so that we can ignore any dependence of $\Delta F^q$ on $p$.
Thus the ``delta corrections'' to be derived and applied to the BCD data
can be represented as simple offset terms, as in equation (1), but now without
a gain term in the equation. The data model that is applied when starting with 
the IRAC BCD frames can be represented as a slight variation on equation (1):
\begin{equation}
D^i = S^{\alpha} + F^p + F^q\ .
\end{equation}
With no gain term, this model has the advantage that
no contrasting data set (e.g. the high zodiacal brightness
field used with the QSO 1700 data) is required to separate
degeneracies between gain and offset. Any true gain errors that
are present in the BCD data will be absorbed in the offset term $F^p$,
by assuming that $S^\alpha$ is constant. The size of the errors made by 
approximating gain errors as offset errors are of order 
$\delta S^{\alpha} \delta G^p$ in a single BCD frame, where 
$\delta S^{\alpha} \equiv S^{\alpha} - \langle S^{\alpha} \rangle$.
The errors are reduced further by the square root of the number of dithered
frames ($\sqrt{N}$) at each location ($N>100$ frames for the deeper 
GOODS and QSO 1700 fields). The fluctuations, $\delta S^{\alpha}$, in 
the dominant zodiacal light and cirrus foregrounds are already 
estimated to be at or below the residual fluctuations (see Figure 1 of KAMM1).
Further reduction of these fluctuations by factors of $\delta G^p$ ($<1\%$, IRAC 
Data Handbook) and $1/\sqrt{N}$ means that the approximation of gain errors 
as offset errors only affects results at levels $\ll1\%$ of the detected signal.
We note that the approximation would be more problematic if we were interested 
in accurate photometry of the brighter resolved sources, for which $\delta S^{\alpha}$
would be very large.

\subsection{GOODS HDFN and CDFS}

The GOODS Legacy program (Program ID numbers 169 \& 194) is designed to obtain very deep,
confusion--limited observations over small ($10'\times 15'$) fields. The
chosen fields are the Hubble Deep Field -- North (HDFN) and the Chandra
Deep Field -- South (CDFS), which is also the location of the Hubble Ultra Deep Field.

These observations also used 200-sec frame times, and were carried out at two epochs
separated by $\sim6$ months. At the first epoch (HDFN-e1, CDFS-e1),
the two IRAC fields of view cover
partially overlapping $10'\times 10'$ fields. At the second
epoch (HDFN-e2, CDFS-e2) the IRAC fields
of view are reversed, thus providing complementary coverage.

For each channel, the BCD frames of each of the $\sim20$ AORs were processed separately to
determine preliminary calibration factors, $F^p_1$ and $F^q_1$. Maps made from these
calibrated data would show large systematic
errors because there is no constraint between the AORs
to produce the same mean sky intensity. Therefore, at each epoch, we also performed the
self--calibration on the entire data set, but after downsizing the data set by
performing $2\times2$ pixel--averaging on each BCD frame (resulting in $128\times128$
pixel frames). The derived calibration parameters $F^p_2$ and $F^q_2$
produce consistency across the entire data set, but with
limited spatial resolution and with averaging over some real temporal variations 
(between AORs) in the detector offsets ($F^p_2$). Thus, we calibrated the frames of each AOR
using $F^p = F^p_1 + \nabla(F^p_2 - F^p_1)$ and $F^q = F^q_2$,
where $\nabla(F^p_2 - F^p_1)$ is the 2--dimensional linear gradient in the difference
between the derived detector offsets. 
This combines the individual detail of the
detector offsets ($F^p_1$) derived for each AOR, with the overall consistency 
provided (via $F^q_2$) by simultaneous self--calibration of all AORs.

\subsection{Extragalactic First Look Survey}

The Extragalactic First Look Survey (FLS; Program ID number 26) is a shallow survey
covering a $2\arcdeg\times2\arcdeg$ field. Observations used 12--sec frame times with
a depth of 5 exposures dithered with the small--scale Gaussian pattern.
These data were examined to explore larger spatial scales than the deeper data sets,
despite the fact that the depth and the dithering are not especially
well--suited for self--calibration.

\subsection{Final Images}

For each of the data sets described above we mapped the
artifact--corrected and self--calibrated BCD frames into final
mosaics. The mapping procedure we used is an interlacing
algorithm, where each pixel of the BCD frame is mapped into the
pixel in the sky map that contains the center of the BCD pixel.
This is similar to a drizzle algorithm with the ``pixfrac''
parameter set to zero (Fruchter \& Hook, 2002).
A desirable aspect of this mapping
procedure is that it does not induce any pixel-to-pixel
correlations in the noise, which does occur with procedures that
map the flux of a single input pixel into multiple sky map pixels.
Another asset of this procedure is that it can easily create sky 
maps with pixel scales and orientations that are independent
of the scale and orientation of the detector pixels.
In general we prepared several variations of the final images. The
most basic images are generated by mapping the entire data set
into images with a scale of $1\farcs2$/pixel (the detector pixel
scale). For the deeper data sets, we also produced images with
scales of $0\farcs6$/pixel. This allows slightly better
discrimination of resolved point sources. For the GOODS and EGS
data sets, the data from each epoch ($\sim6$ months apart) were
mapped into separate images. These images are useful as a check
for systematic errors. Finally, for all data sets we created ``A''
and ``B'' images by mapping all the even numbered frames from the
sequence of exposures into the ``A'' image, and the odd numbered
frames into the ``B'' image. Any systematic errors should be
manifested very similarly in the A and B images, and thus the
(A-B)/2 difference images provide a useful means of characterizing
the random (noise) properties of the data sets.

%%%%%%%%%%%%%%%%%%%%%%%%%%%%%%%%
\section{Fluctuation Analysis}

In this section we present the analysis procedures that are applied to
the mosaicked IRAC images to derive the power spectra of the background. 
The analysis consist of two main parts: (1) removal of individually 
resolved sources via modeling and masking, and (2) calculation of the 
power spectra of the remaining background. These stages are 
shown in the overall data processing and analysis
schematic flow chart in Figure \ref{fig:flow}.
For clarity, this section is restricted to a direct description of the processes.
There are several aspects of the analyses which can have significant 
effects on the results. Tests of the effects that these processing
details have on the final derived power spectra will be presented
in \S4. 

\subsection{Source Removal}

In order to study the spatial fluctuations of the unresolved
extragalactic background emission, we must have a means of
removing or ignoring the influence of the brighter, resolved
galaxies and foreground stars. One such method is that the sources
can be individually fit with a model and subtracted. Practical
difficulties with this approach are (a) limitations in the
accuracy of the point response 
functions\footnote{The PRF includes the sampling of the detector pixels
as well as the point spread function (PSF) which describes the 
light incident at the surface of the detector. Description of the IRAC
PRF, and the most current PRFs, are found at: http://ssc.spitzer.caltech.edu/irac/psf.html.}
 (PRFs), and (b)
limitations in the modeling of sources that have resolved extended
structure. Small fractional errors in the PRF or source model can
result in large residuals at very bright sources. Furthermore, the
power spectrum of the residuals can exhibit different behavior
than the power spectra of the original sources or the PRF. A
complementary approach is to mask the bright resolved sources in
the images. Depending on the type of analysis to be performed, the
masked regions can either be ignored (e.g. when computing
correlation functions) or else filled with zeros or noise at the
appropriate level (e.g. when computing power spectra). The
difficulties in the masking approach are (a) for deep
observations, masking all resolved sources including their
extended wings (both due to the PRF and any extended emission) can
leave little or no data left for analysis, and (b) the masking
will likely alter the shape of the calculated power spectrum of
the image.

In our studies, we apply both techniques. A source modeling procedure is used
to ensure that the faintest resolved point sources and extended sources are removed
from the images. Masking is then applied to eliminate artifacts in the modeling and
subtraction of only the brighter emission, thus minimize the influence of the masking
on the power spectra. These steps are shown schematically in Figure \ref{fig:flow}.

\subsubsection{Source Modeling}
Our source modeling procedure is conceptually similar to the CLEAN
algorithm, which is used to remove the effects of beam sidelobes
in radio images (H\"ogbom 1974). We start with the original image
and a corresponding model image which is set to zero. The first
step is to locate the brightest pixel in the original image. At
that location, we subtract the IRAC PRF, normalized such that the
peak is a specified fraction $f$ of the pixel intensity. We {\it
add} the same scaled PRF to the corresponding location in the
model image. This process is iterated by locating the brightest
pixel in the modified image. The scaled PRF is again subtracted
from the image and added to the model. 
Because only a fraction $f$ of a source (a ``component'') 
is subtracted at each iteration,
even a point source is modeled by multiple components. 
The residual flux of an ideal point source, matched by the PRF, 
will be proportional to $(1-f)^n$ after subtraction of $n$ components.
For our analysis we used $f = 0.5$, as a compromise between speed 
(high values of $f$) and insensitivity to any PRF errors (low values 
of $f$; discussed below). So for the faintest sources, the 
residual emission of the point source is lost in the noise with $n = 2-3$.
Whereas for bright sources, several dozen components may be needed 
to reach the same level of residual emission.
The loop of finding and subtracting components is repeated on
order of 10$^4$ times, depending on the number of resolved
sources, the size of the image, and the value of $f$. The total
number of iterations is chosen so that the brightest pixels left
in the image are approximately at the 3 $\sigma$ noise level. We
save the model after every $\sim10^3$ iterations, so that we have
a series of $\sim10$ models at various depths. These can be
examined afterwards to determine how the model-subtracted image
varies as a function of model depth, and determine the optimal
model depth.

There are several important details to be noted in the modeling
procedure. First is that the noise level is not completely uniform
across the original image. Therefore, we actually model an image
that is weighted by the exposure depth, which produces an image
with flat noise properties. This is equivalent to searching for
the most significant, rather than the brightest, pixel in the
image at each iteration. The model thus produced needs to be
deweighted before subtraction from the data. The second important
detail is that the choice of PRF can be important. If the model
PRF is sharper (narrower) than the actual PRF, then a point source
will behave as an extended source, requiring subtraction of
multiple components at slightly different positions, and thus the
overall number of iterations would need to be increased. Despite
being slower, a good result should still be attained. If the model
PRF is too wide, however, the emission from point sources will be
oversubtracted immediately around the source. This error is not
recoverable, as we only are fitting components to the most
significant positive pixels. Because we are interested in faint
background fluctuations, it is also important that our modeling
procedure uses PRFs that map the IRAC beam out to large angular
distances, i.e. that it include the extended wings of the PRF. If
the wings are not included in the model PRF, then the actual wings
of sources in our image will not be modeled and removed, and may
provide an undesired contribution to the power spectrum. 
The PRF used in these studies was described in the in the Supplementary 
Information of KAMM1. It consists of the core PSF (measured out to a 
radius of $12''$) which was made
available by the SSC following the in-orbit checkout (IOC), combined
with the broad wings (measured out to $\sim150''$) that were 
observed in long frame time observations of Fomalhaut (AORID = 6066432). 
Tests on the sensitivity of the power spectra to 
details of the adopted PRF are presented in \S4.2.
The third detail of the procedure is that it is important to set the correct
background level in the original image. If the background level is
set too low, then the apparent brightness of the sources will be
set too high and sources may be oversubtracted, unless the
parameter $f$ is set to a relatively small value. If the
background is set too high, then sources will tend to be
undersubtracted. In this case, combined with relatively small
values of $f$, the model-subtracted data will appear to have had
sources removed by truncation at a particular brightness level. 

\subsubsection{Source Masking}
The source masking of KAMM1 was calculated iteratively from the original image.
The mask is initially defined as all pixels with intensity more than $N_{clip}\sigma$ above the
mean intensity, and all pixels surrounding these within a square $N_{mask}\times N_{mask}$ window. (The primary results of KAMM1 used $N_{mask} = 3$.)
The process is then repeated with $\sigma$ being replaced by $\sigma_{unmasked}$
(derived only from the unmasked data),
and newly identified pixels being added to the mask. After several iterations the
procedure will converge and no unmasked pixels with intensities $> N_{clip}\sigma_{unmasked}$
remain. The final result is very similar to masking the image at a fixed surface brightness
threshold, and then expanding (dilating or growing) the mask to include the
$N_{mask}$ neighboring pixels.

An additional detail of the masking procedure is that we also construct masks from
the models (described above), and then apply the union of both masks to the data analysis.
This is done primarily to eliminate artifacts from ghost images. Ghost images are 
generally weak in our images because of dithering combined with the fact that 
the position of the ghost image will shift as a function of the location of the source 
on the detector array. Using the model to help insure they are 
masked increases the masked area by $\sim2\%$.

Throughout the study we compute the power spectrum from Fourier
transforms for fields in which $\sim 20-25\%$ of the pixels
are masked and set to 0.0.

\subsection{Power Spectra and Fluctuation Spectra}
As presented here, the two dimensional power $P(u,v)$ of the model-subtracted and masked
background intensity $\delta F(x,y)$ is simply derived
from the discrete fast Fourier transform (FFT) of the image
\begin{equation}
P(u,v) = |FFT[\delta F(x,y)]|^2 =
\left|\frac{1}{M}\frac{1}{N}\sum_{x=0}^{M-1}\sum_{y=0}^{N-1}\delta
F(x,y)\exp[-2\pi i(ux/M+vy/N)]\right|^2.
\end{equation}
This is reduced to a power spectrum,
$P(q)$, where $q = 2\pi[(u/N)^2+(v/M)^2]^{0.5}/\theta_{pixel}$,
by averaging $P(u,v)$ in bins with spatial frequencies in the
ranges $[q,q+\delta q]$, where the bin width $\delta q =
2\pi/[\theta_{pixel}\max{(N,M)}]$. For the binned power spectra,
uncertainties estimated for $P(q)$ are calculated as the
standard deviation of the mean for each bin. At the largest
spatial scales, both the power and its estimated uncertainty are
subject to large errors due to the small number (sometimes only 2)
of independent measurements on these scales. We divide our power
spectra by the fraction of pixels in the image that have not been
masked (masked areas are set to zero intensity). Masking in the image 
domain corresponds to a convolution in the Fourier transformed domain.
If the power spectrum is a flat function of frequency, 
then its convolution with the FFT of the mask will also be flat
and unchanged (after rescaling for the fractional area masked).
For power spectra that are strongly peaked at low frequencies (large spatial scales),
the convolution produced by the masking shifts some of the power to higher
frequencies, leading to underestimates of the large scale power. 
In \S4.2 we show that this shift in power does not qualitatively affect 
our results, with the largest change being a reduction in power at the largest spatial scales 
by approximately the same fraction as the masked area of the image.
Our power spectra are plotted as a function of $2\pi/q$ which 
is the spatial wavelength.

\begin{figure}[ht] 
\includegraphics[angle=90, width=6.5in]{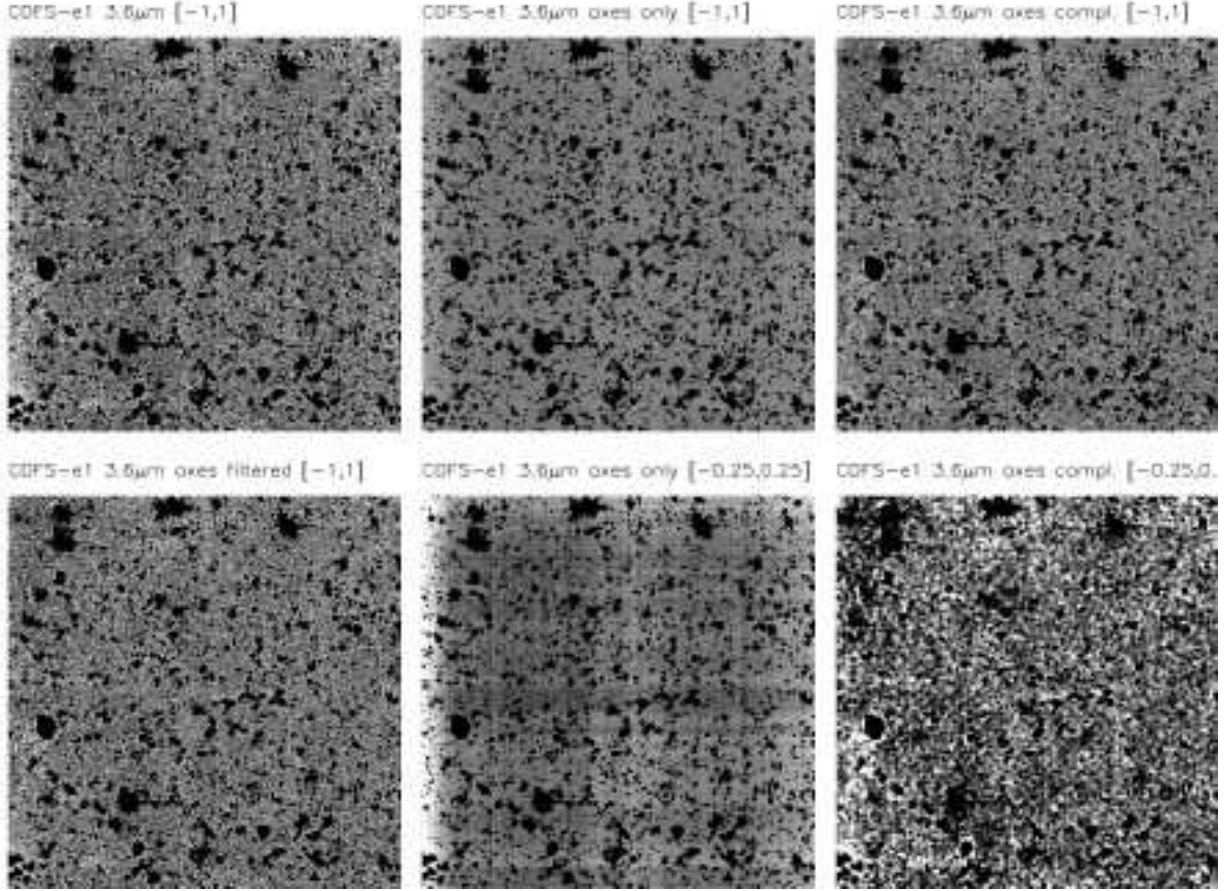}
   \caption{Illustration of systematic effects induced by artificial power on the axes
   of the Fourier transformed images. (top left) The blanked and model-subtracted
   CDFS epoch 1 field at 3.6 $\micron$, scaled from [-1,1] nW m$^{-1}$ sr$^{-1}$.
   (bottom left) The same field after subtraction of power along the FFT axes and at scales $>9.6''$.
   (top center) Image of the FFT power at scales $>9.6''$ and on the axes of the FFT (i.e. the difference
   of the two images at left).
   (bottom center) Same image but on a narrower display range to better show structure related to the
   coverage of the observations.
   (top right) Image of the FFT power at scales $>9.6''$, but {\it excluding} the axes of the FFT
   (i.e. this is the ``complement'' of the figure at top center).
   (bottom right) Same image as above, but on a narrower display range to show that the off axis power
   has little or no resemblance to the coverage or known detector artifacts.}
   \label{fig:zero_axes}
\end{figure}

Because of the constrained detector orientation (position angle) during any given set
of {\it Spitzer} observations, the self--calibration procedure cannot distinguish between
strictly linear gradients in the sky and correlated gradients in the detector offsets $F^p$
and $F^q$. Such gradients can be caused by calibration errors, zodiacal light, or the
true astronomical background. In any case, to ensure that they have no effect, we omit
the data along the $x$ and $y$ axes of the Fourier transformed image when calculating
$P(q)$. This is done by simply omitting measurements $P(u,v)$ where $|u| \leq \delta q$ or $|v| \leq \delta q$
when constructing $P(q) = \langle P(u,v) \rangle_{[q,q+\delta q]}$ as described above.
Doing so means that we obtain no result for the largest spatial scale (smallest $q$) 
that could be measured in principle, and the results at
other large spatial scales are made slightly more uncertain. Omitting the power measured
along the axis also eliminates the effects of many systematic errors, which tend to
be aligned with the detector and thus are preferentially found along the axes of the
2-dimensional Fourier transform of the image. This is illustrated in Figure \ref{fig:zero_axes}. 
Unless otherwise noted, all power spectra shown in this paper do
not include power on the axes.

For comparison with the original images and the brightnesses of the resolved sources
in the images, it is sometimes convenient to display the fluctuation spectrum,
which is defined as $[q^2P(q)/(2\pi)]^{0.5}$.

\section{Test Cases and Results}

In this section we present several tests aimed at identifying
possible problems in the analysis of the clipped and
model-subtracted background fluctuation. Table \ref{tab:checks} is
provided as a summary of the tests presented below and in our
prior reports.

\subsection{Self-calibration vs. GOODS processing}
\begin{figure}[ht] 
   \plotone{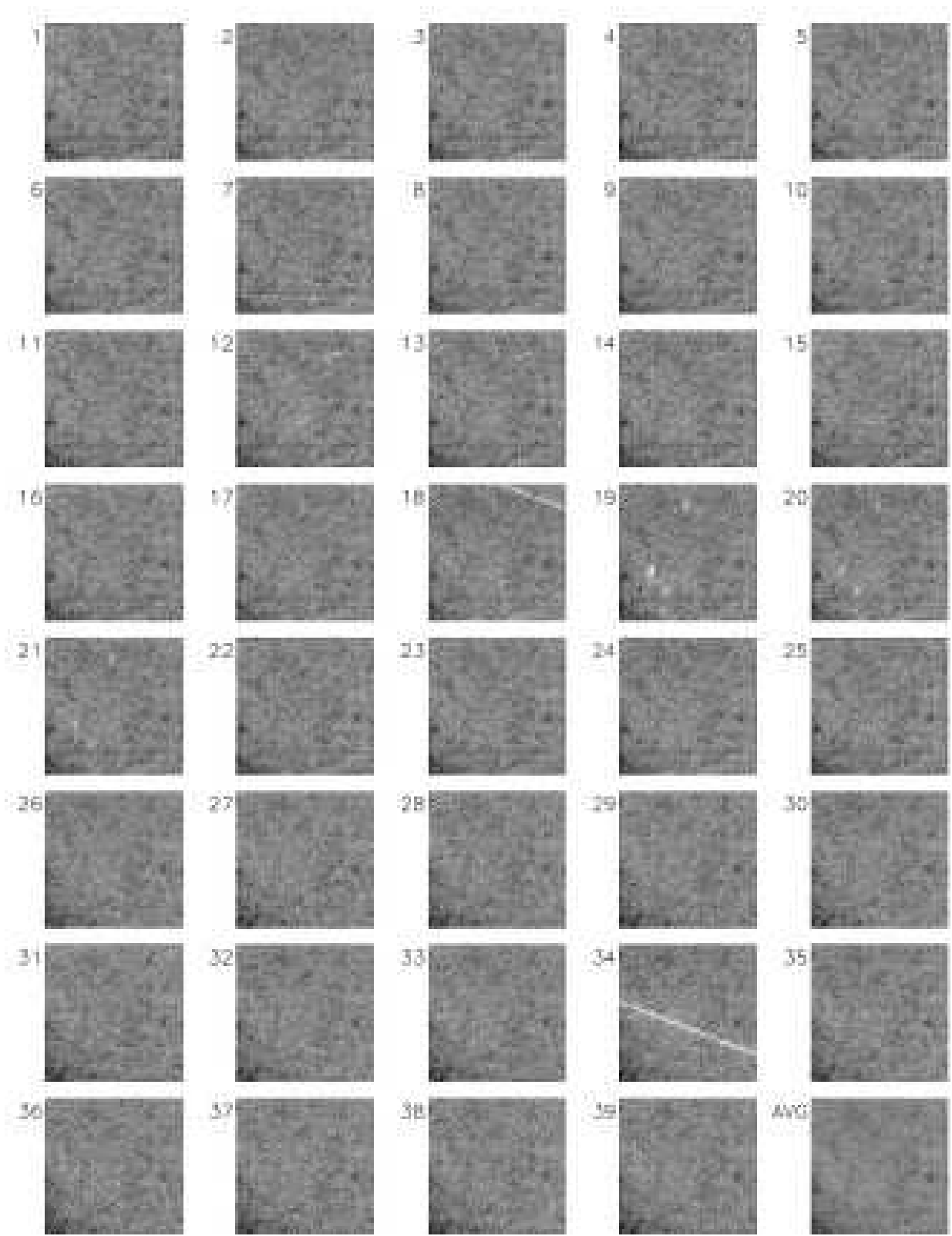}
   \caption{Array offset terms $F^p$ at 3.6 $\micron$ derived for each AOR of CDFS-e1 from self-calibration.
   Note that some AORs (e.g. 19 -- 21) contain latent images from intermixed observations of bright sources, or tracks from slewing across bright sources (e.g. 18, 34).
   Images are shown on a linear (black to white) stretch from [-0.01,0.01] MJy sr$^{-1}$
   (or equivalently [-8.33,8.33] nW m$^{-2}$ sr$^{-1}$).}
   \label{fig:offsets}
\end{figure}

An assessment of the usefulness of the self-calibration can be made by examining images
of the derived array offsets $F^p$ and the temporal trends of the variable offsets $F^q$.
Figure \ref{fig:offsets} shows the array offsets $F^p$ derived at 3.6 $\micron$ for each of the AORs
of the CDFS--e1 observations. These offsets display a relatively constant pattern of dark
features. These represent long--term changes in the detector response compared to
the standard gains and offsets applied by the BCD pipeline. There are both spotty
features that likely represent long term latent images from previous observations,
and a horizontal linear feature (about 64 pixels from the bottom of the detector array) that is
more directly related to the hardware.
Additionally, there are short--term detector changes that appear as white spots and
lines. These are caused by staring at or slewing over bright sources in the time
preceding the AOR where they appear. These features decay relatively quickly, affecting
no more than 3 consecutive AORs. The detectors at 4.5 and 5.8 $\micron$ are not
strongly affected by latent images, but at all wavelengths the self--calibration
does find array offsets with fairly fixed patterns along with variable features that
can change from one AOR to another. Thus with the self--calibration,
we remove these artifacts as appropriate for each AOR.

\begin{figure}[ht] 
   \plotone{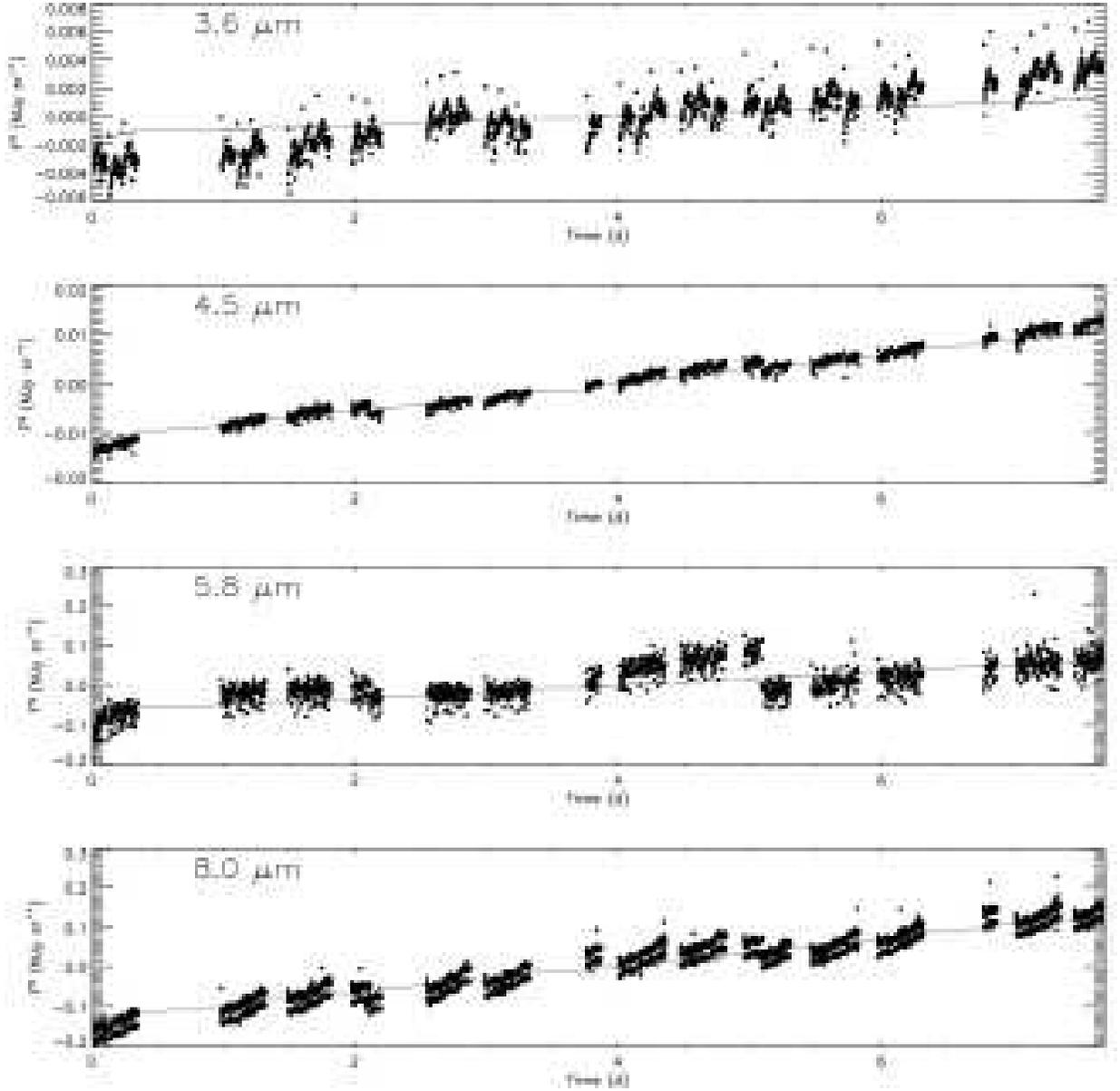}
   \caption{Variable offset terms $F^q$ (dots) per frame derived by the self-calibration 
   are plotted as a function of time since the start of the CDFS-e1 observations.
   The general trend of the changes in $F^q$ correlates well with that expected for 
   the zodiacal light as estimated by the ZODY$\_$EST keyword
   values from the BCD headers (shown as solid lines). Smaller variations and jumps 
   with respect to this trend are due to instrumental changes. The infrequent and nearly 
   periodic outliers are the result of changes in IRAC's dark frame (the ``first frame effect'').}
   \label{fig:quads}
\end{figure}

The self--calibration also derives variable offset terms $F^q$. Example of these
offsets at each wavelength are shown for the CDFS-e1 observations in Figure \ref{fig:quads}. 
In this figure, each of the four $F^q$ values per frame is plotted (dots) as a function of 
time (in days) since the start of the observations of this field. The self-calibration assumes
that the sky is a stable calibration source. However, since the zodiacal light intensity does 
change on a time scale of days, this variation is absorbed by the $F^q$ term in the self-calibration.
Thus, the figure shows a steady drift in $F^q$ which is well--correlated with the change
in the zodiacal light intensity (solid line). The infrequent but nearly periodic outliers are evidence of
incomplete correction of the offsets in the initial frame of each observing sequence 
(the ``first frame effect''). Other smaller scale ($\lesssim 1$d) drifts and jumps 
with resect to the zodiacal light trend are likely caused by instrumental changes, 
because their strength does not vary with wavelength as would be expected if they 
were caused by short-term variations in the zodiacal light. The clustering of points in time 
simply reflects the scheduling of the observations. 

\begin{figure}[ht] 
   \plotone{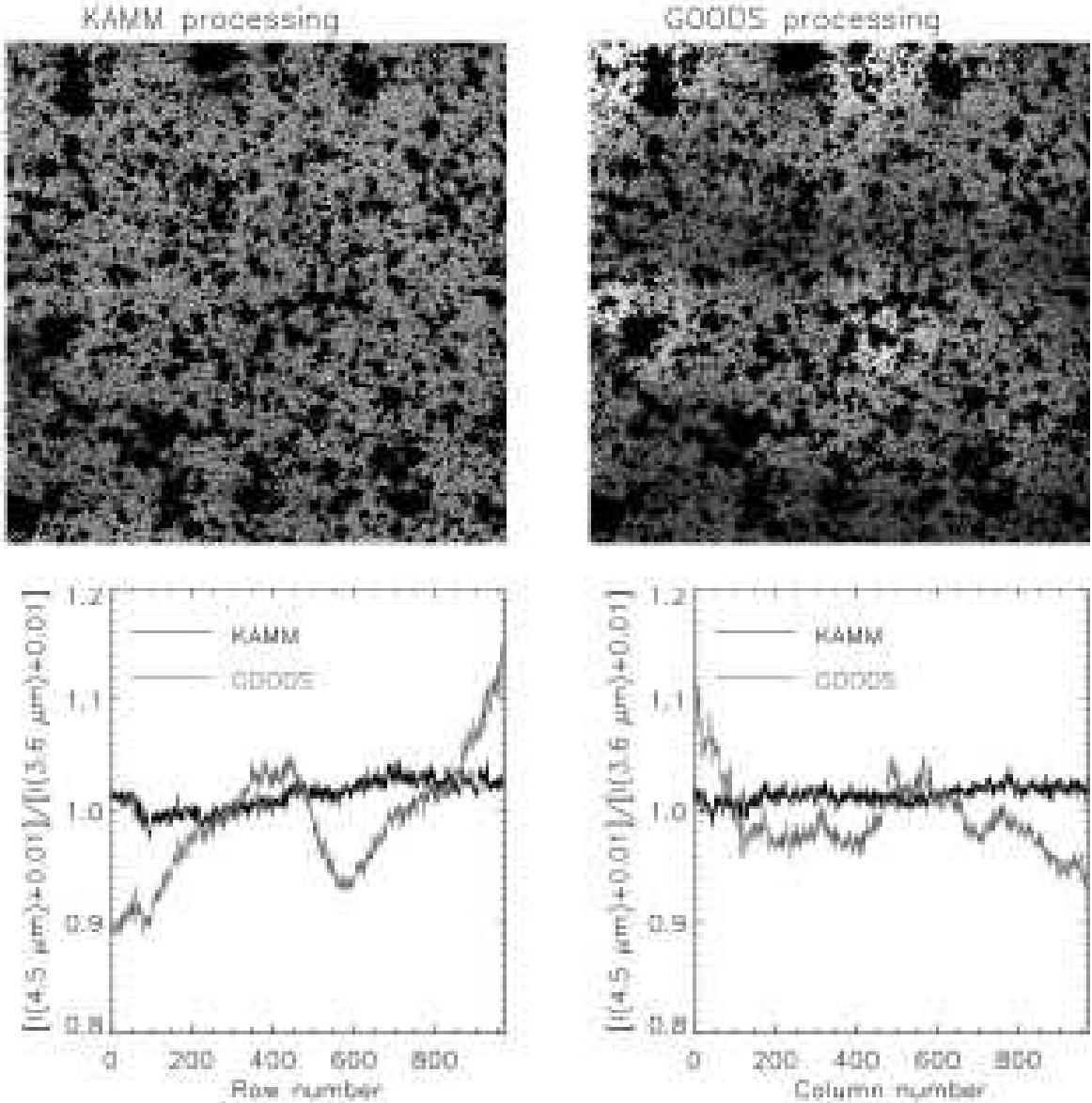}
   \caption{Ratio of 4.5 $\micron$ / 3.6 $\micron$ CDFS-e1
   images for the KAMM processing and the
   GOODS processing (v 0.30). Bright sources in the images have been masked
   identically, but no model has been subtracted from either. One or both
   channels of the GOODS data contain a
large scale artifact that reveals the 2x2 mosaicked coverage of
the field. The lower panels compare median intensities across each
ratio image as a function of row and column. Small offsets are
added to the ratios so that the ratios are always positive with a
mean near 1. The images on the right are clearly problematic in
uncovering faint diffuse signal but were used in CIB analysis of
Cooray et al. (2007). The pattern seen in the GOODS processing is 
related to the calibration of the detector offsets, as shown in Fig. 
\ref{fig:dither_offsets} and discussed in \S4.3.}
   \label{fig:compare12}
\end{figure}

\begin{figure}[ht] 
   \plotone{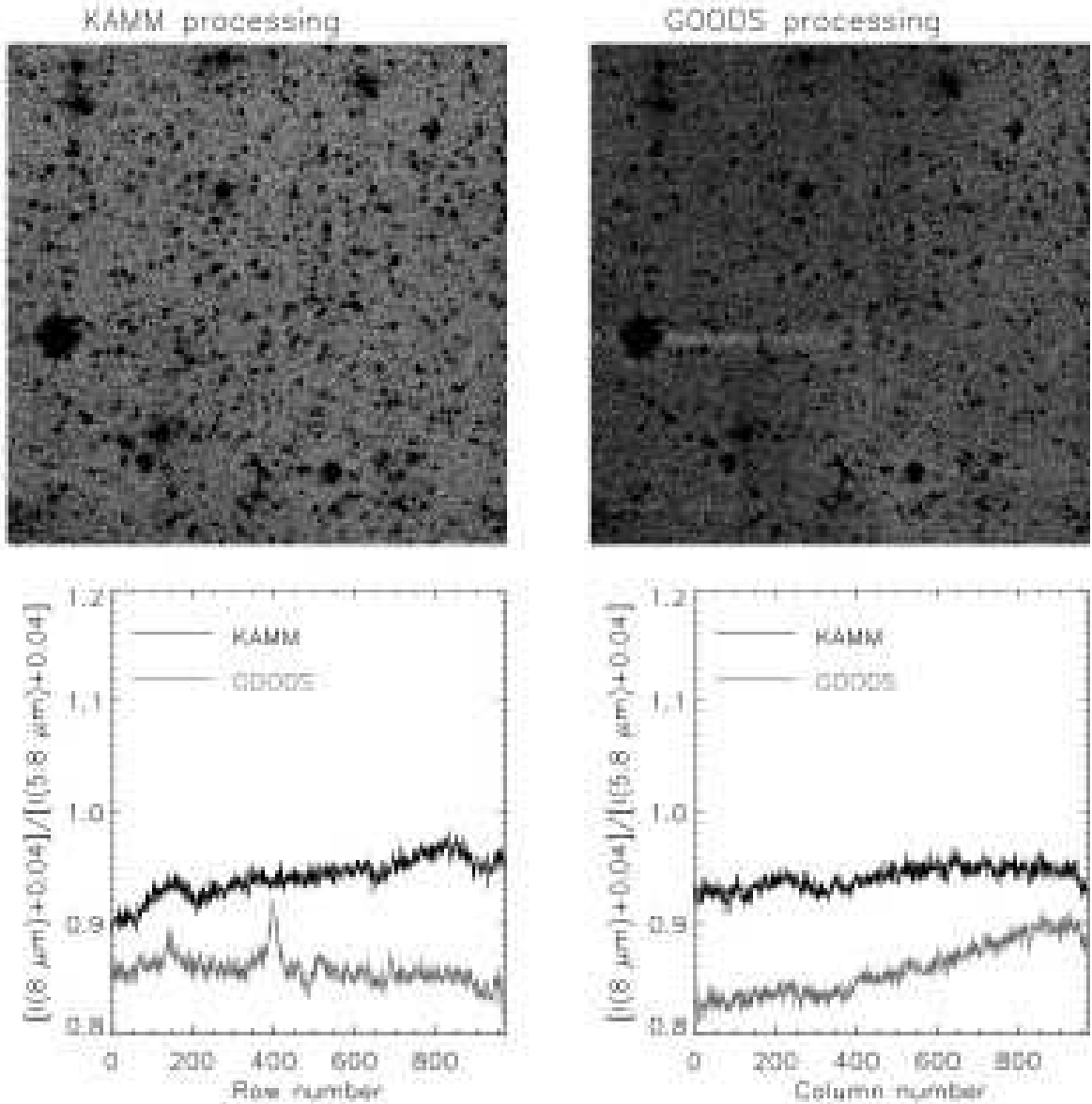}
   \caption{Ratio of 8.0 $\micron$ / 5.8 $\micron$ CDFS-e1
   images for the KAMM processing and the
   GOODS processing (v 0.30). Bright sources in the images have been masked identically, but no model has been subtracted from either. The lower panels compare median intensities across each ratio image as a function of row and column. Small offsets are added to the ratios so that the ratios are always positive.}
   \label{fig:compare34}
\end{figure}

The differences made by application of the self-calibration to the data are often small
compared to the brightness of typical sources in the images. Therefore, it is difficult to
see the effect of self-calibration on a full-intensity image. However, the effects
become very evident when examining certain processed results in which the appearance
of the point sources is minimized. Ratio images between our self--calibrated mosaics,
and those prepared by the GOODS team show evident differences, but in such images it
can be difficult to determine which of the original images is causing which artifacts
in the ratio. A more decisive comparison can be made by examining the ratio of
mosaics at two wavelengths for the self-calibration, and the corresponding ratio for the
GOODS pipeline processed data. Artifacts in these ratio maps are definitely the
fault of the corresponding data reduction, although here it may be ambiguous as to
which wavelength (if not both) contains the flaws. In constructing these ratios,
we add small offsets ($\sim3$ times
the noise level) to the data such that the ratios are always positive and the noise
does not dominate the appearance. Figure \ref{fig:compare12} shows the ratio images
of 4.5 / 3.6 $\micron$ mosaics for our self-calibrated data, and for the GOODS
processed data (v0.3). Also shown are the median intensities of each ratio, taken across
rows and columns. Masking is applied to the bright sources to eliminate the distraction
of the intrinsic color variations of some of these sources. The comparison shows that
while neither ratio is perfectly flat, the self-calibrated data show significantly
less large--scale structure.
Here it is very clear that the GOODS result contains an artifact related to the
coverage of the field, whereas the self-calibrated result is much flatter. This difference
can be traced to a gradient in the detector offset that we identify and remove
through the self-calibration process (see \S 4.3).
The corresponding ratio images for 8 / 5.8 $\micron$ are shown in Figure \ref{fig:compare34}.

\subsection{Modeling and clipping}
\begin{figure}[ht] 
   \plotone{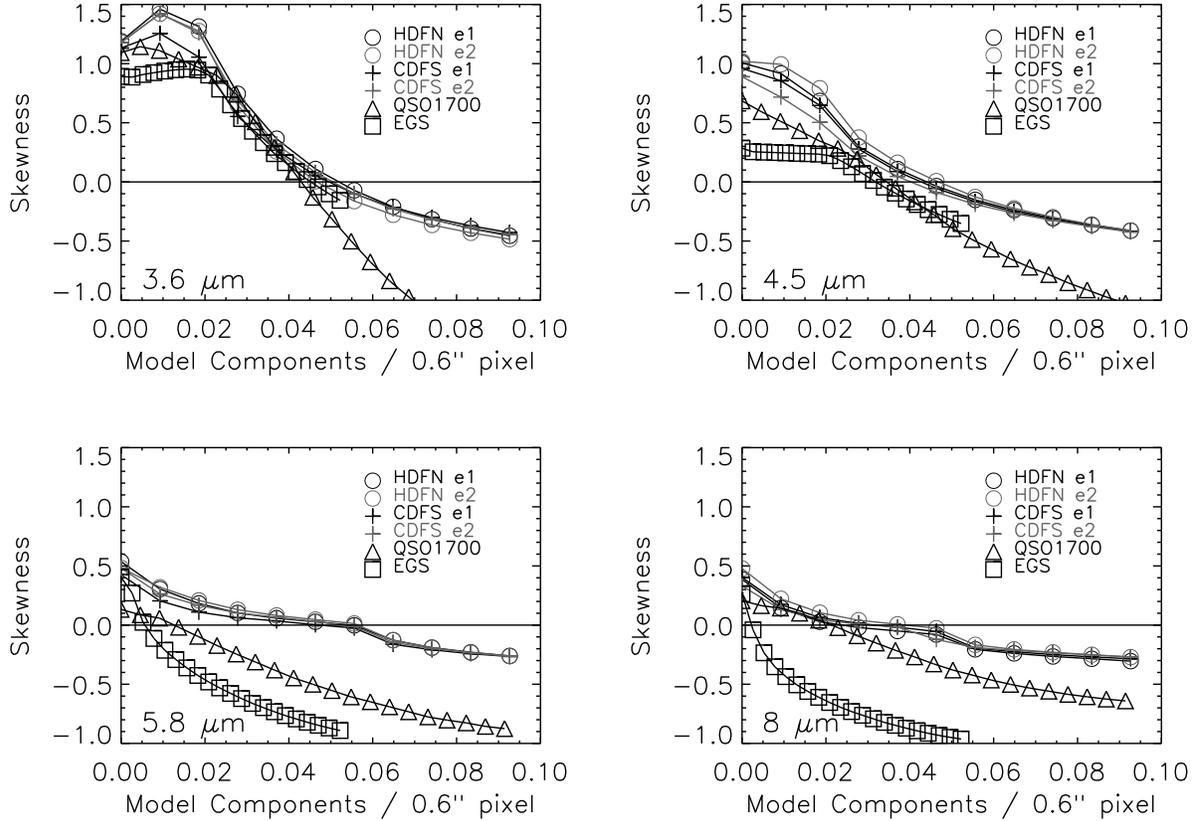}
   \caption{The skewness of the distribution of pixel intensities for the model-subtracted
   fields, as a function of the density of model components (the number of components 
   subtracted by the model divided by the area of the field). The symbols 
   denote intervals of 3000 model components for the QSO 1700 and EGS fields, 
   and 120000 components for the other fields. Models that yield negative skewness 
   are likely too deep, and are increasingly attacking random noise
   rather than actual sources.}
   \label{fig:skewness}
\end{figure}

An important aspect of the source modeling procedure (\S 3.1.1) is
the determination of the optimal depth of the model. For the
results presented here, we have chosen the optimal depth to be
that where the residual intensities (after clipping and
subtraction of the model) exhibit zero skewness [i.e. the
normalized third moment of the distribution: $\langle(x -
\bar{x})^3\rangle/\sigma^3$]. This is because any true sky sources
contribute to the positive tail of the distribution, whereas the
noise is expected to have zero skewness. We note, however, that
the final results are very similar if we use our prior criteria,
such as either the iteration where negligible correlation with
removed emissions is reached, or when the shot noise from the
remaining sources is sufficiently larger than the $A-B$ estimate
of the instrument noise so that no significant amount of the
instrument noise is removed in the modeling. Figure
\ref{fig:skewness} shows the change in skewness of the pixel
intensities for our modeled fields as function of the mean density
of components. 
The density of components is the number of components 
subtracted by the model, divided by the area of the field. 
This quantity is related to the density of sources in the field, 
however because multiple components are required to model
each source (\S 3.1.1), the actual density of sources in the images is
several times lower. If the modeling is not sufficiently deep,
remaining point sources leave a positive skew in the distribution
of the residual intensities. If the model is too deep, the highest
noise peaks begin to get subtracted, and the residual intensities
develop a negative skew.

\begin{figure}[ht] 
   \plotone{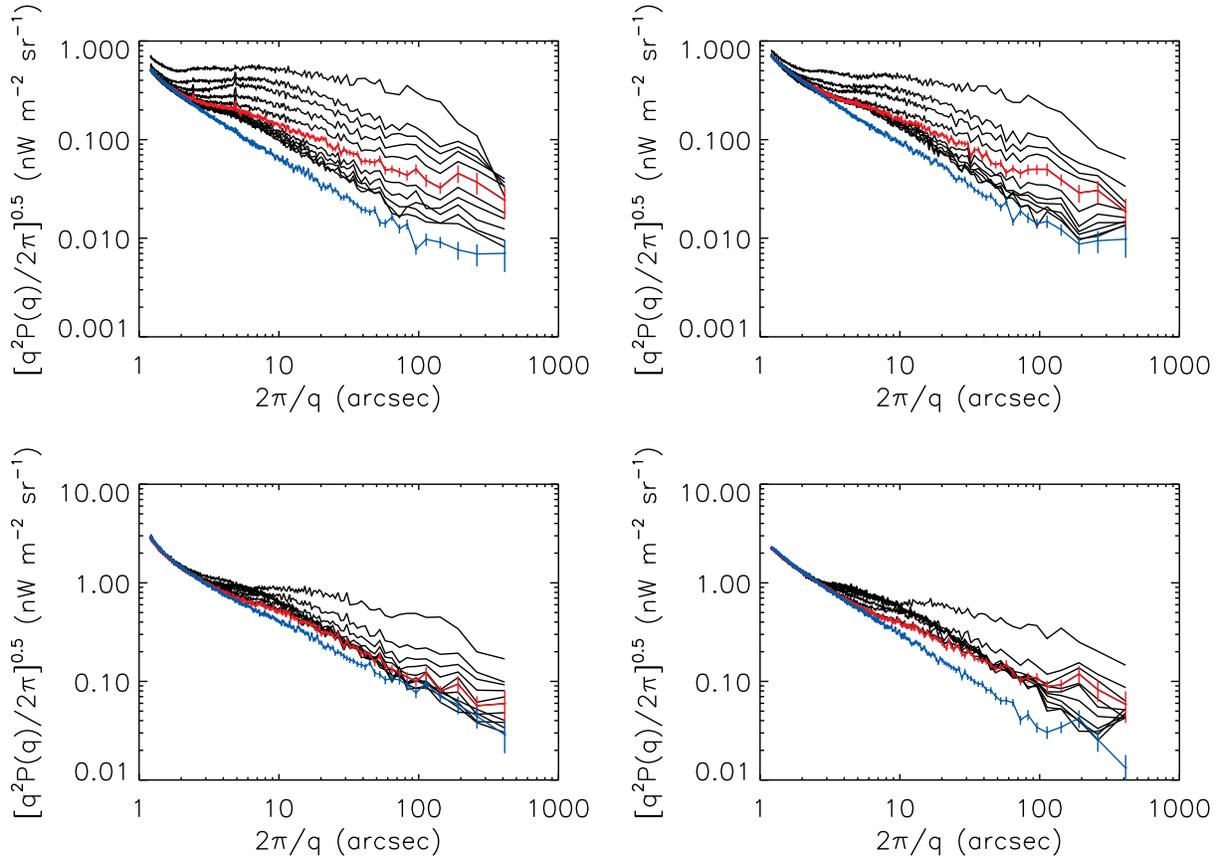}
   \caption{Fluctuation spectra as a function of model depth for the CDFS epoch 1 field.
   The red line indicates the optimal (zero skewness) model. The blue line indicates
   the (A-B)/2 noise fluctuations. The relative uncertainties at each model depth are similar to
   those that are depicted for the optimal model. Left to right and top to bottom are results
   for 3.6, 4.5, 5.8, and 8 $\micron$ respectively.}
   \label{fig:figure5_CDFS_ep1}
\end{figure}

\begin{figure}[ht] 
   \plotone{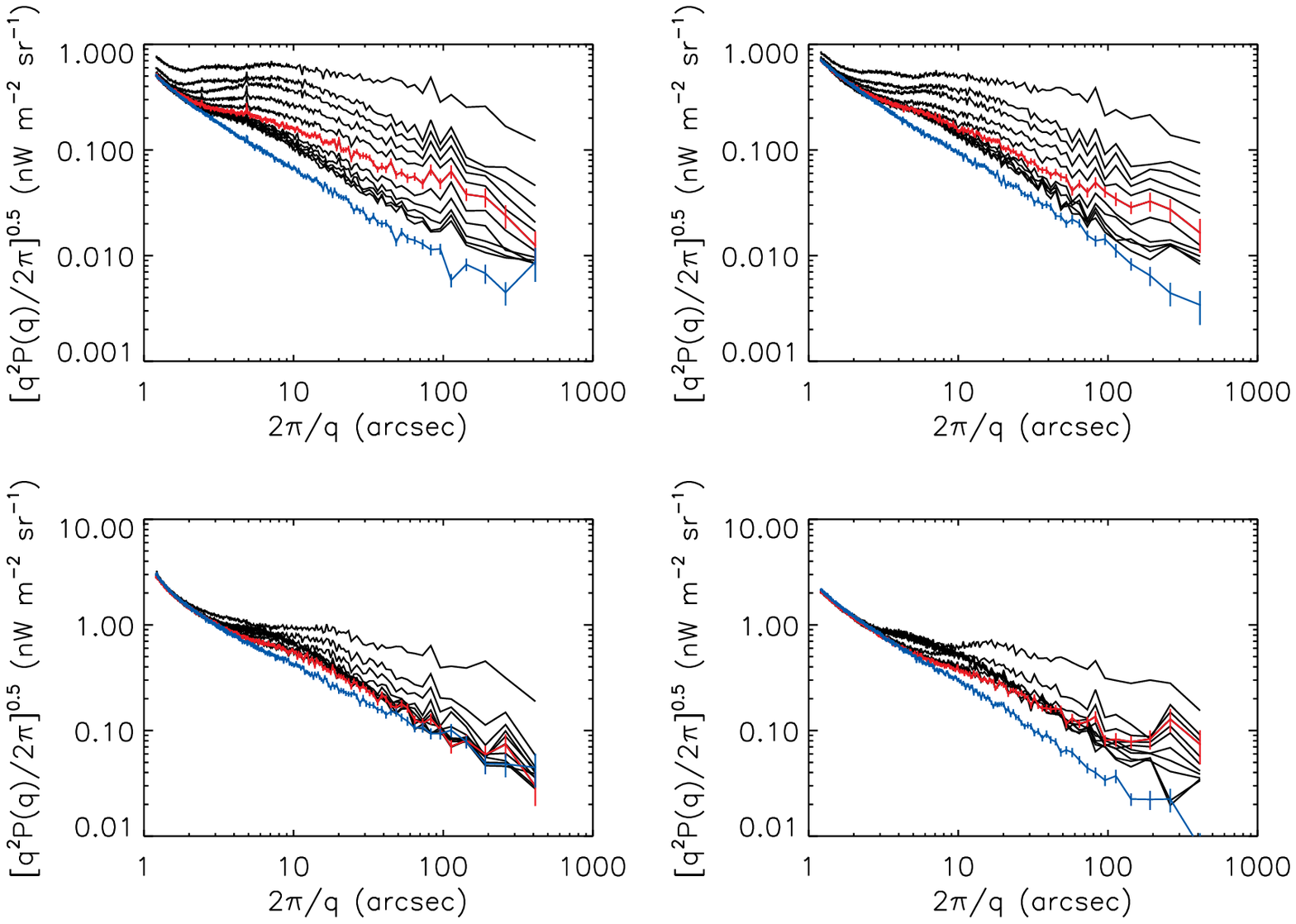}
   \caption{Same as Fig. \ref{fig:figure5_CDFS_ep1}, except for the CDFS epoch 2 field.}
   \label{fig:figure5_CDFS_ep2}
\end{figure}

\begin{figure}[ht] 
   \plotone{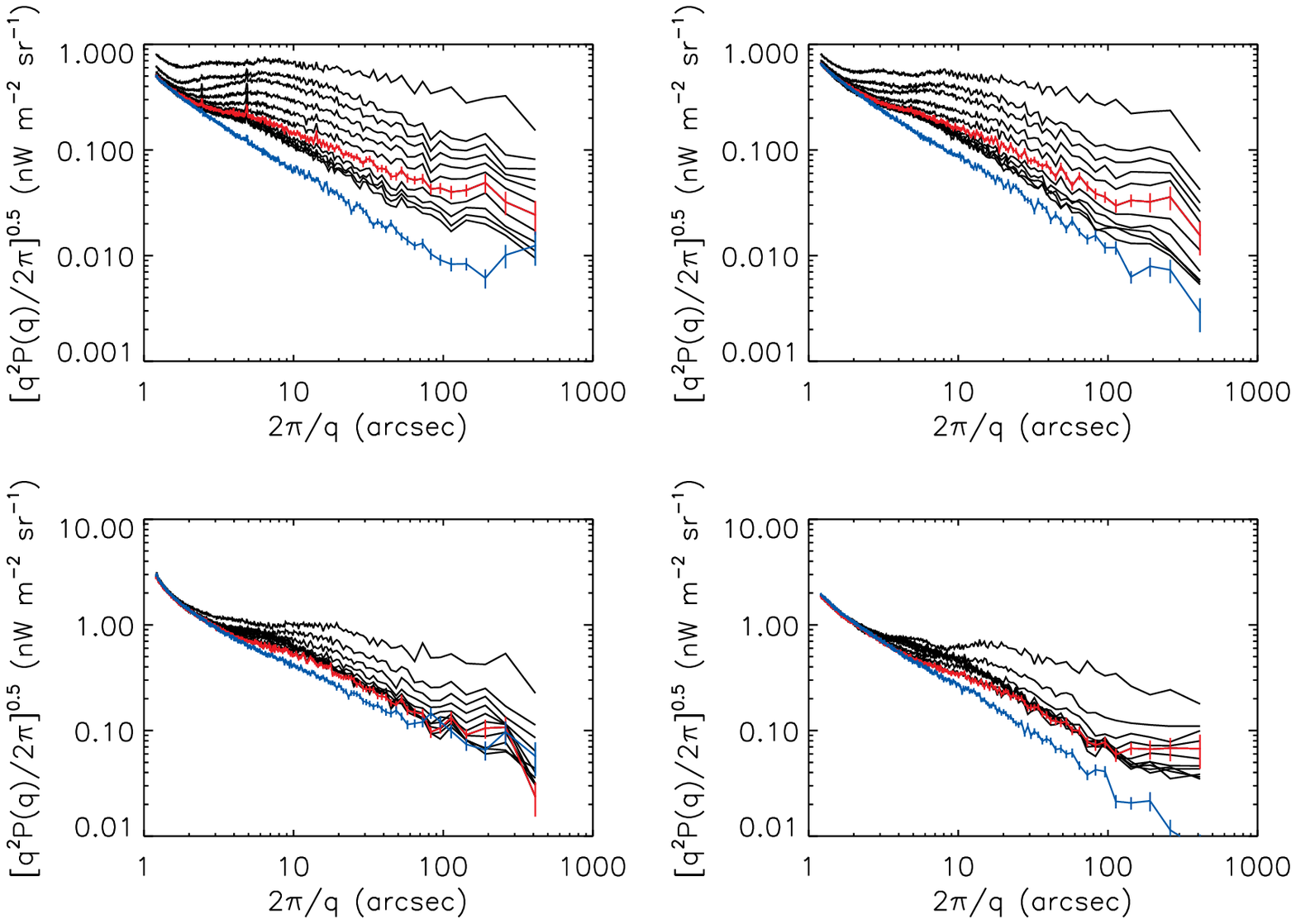}
   \caption{Same as Fig. \ref{fig:figure5_CDFS_ep1}, except for the HDFN epoch 1 field.}
   \label{fig:figure5_HDFN_ep1}
\end{figure}

\begin{figure}[ht] 
   \plotone{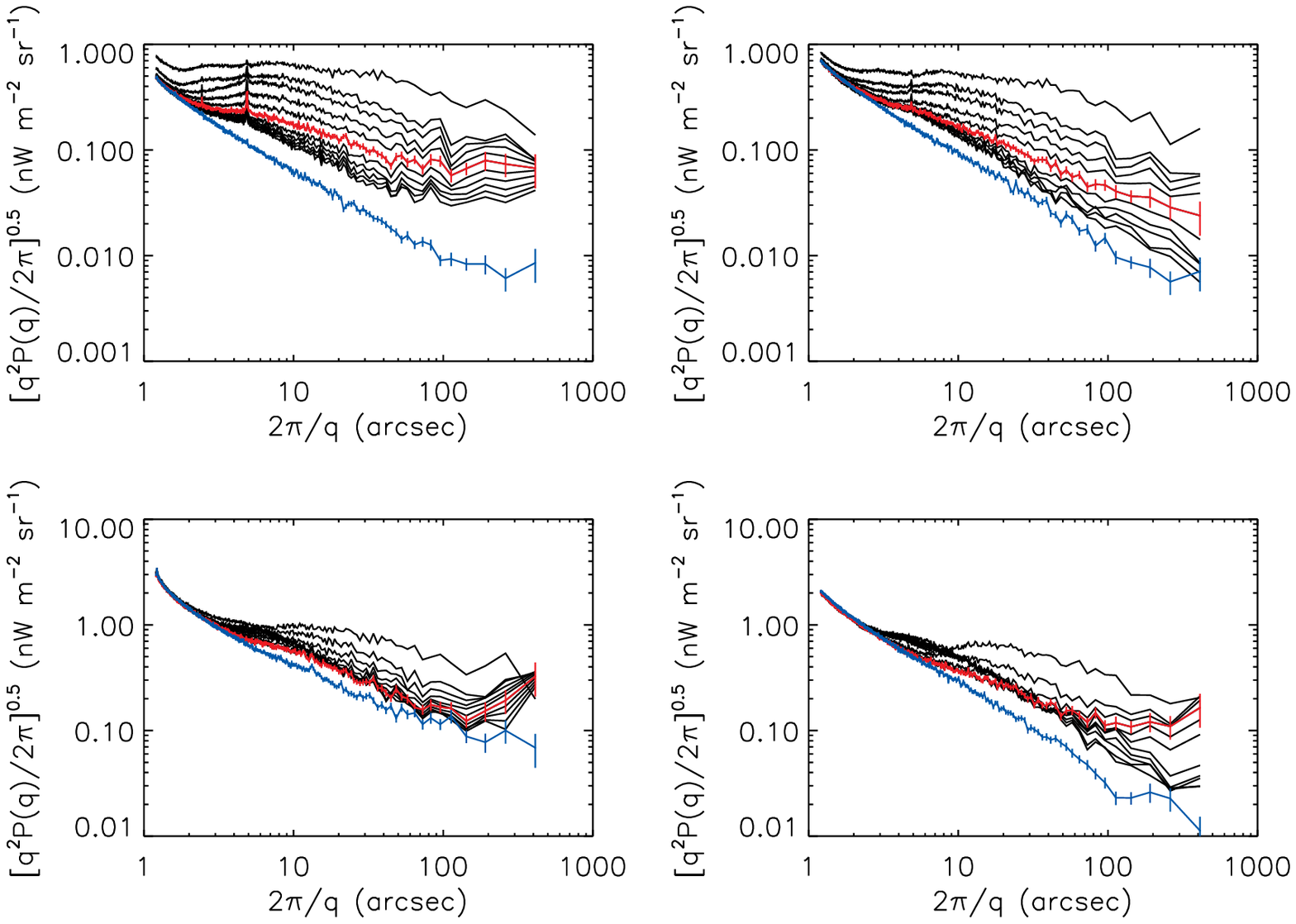}
   \caption{Same as Fig. \ref{fig:figure5_CDFS_ep1}, except for the HDFN epoch 2 field.}
   \label{fig:figure5_HDFN_ep2}
\end{figure}

\begin{figure}[ht] 
   \plotone{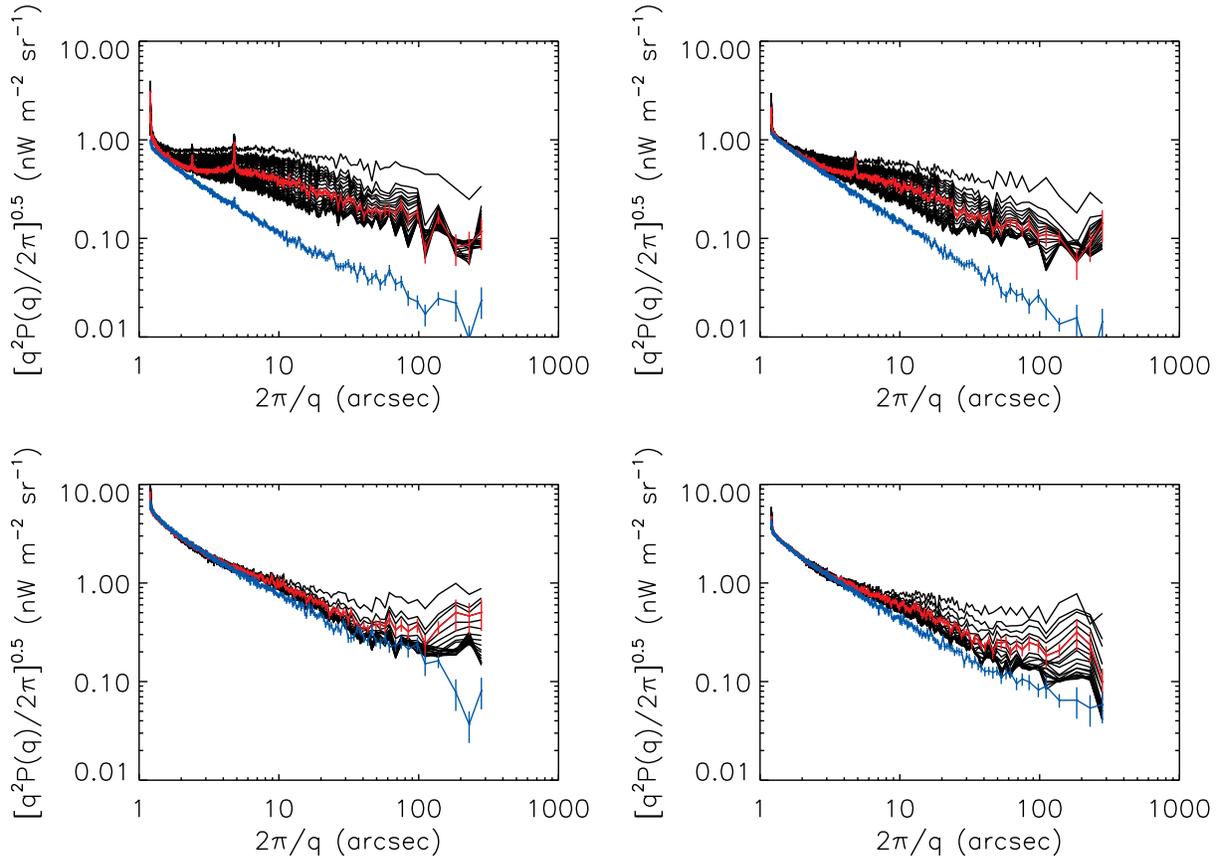}
   \caption{Same as Fig. \ref{fig:figure5_CDFS_ep1}, except for the QSO 1700 field.}
   \label{fig:figure5_QSO1700}
\end{figure}

\begin{figure}[ht] 
   \plotone{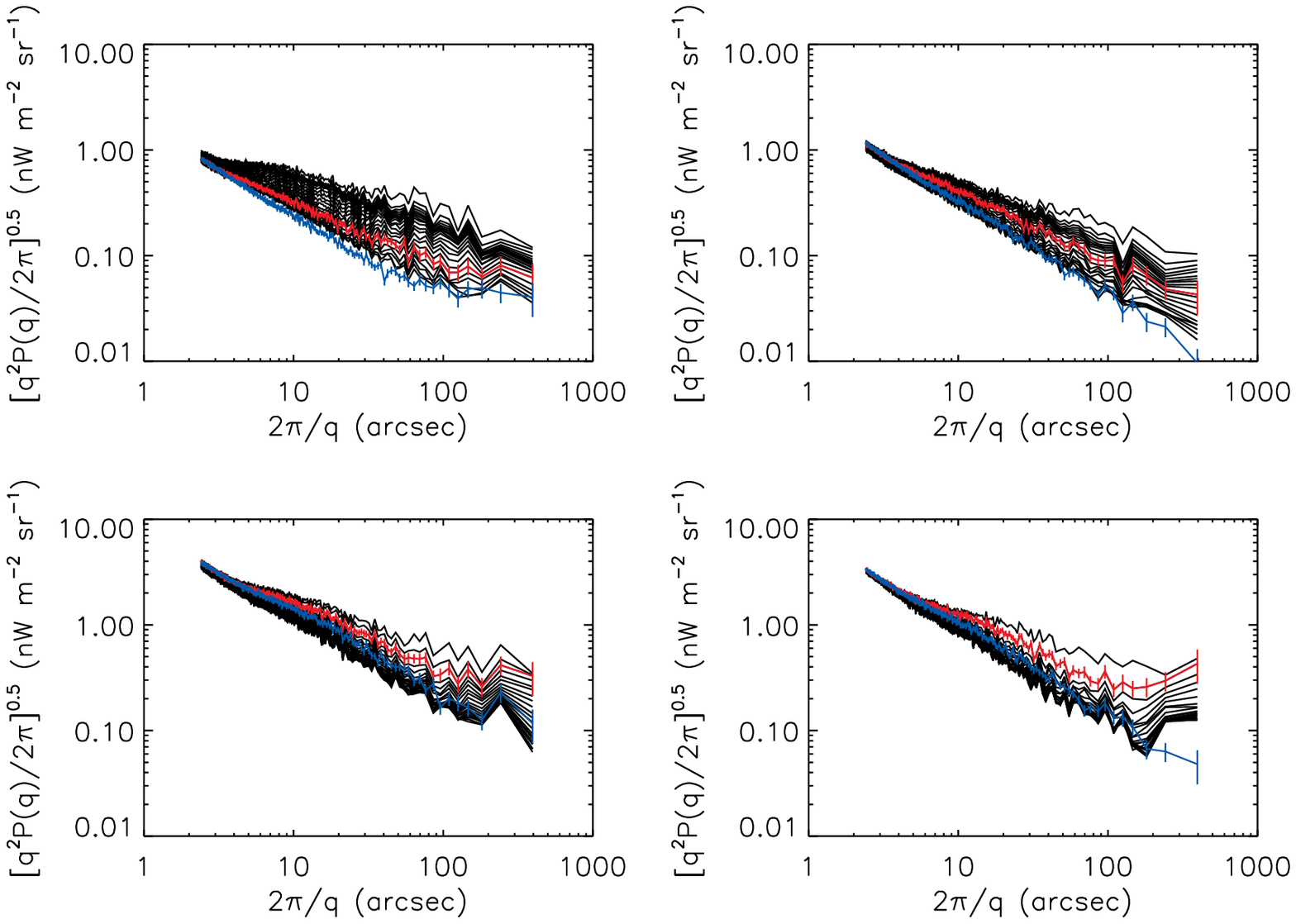}
   \caption{Same as Fig. \ref{fig:figure5_CDFS_ep1}, except for the EGS field.}
   \label{fig:figure5_EGS}
\end{figure}

Figures \ref{fig:figure5_CDFS_ep1}-\ref{fig:figure5_EGS} show the
changes in the fluctuation spectra as a function of model depth
for several fields. At 3.6 and 4.5 $\micron$, there are large
changes in the residual fluctuation as a function of model depth
on medium and large scales. Changes are small at the smallest
angular scales ($\lesssim4''$), which are instrument noise
dominated. At these wavelengths it is relatively difficult for the
models to fit the fluctuations down to the noise level on all
scales. This indicates significant structure in the images, but
does not reveal whether the structure is astronomical, or rather
an instrument or data artifact. At 5.8 and 8 $\micron$ changes in
the fluctuations with model depth are less pronounced. The noisier
fields (EGS and QSO 1700) can be modeled down to the (A-B)/2 noise
level (\S 2.5) and lower. In the GOODS fields, after approximately
zero skewness is reached, power at intermediate scales ($4-10''$)
begins to rise, as the model starts to imprint a (negative) PRF
into a formerly random unstructured noise background.

We tested several other criteria for selection of the optimal
model depth. These included the correlation coefficients between
(a) the model $M_i$ at iteration $i$ and the residual intensity
$\delta F_i$, (b) the change in the models $M_i - M_{i-1}$ and the
residual intensity $\delta F_i$, (c) the model $M_i$ at iteration
$i$ and the original intensity $F$, (d) the change in the models
$M_i - M_{i-1}$ and the original intensity $F$. The former two
correlation coefficients generally change from positive to
negative at approximately the same depth as the
skewness. However, these correlations are more strongly affected
than the skewness by the initial zero level used in the modeling
procedure. The latter two criteria are less suitable as they tend
to asymptotically approach 1 and 0 respectively. We emphasize
again that for the final iterations there are negligible
correlations between the modeled sources and the source-subtracted
maps.

\begin{figure}[ht] 
   \plotone{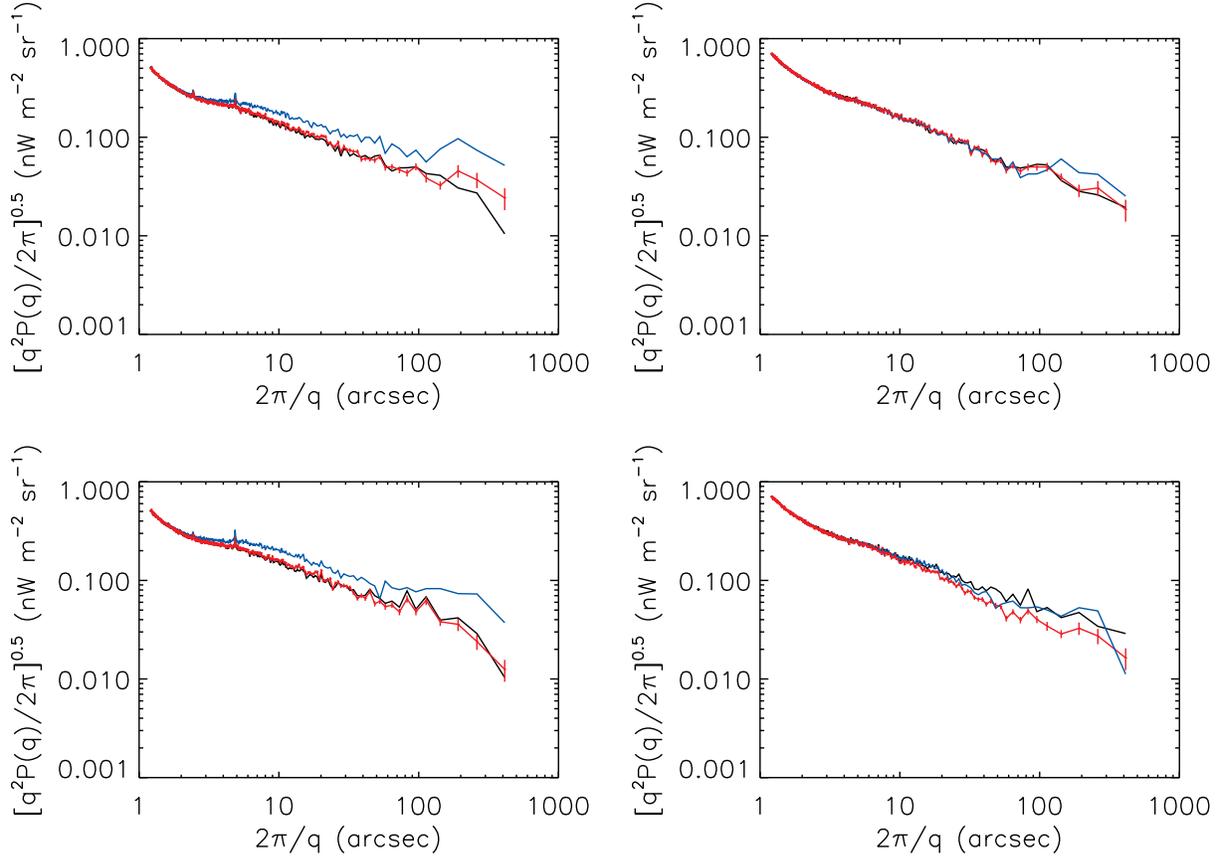}
   \caption{Fluctuation spectra as a function of PRF for the CDFS fields.
   The epoch 1 fields are shown in the top row, while epoch 2 are shown in the bottom.
   The left column shows 3.6 $\micron$ results, with 4.5 $\micron$ results in the right column.
   The red line indicates the optimal (zero skewness) model. The blue line indicates
   the wider PRF ($PRF^{0.95}$). The black line indicates the narrower PRF ($PRF^{1.05}$).
   A wider PRF would have been less effective at modeling and removing resolved sources at 3.6 $\micron$.}
   \label{fig:figure6_PRF}
\end{figure}

Another important aspect of the modeling procedure is that the correct PRFs are used.
If the PRF core or wings are too broad, point sources will be poorly fit, generally
oversubtracted in the outer portions and undersubtracted in the inner portions.
A too narrow PRF core is less of a problem, but will require a larger number of
model components to fit each source. If the PRF wings are too weak compared to the
true PRF, then it will be impossible to remove faint large-scale structure of bright
point sources. To investigate the sensitivity of the residual emission to the PRF
shape, we constructed models for the CDFS fields at 3.6 and 4.5 $\micron$ using
PRFs that are raised to the 0.95 and 1.05 powers to effectively widen and narrow
the PRF respectively. Figure \ref{fig:figure6_PRF} shows that using the narrower PRF has little effect on the
results, but does require additional model iterations ($\sim20\%$ more). 
At 3.6 $\micron$, the models using the
wider PRF require ($\sim20\%$) fewer iterations, but leave an increased level of fluctuations
in the residual intensity. However, at 4.5 $\micron$ the wider PRF produces similar
results to those obtained with the nominal PRF. This
indicates that our nominal 4.5 $\micron$ PRFs are slightly
too narrow and/or have somewhat weaker wings than the true 4.5 $\micron$ PRF.

\begin{figure}[ht] 
   \plotone{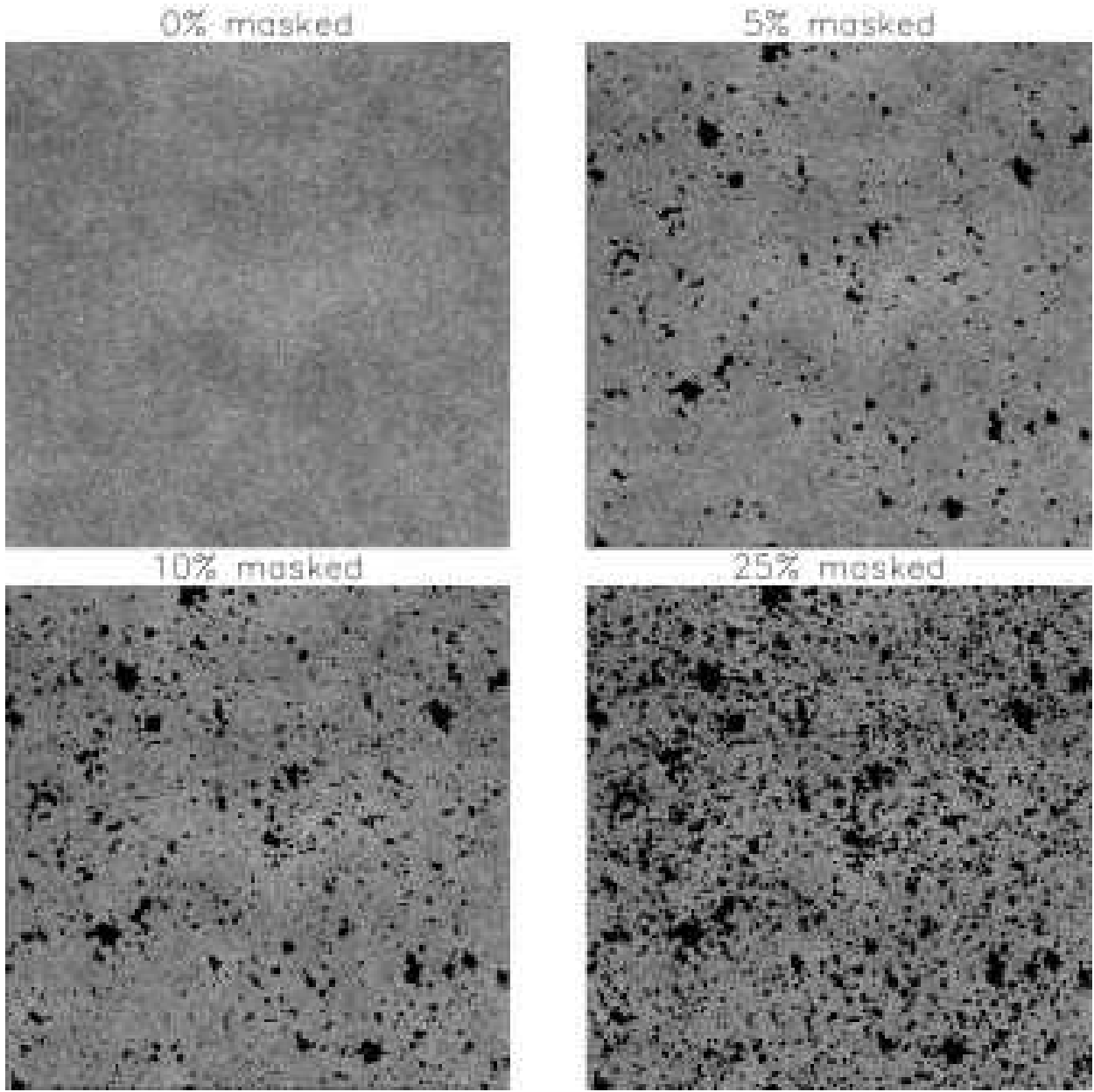}
   \caption{Simulated images containing large scales structure, shot noise, and instrument noise
   (but not with individually resolved sources), shown without any masking and with masking 
   at 5, 10 and 25\%. The 25\% mask is from the actual observations, and the other masks are 
   derived from it.}
   \label{fig:figure_sim_images}
\end{figure}

\begin{figure}[ht] 
   \plotone{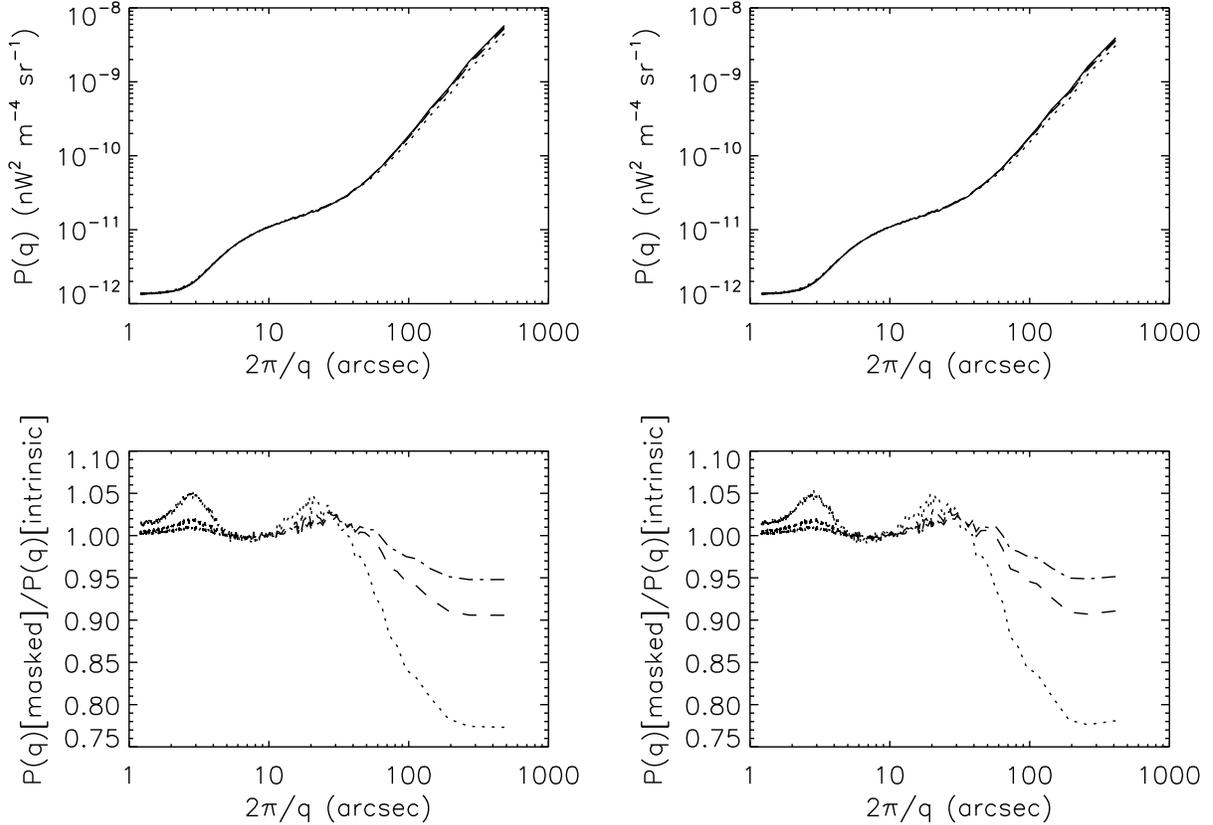}
   \caption{Mean power spectra of 160 simulated images with various amounts of masking 
   (0\% masked = solid lines, 5\% masked = dot-dashed lines, 
   10\% masked = dashed lines, 25\% masked = dotted lines). The power spectra on the left 
   include all power, while those on the right exclude power along the axes of the Fourier 
   transforms. The lower panels reveal finer detail by normalizing the masked power 
   spectra by the unmasked power spectra.
   At much higher levels of masking, a correlation function analysis would be immune from 
   the spurious drop in large scale power that would occur in the power spectrum calculation
   (e.g. Kashlinsky 2007).}
   \label{fig:figure_mask_test2}
\end{figure}

To investigate the effect that the source masking has on the derived power,
we constructed a set of 160 simulations of fields the same size as the HDFN and CDFS.
The simulated fields include: (1) a flat instrument noise component, (2) a shot noise
component (flat at large scale and rolled off at small scale by the PRF) 
representing faint unresolved sources, and (3) a structured background (a power
law at large scales, but also rolled off by the PRF at small scales). 
Individually resolved sources were omitted from the simulation so that the effects
on the structured background would be more clearly displayed. One example of a simulated
image is shown in Figure \ref{fig:figure_sim_images}. The figure also shows the same
simulation after masking roughly 5, 10 and 25\% of the area. The latter mask is taken
from the actual data. The other masks are versions that are processed to reduce to 
masked area. These masks were applied to each of the 160 simulations, and the power
spectra were calculated. The mean of these power spectra are shown in Figure 
\ref{fig:figure_mask_test2}. When the power spectra are plotted on logarithmic axes,
the effects of the masking appear to be very minor. When the power spectra of the masked
simulations are normalized by those of the unmasked simulations, the differences
become more apparent. Increased masking reduces the power at the largest scales
by approximately the same fraction as the masked area. There is a less significant 
increase in the power at small spatial scales. The figure also shows that the masking 
is not the cause of any excess power along the axes of the Fourier transformed images.

\begin{figure}[ht] 
   \plotone{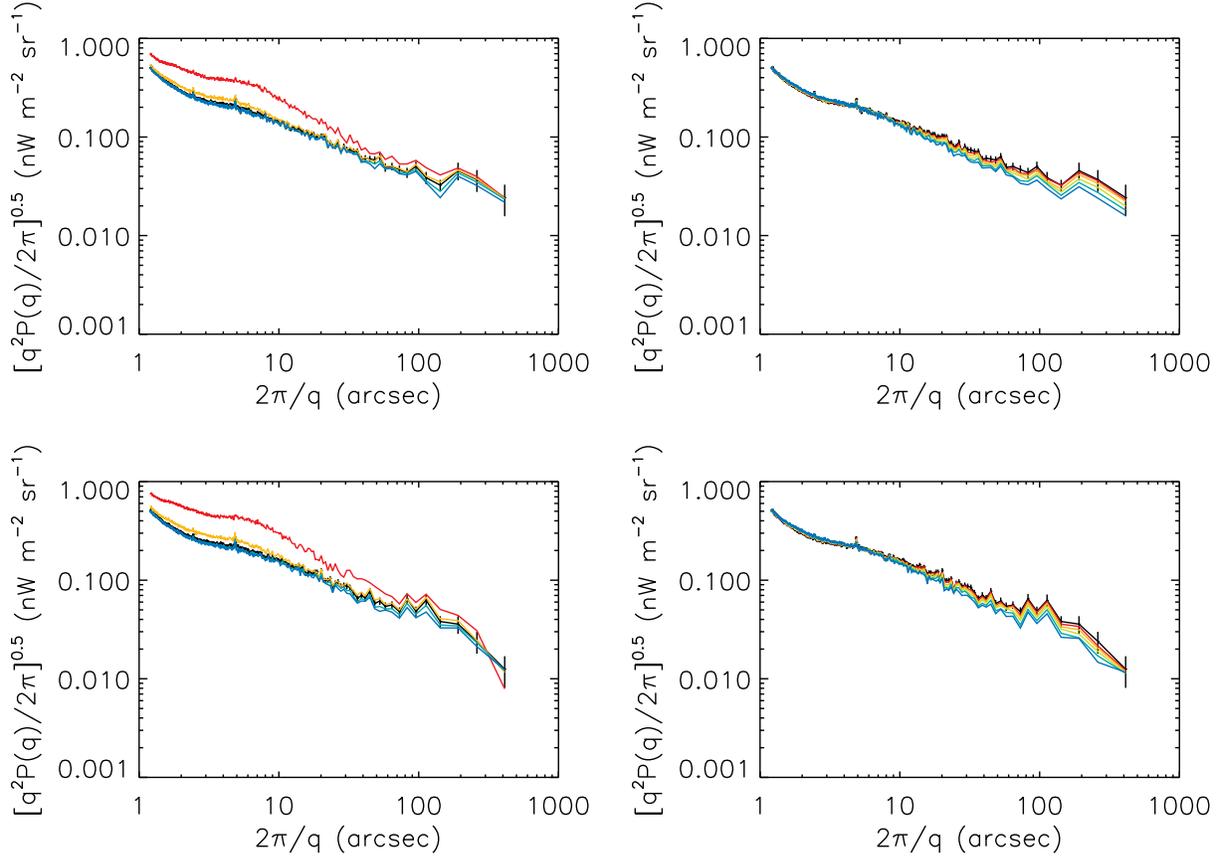}
   \caption{Fluctuation spectra of the CDFS field
   at 3.6 $\micron$ for various modifications of the source masking.
   The epoch 1 fields are shown in the top row, while epoch 2 are shown in the bottom.
   In the left column the black line
   represents the standard result. The orange and red lines indicate results where the
   clipping mask has been eroded by 1 and 2 pixels respectively, (i.e. decreasing amounts
   of clipped data). The green and blue lines indicate results where the
   clipping mask has been dilated by 1 and 2 pixels respectively, (i.e. increasing amounts
   of clipped data). In the right column the black line again represents the standard result.
   The red, orange, yellow, green and blue lines indicate results when an additional
   randomly placed 3$\times$3 pixel masks are applied covering 10, 20, 30, 40, and 50 \%
   of the total area respectively.}
   \label{fig:figure_mask1}
\end{figure}

\begin{figure}[ht] 
   \plotone{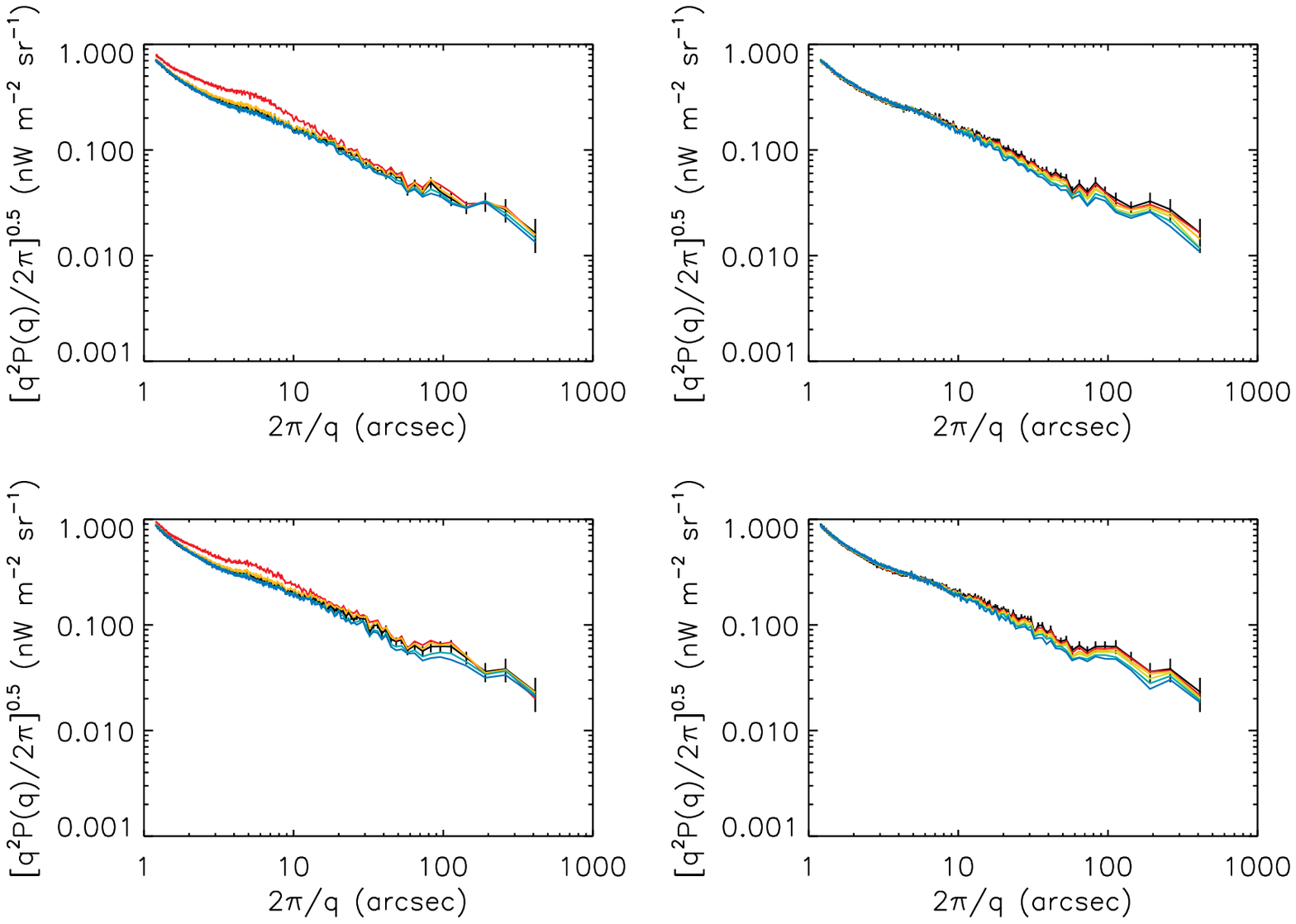}
   \caption{Same as Fig. \ref{fig:figure_mask1}, except for 4.5 $\micron$.}
   \label{fig:figure_mask2}
\end{figure}

\begin{figure}[ht] 
   \plotone{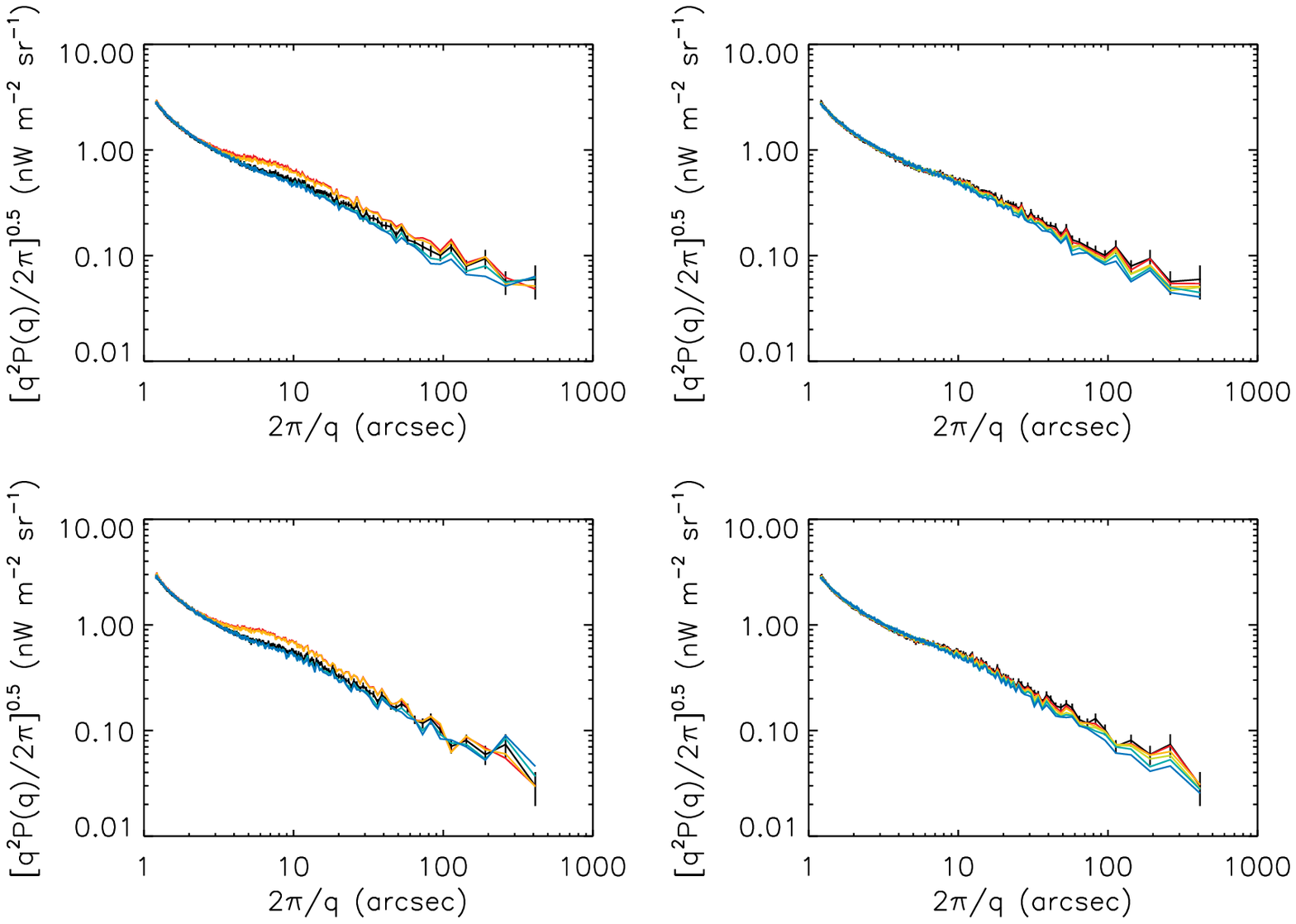}
   \caption{Same as Fig. \ref{fig:figure_mask1}, except for 5.8 $\micron$.}
   \label{fig:figure_mask3}
\end{figure}

\begin{figure}[ht] 
   \plotone{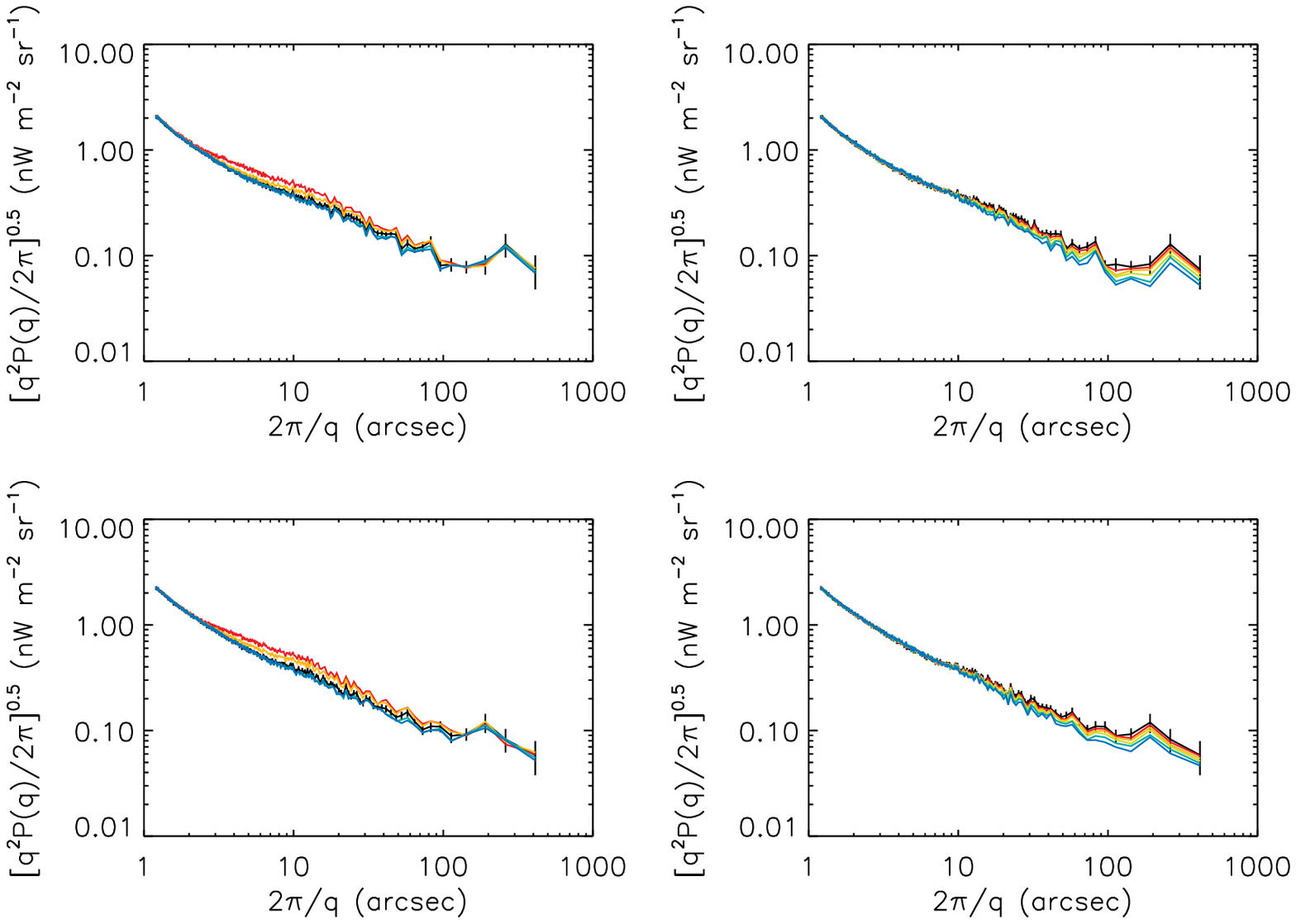}
   \caption{Same as Fig. \ref{fig:figure_mask1}, except for 8 $\micron$.}
   \label{fig:figure_mask4}
\end{figure}

Other tests of the effects of the clipping were performed using the actual images. 
Figures \ref{fig:figure_mask1}-\ref{fig:figure_mask4}
show the fluctuation spectra for the CDFS epoch 1 and
epoch 2 fields, for several variations of the clipping mask at a fixed model depth.
When the clipping mask is expanded by one or two pixels (i.e. the clipped regions are
increased in size) using a mathematical erosion operator, there is very little change in
the fluctuation spectrum. In most cases the largest changes are less than the 1$\sigma$
uncertainties. The changes are similarly small when the clipping mask is reduced in size
by 1 pixel using a mathematical dilation operator, with the exception of a small but
significant increase in the fluctuation amplitude at scales $\lesssim10''$.
When the clipping mask is reduced in size by 2 pixels, many faint sources are no longer
subject to any blanking at all. This creates large increases in the signal at
the smaller angular scales, especially at 3.6 and 4.5 $\micron$ where the faintest
sources blanked by the standard masking are well above the instrument noise limits.

The final masking test involved the additional masking by randomly located $3\times3$ 
pixel patches. Such masks were generated in which 10, 20, 30, 40, or 50\% of the 
pixels were masked. These masks were applied in addition to the standard clipping masks.
The results (shown in the right-hand panels of Figures \ref{fig:figure_mask1}-\ref{fig:figure_mask4}) 
are similar to those of the simulations above. As the total are masked is increase,
the power at the largest spatial scales decreases proportionally with the 
fractional area that is not masked, while the power at small spatial scales remains unaffected.

\subsection{Constructed tests}
As a check on the possible systematic errors in the power (or fluctuation) spectra of the
data, we have calculated power (or fluctuation) spectra for various artificial images 
that are related to different aspects of the analysis. Similarities between the 
measured and artificial power spectra can indicate possible errors.

\begin{figure}[ht] 
   \plotone{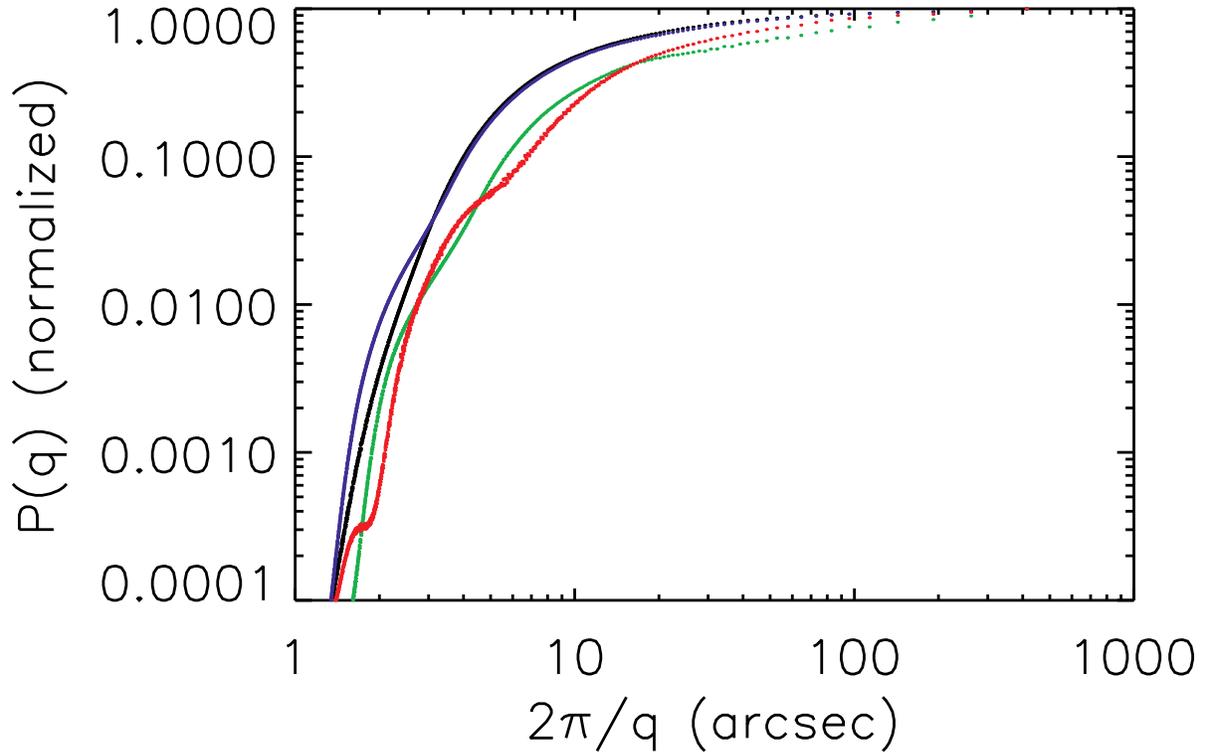}
   \caption{Power spectra of the IRAC PRFs. Black, blue, green, and red symbols
   represent 3.6, 4.6, 5.8 and 8 $\micron$ PRFs respectively.
   Celestial components of the power spectrum will be
   modulated by these functions; other (e.g. instrumental) components of the power spectrum will not.}
   \label{fig:beams}
\end{figure}

In Figure \ref{fig:beams}, we present the power spectra of the IRAC PRFs that we used
in the source modeling procedure. Any real signal from the sky will be convolved by
the PRF and thus its power spectrum will be multiplied by those shown here. The PRF will
reduce power by factors larger than 2 on scales $\lesssim10''$ at 3.6 $\micron$ to
$\lesssim20''$ at 8 $\micron$. Power arising from other sources (e.g. instrumental noise)
will not be modulated (multiplied) by the PRF power spectra.

\begin{figure}[ht] 
   \plotone{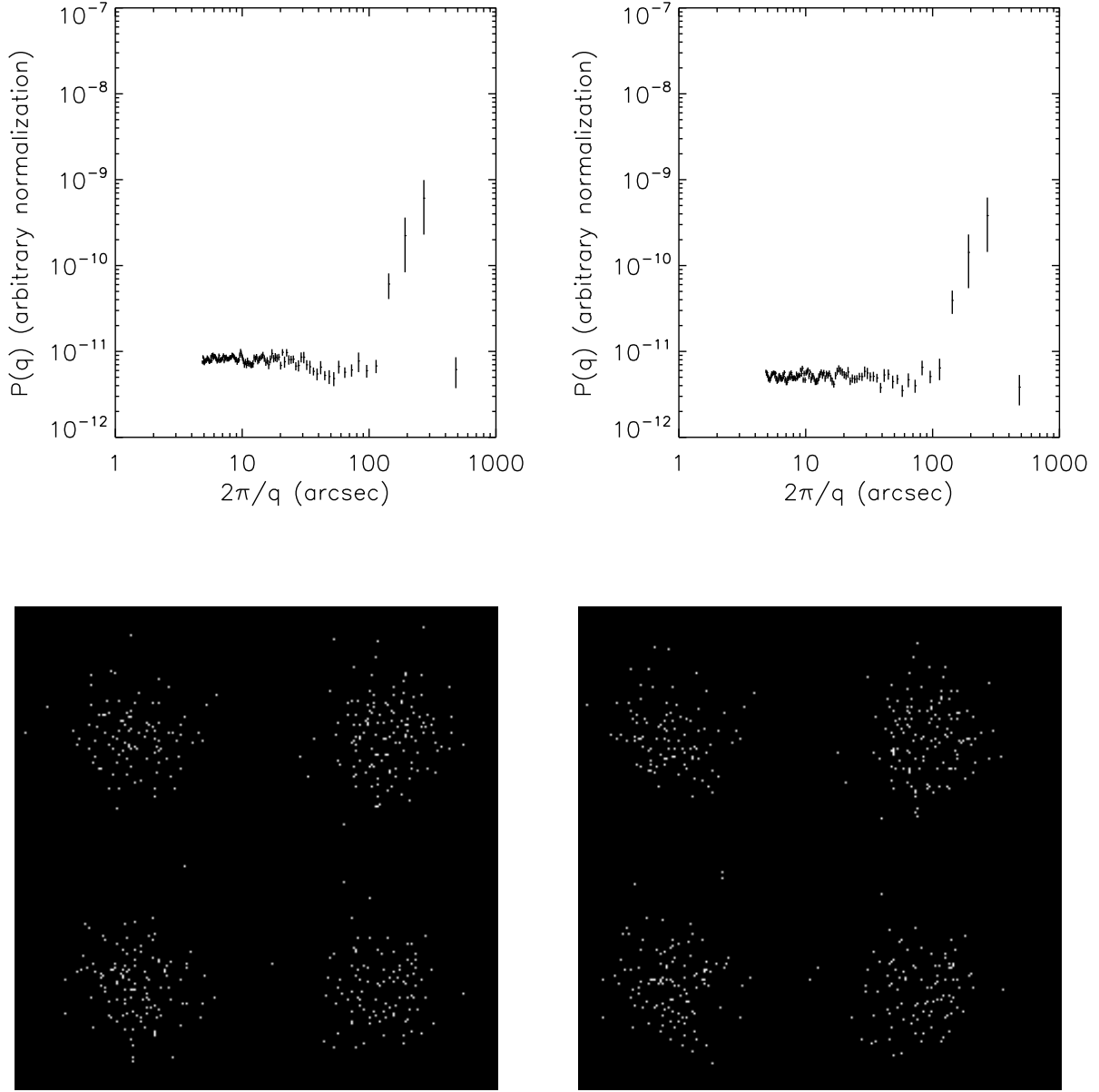}
   \caption{Power spectra of the CDFS-e1 dither patterns at 3.6 (left) and 4.5 (right) $\micron$.
   Excess power appears only at scales $150 - 330''$, which is caused by the clustering
   of pointings within the $2\times2$ instrument fields of view required to mosaic the entire field.
   The lower panels depict the net dither patterns for all AORs covering this field.}
   \label{fig:dither}
\end{figure}

\begin{figure}[ht] 
   \plotone{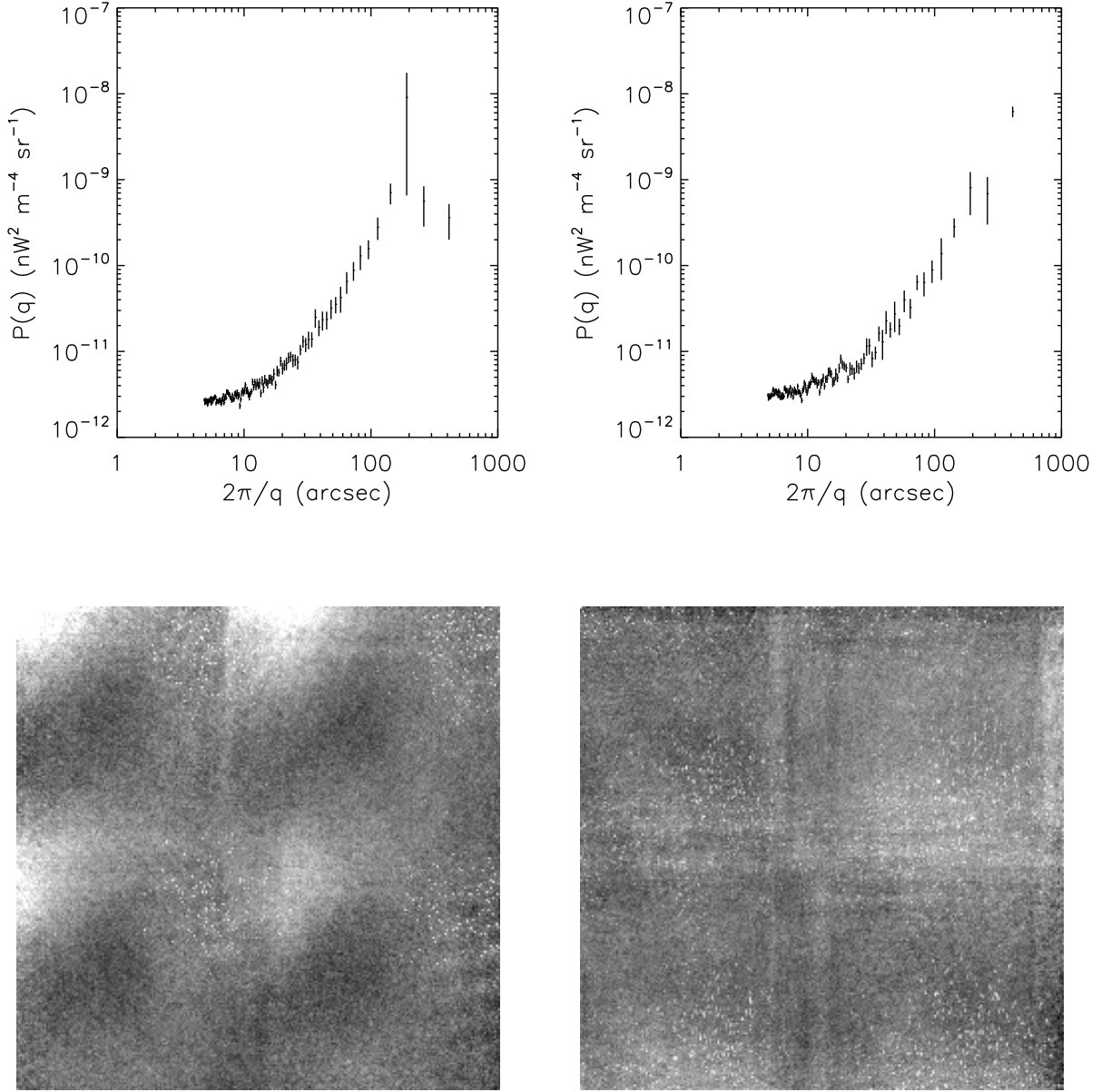}
   \caption{Power spectra of the CDFS-e1 offsets at 3.6 (left) and 4.5 (right) $\micron$, as reprojected
   onto the sky by the net dither pattern (Fig. \ref{fig:dither}). These artificial images are shown in
   the bottom row, with the corresponding power spectra at the top (range = [-1,1] nW m$^{-1}$ sr$^{-2}$). 
   Note the strong similarities with the GOODS-processed image in Fig. 5.}
   \label{fig:dither_offsets}
\end{figure}

Figure \ref{fig:dither} shows artificial images constructed by
distributing delta functions in the same pattern as the dithering
and the (2$\times$2) raster mapping used for the CFDS-e1 field at
3.6 and 4.5 $\micron$. If power along the axes were included, the
corresponding power spectra of these images would show a strong
feature at $\sim100-300''$, corresponding to the offset between
the 4 raster pointings of the map. When the power on the axes is
excluded (as in the figure), excess power is eliminated at
$\sim300''$, and is reduced at smaller spatial scales. This sort
of feature is not directly present in the power spectra of any
field. However, most instrumental errors will be much more highly
structured than a delta function. As an example of such an error
we took the detector offsets ($F^p$ and $F^q$) derived from the
self-calibration of the full data set at half-resolution, and
repeatedly added them to a blank sky in the same pattern as the
dithering. This creates images representing the errors that would
be present had we not removed the detector offsets via the
self-calibration. These are shown in Figure
\ref{fig:dither_offsets} along with the corresponding power
spectra. Both power spectra can be approximated as the sum of a
flat white--noise component and a steeply rising power law
component with an index of 2.3 - 2.4. This rising component is
steeper than that seen in the power spectra of the actual sky. At
4.5 $\micron$, the turnover at $200''$ is hidden by large scale power 
represented in the $F^q$ offsets. The amplitudes of these power spectra are large
enough that they would contribute significantly to the result if
we had not self-calibrated the data. In fact the pattern seen here
in the 3.6 $\micron$ offsets is largely responsible for that seen
in the ratio of the 3.6/4.5 $\micron$ data reduced by the GOODS
team (see Figure \ref{fig:compare12}).

\begin{figure}[ht] 
   \plotone{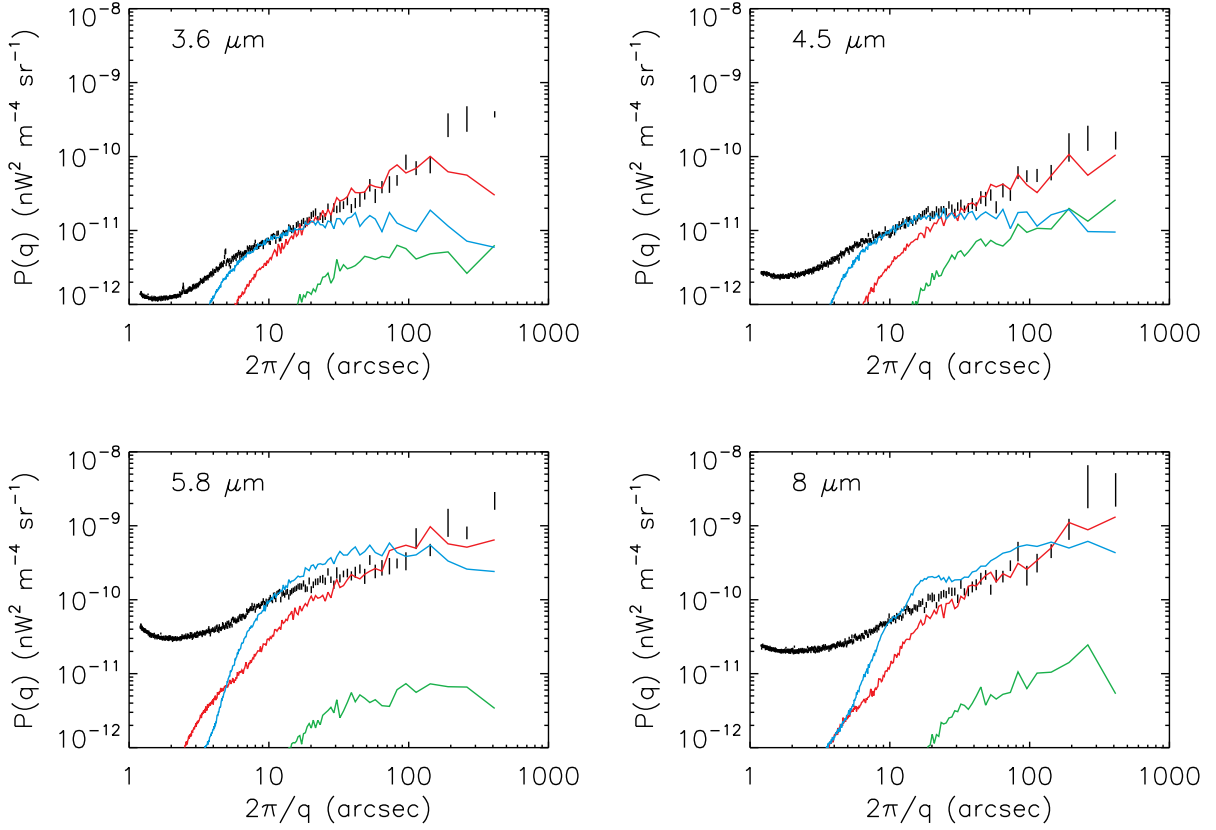}
   \caption{Power spectra of artificial CDFS-e1 images compared to that of the actual CIB.
   The black points with 1 $\sigma$ error bars are the CIB power spectrum.
   The red line indicates the power spectrum of the clipping mask,
   arbitrarily normalized to match at scales $2\pi/q > 30''$.
   The cyan line indicates the power spectrum of the sources that are masked
   (i.e. using the inverse of the nominal clipping mask),
   arbitrarily normalized to match at scales $8'' < 2\pi/q < 15''$.
   The green line indicates the power spectrum of the ``halo'' test
   image, based on the clipping mask. In this case, the power is appropriately
   scaled according to the (small) correlation between the halo image and the actual
   CIB fluctuations. }
   \label{fig:power_mask_CDFSe1}
\end{figure}

\begin{figure}[ht] 
   \plotone{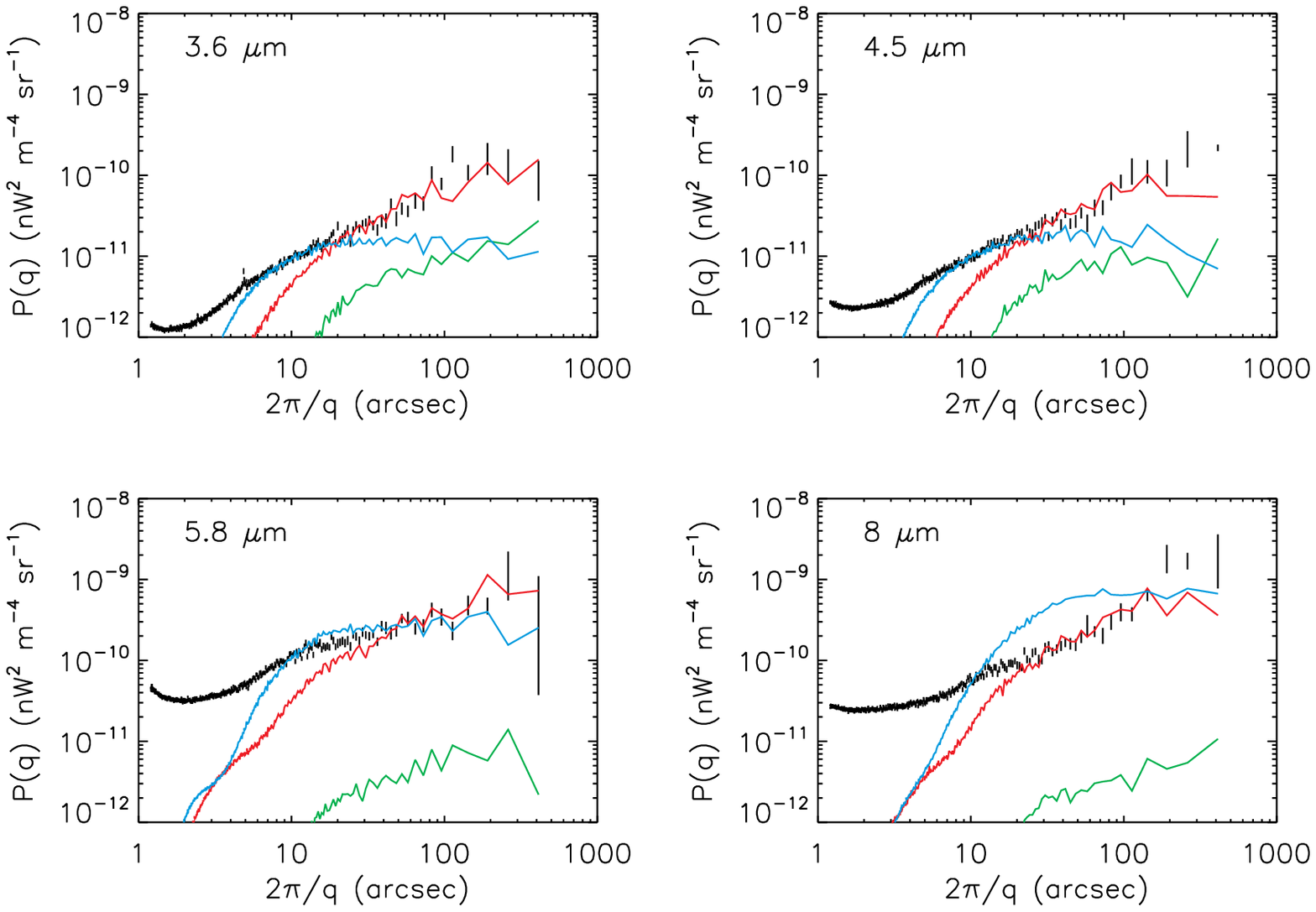}
   \caption{Same as Fig. \ref{fig:power_mask_CDFSe1}, except for the CDFS-e2 field.}
   \label{fig:power_mask_CDFSe2}
\end{figure}

\begin{figure}[ht] 
   \plotone{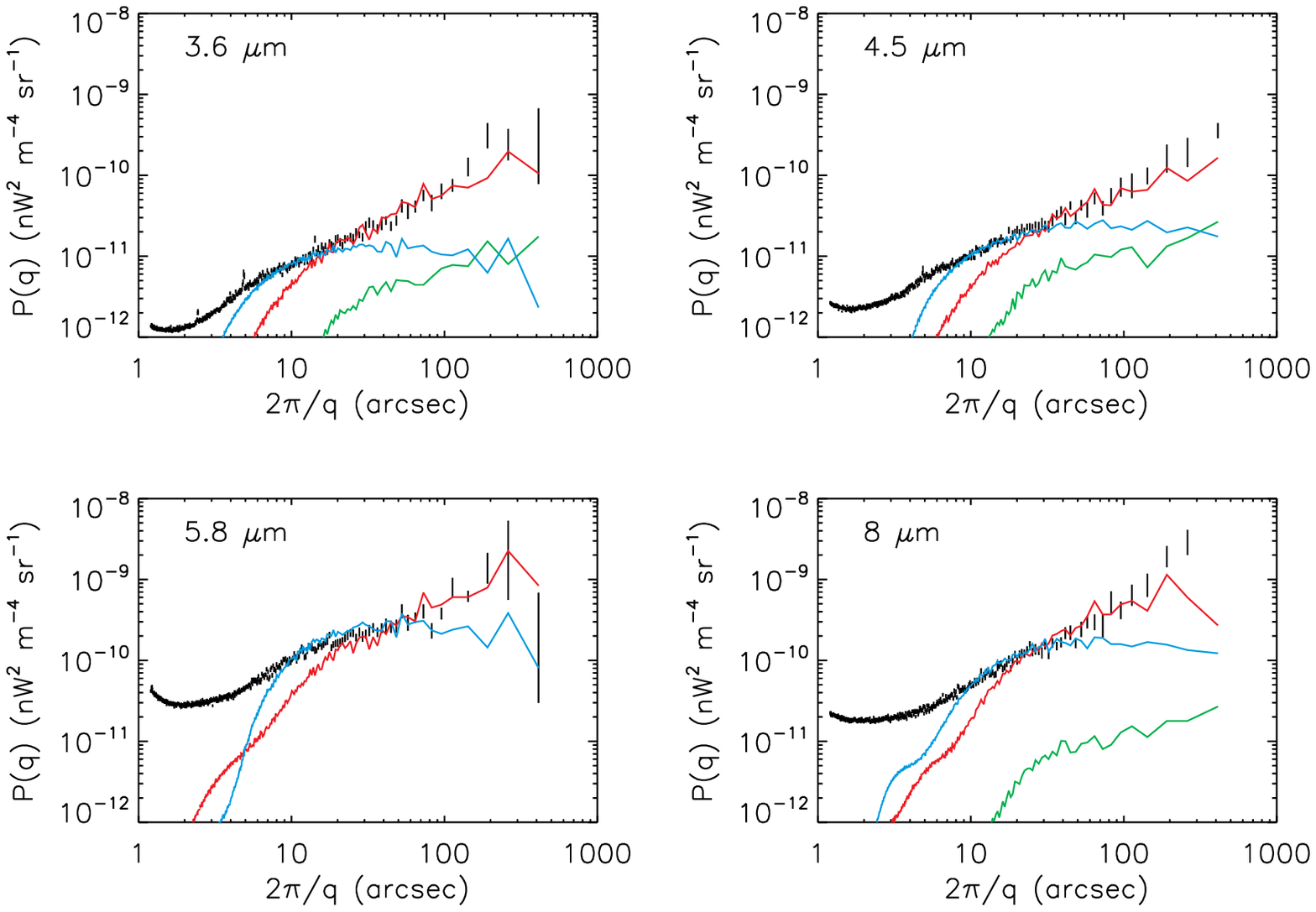}
   \caption{Same as Fig. \ref{fig:power_mask_CDFSe1}, except for the HDFN-e1 field.}
   \label{fig:power_mask_HDFNe1}
\end{figure}

\begin{figure}[ht] 
   \plotone{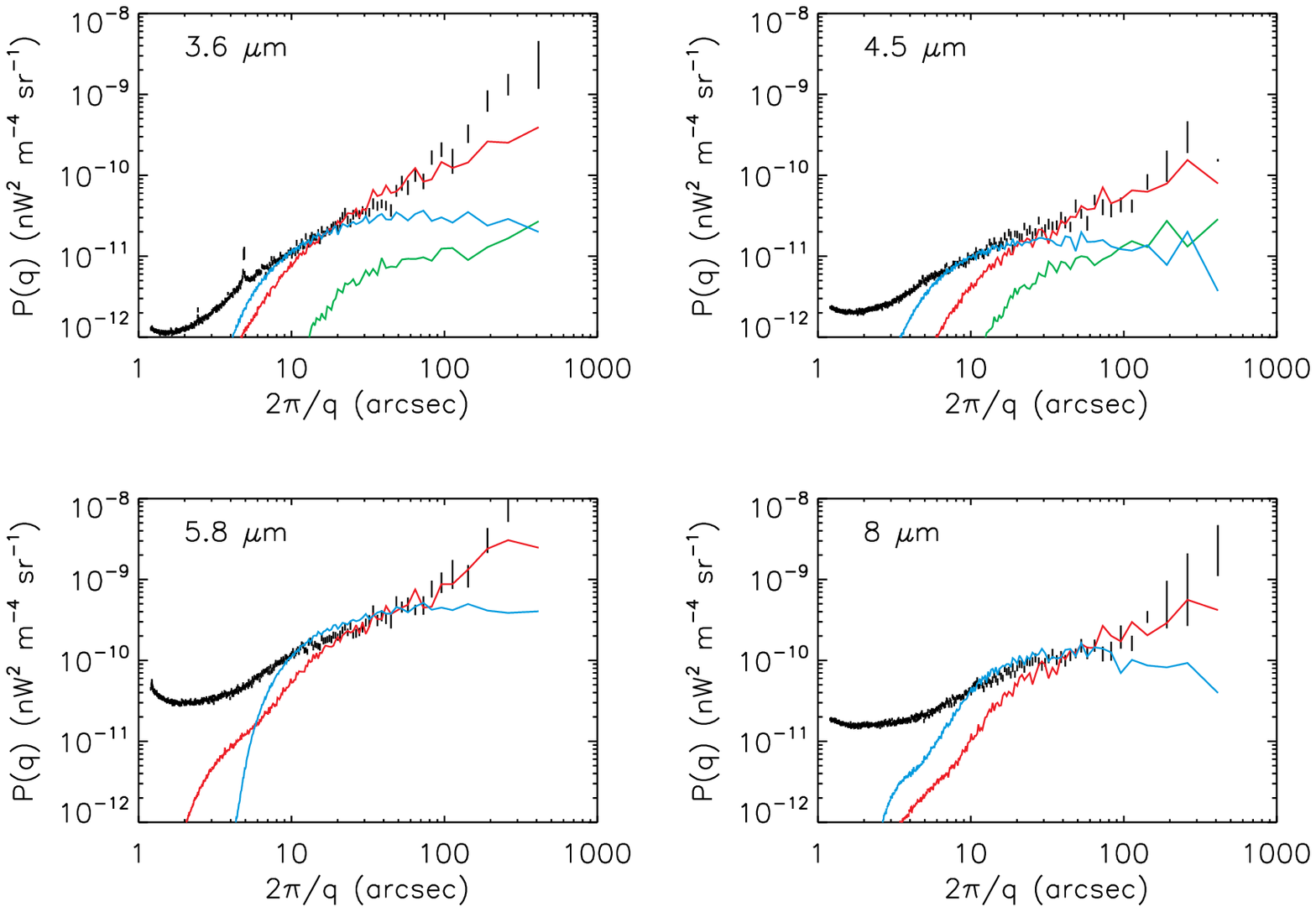}
   \caption{Same as Fig. \ref{fig:power_mask_CDFSe1}, except for the HDFN-e2 field.}
   \label{fig:power_mask_HDFNe2}
\end{figure}

\begin{figure}[ht] 
   \plotone{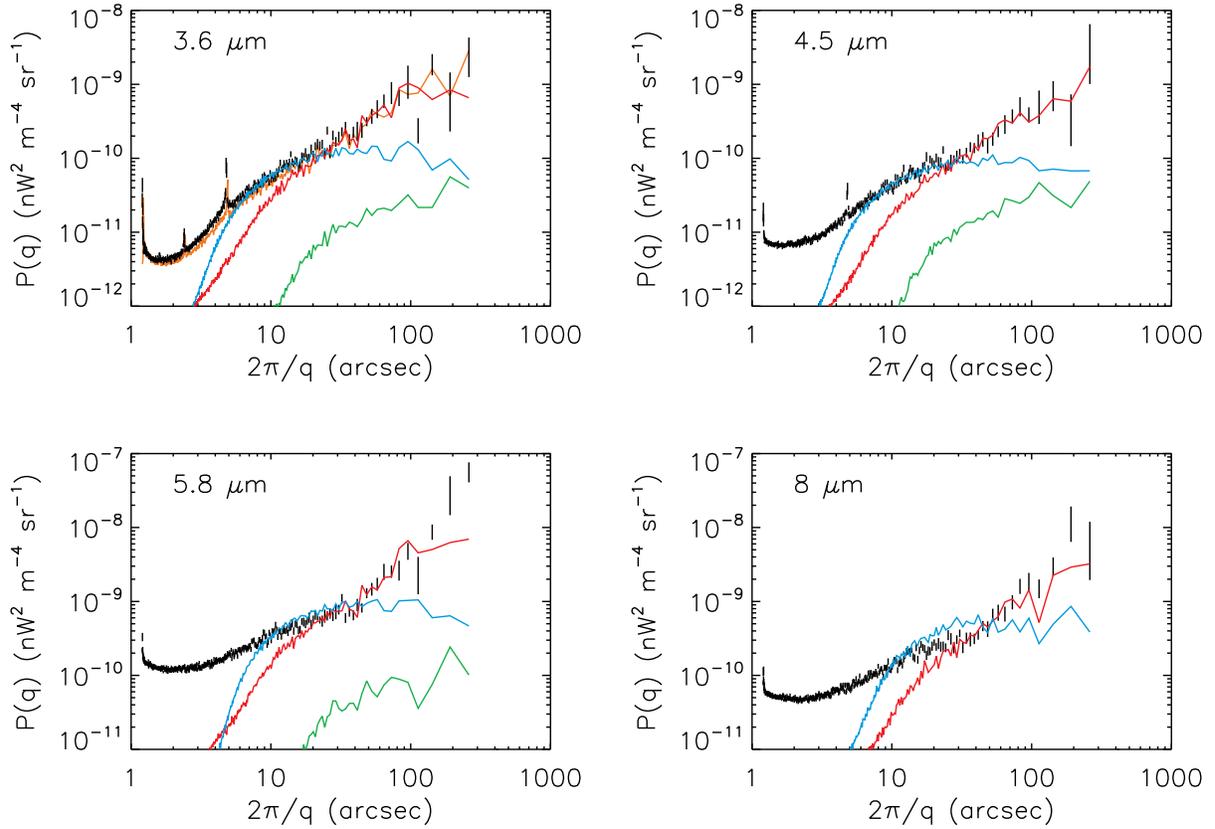}
   \caption{Same as Fig. \ref{fig:power_mask_CDFSe1}, except for the QSO 1700 field.
   The additional orange line in the 3.6 $\micron$ panel shows the power spectrum 
   derived when the processing starts with BCD frames (as for the other fields) rather
   than the raw data. Differences are generally $< 2 \sigma$, except at scales
   $2'' < 2\pi/q < 10''$.}
   \label{fig:power_mask_QSO}
\end{figure}

\begin{figure}[ht] 
   \plotone{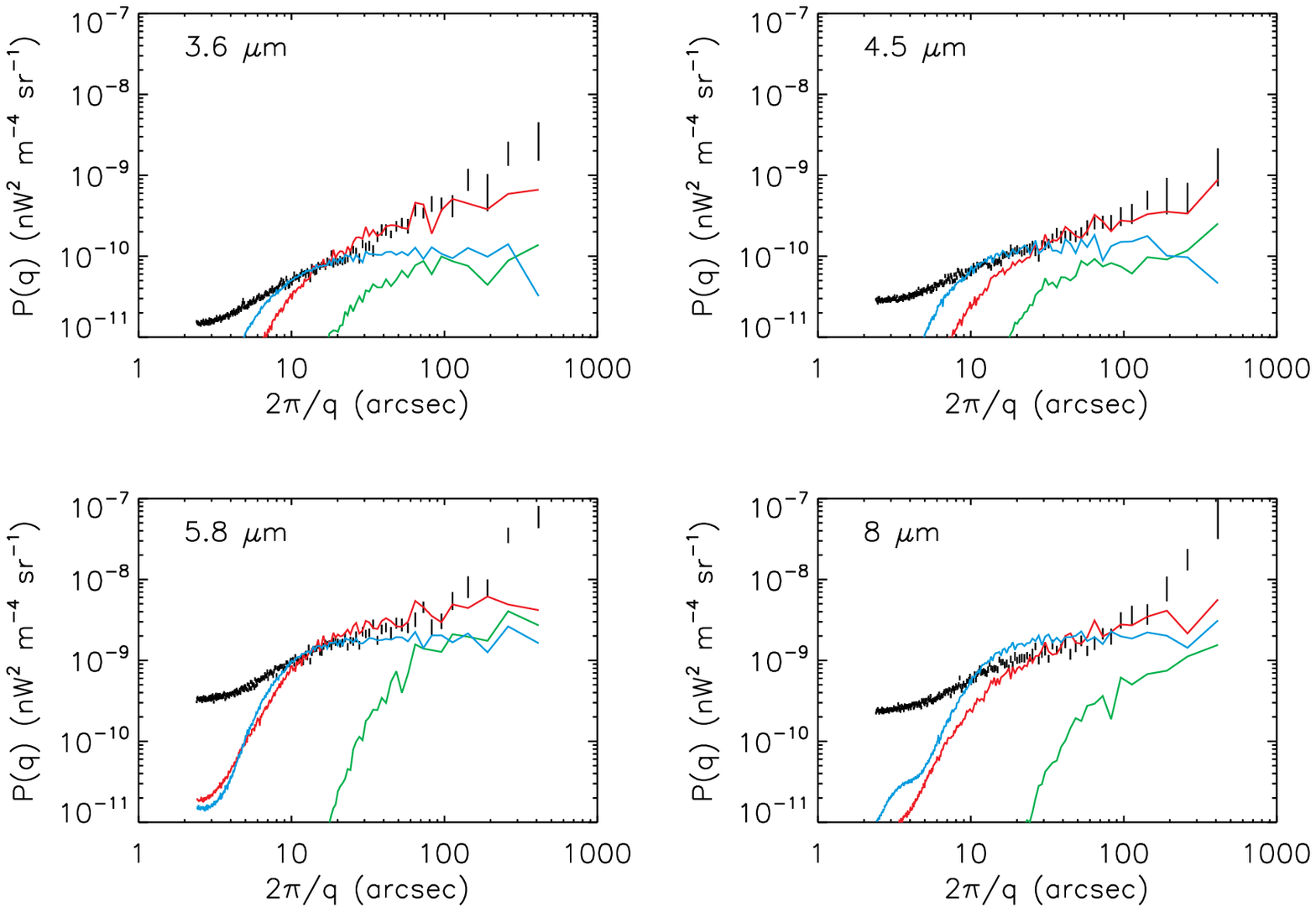}
   \caption{Same as Fig. \ref{fig:power_mask_CDFSe1}, except for the EGS field.}
   \label{fig:power_mask_EGS}
\end{figure}

Other artificial images that we have examined are based on the actual sky instead of
the instrument and observing strategy. The first of these test images is constructed
by applying the complement of the clipping mask to the original sky map. This creates
an image consisting of only the bright sources, with the background set to 0.0 between
them. The power spectrum of this image serves as a check on any large scale structure
that may be intrinsic to the distribution of the bright sources. The results are shown
in Figures \ref{fig:power_mask_CDFSe1}-\ref{fig:power_mask_EGS} (blue lines),
where we have renormalized these power spectra to match the observed spectra at $8 - 15''$.
In all cases, there is no excess power at large spatial scales. The second sky-based test
is to calculate the power spectrum of the mask itself. This is similar to the previous test,
but it removes the effective weighting with source brightness, which is present in the
previous test. The results are shown as the red lines in Figures
\ref{fig:power_mask_CDFSe1}-\ref{fig:power_mask_EGS}, where they have been
arbitrarily normalized to the observed power spectra at angular scales $>30''$.
While the power spectrum of the mask has a large scale shape similar to the data,
the mask is uncorrelated with the residual fluctuations, and subtracting any scaled version of the mask
only {\it increases} the large scale power. A more complex test image was created
by setting the intensity to be the inverse of the distance from the nearest
blanked region of the mask. This ``halo'' image simulates the extended wings of bright
sources that would remain after the application of a simple masking defined strictly by
a surface brightness threshold. Removal of such wings, whether intrinsic to the source
or caused by the PRF, are a large part of the motivation for the model we subtract. This
simulated image can test whether the model over- or under-subtracts such features.
We scaled these test images using the slopes of linear correlations between these and the
model-subtracted images. Correlations were generally weak, though statistically significant.
The power spectra of the test images are shown as green lines in Figures
\ref{fig:power_mask_CDFSe1}-\ref{fig:power_mask_EGS}. In all cases the power is
not more than $\sim10\%$ of the power of the actual background. (In some cases
the power is below the minimum range in the plots.)

One additional check illustrated in Figure \ref{fig:power_mask_QSO} is the 
comparison of the 3.6 $\micron$ power spectrum derived when starting
with the raw data (as described in \S2.2) and the power spectrum derived when the 
processing starts with the latest version (S18.7.0) of the BCD data. 
The two power spectra show only small differences ($\lesssim 2 \sigma$) at 
the smallest angular scales and at large angular scales ($2\pi/q > 10''$). 
More significant, though still small, differences occur at angular 
scales $2'' < 2\pi/q < 10''$. These scales are typically dominated by the 
shot noise (see next section), which is sensitive to the depth of the source
model that is subtracted. The source models were calculated independently for 
the mosaics derived from the raw and the BCD data.

\section{Characterization of Power in Different Fields}

In this section we characterize the final power spectra for each field and each 
wavelength by fitting the power spectra with empirical models with 3 or 4 free
parameters. To the extent that these models provide good fits, the derived 
parameters may provide a simpler means of representing the power spectra.

Our idealized model of the power spectrum includes 
three components: instrument noise, shot noise, and a power law.
For this model we assume instrument noise has a flat spectrum, with the normalization as
its only free parameter. This is expected if there are no correlations in the 
response of the detector pixels. The shot noise is intended to represent
the random Poisson statistics of sources below the confusion
limit. Intrinsically this component is also flat. However, as the
observed sky is unavoidably convolved by the PRF, this component
is correspondingly modulated by the power spectrum of the PRF,
which greatly reduces power at small angular scales. This
component also has only its normalization as a free parameter. The
power-law component is included to represent any excess power at
large angular scales. It is also modulated by the power spectrum
of the PRF, and contains two free parameters: a normalization and
the power law index. This flat noise model can be expressed as:
\begin{equation}
P(q) = a_0(2\pi/q/100'')^{a_1}P_{PRF}(q) + a_2P_{PRF}(q) + a_3.
\end{equation}

A slightly different model was also fit, in which the flat noise was replaced by the
measured A-B noise spectrum. In this case the normalization is fixed, and there is
no free parameter associated with this component.
\begin{equation}
P(q) = b_0(2\pi/q/100'')^{b_1}P_{PRF}(q) + b_2P_{PRF}(q) + P_{A-B}(q)
\end{equation}

Using the A-B noise spectrum instead of a flat instrument noise 
(assumed above) is an improvement if the structure in the A-B spectrum
is purely instrumental, and does not include any astronomical component 
or systematic errors.

\begin{figure}[ht] 
   \includegraphics[height=6in]{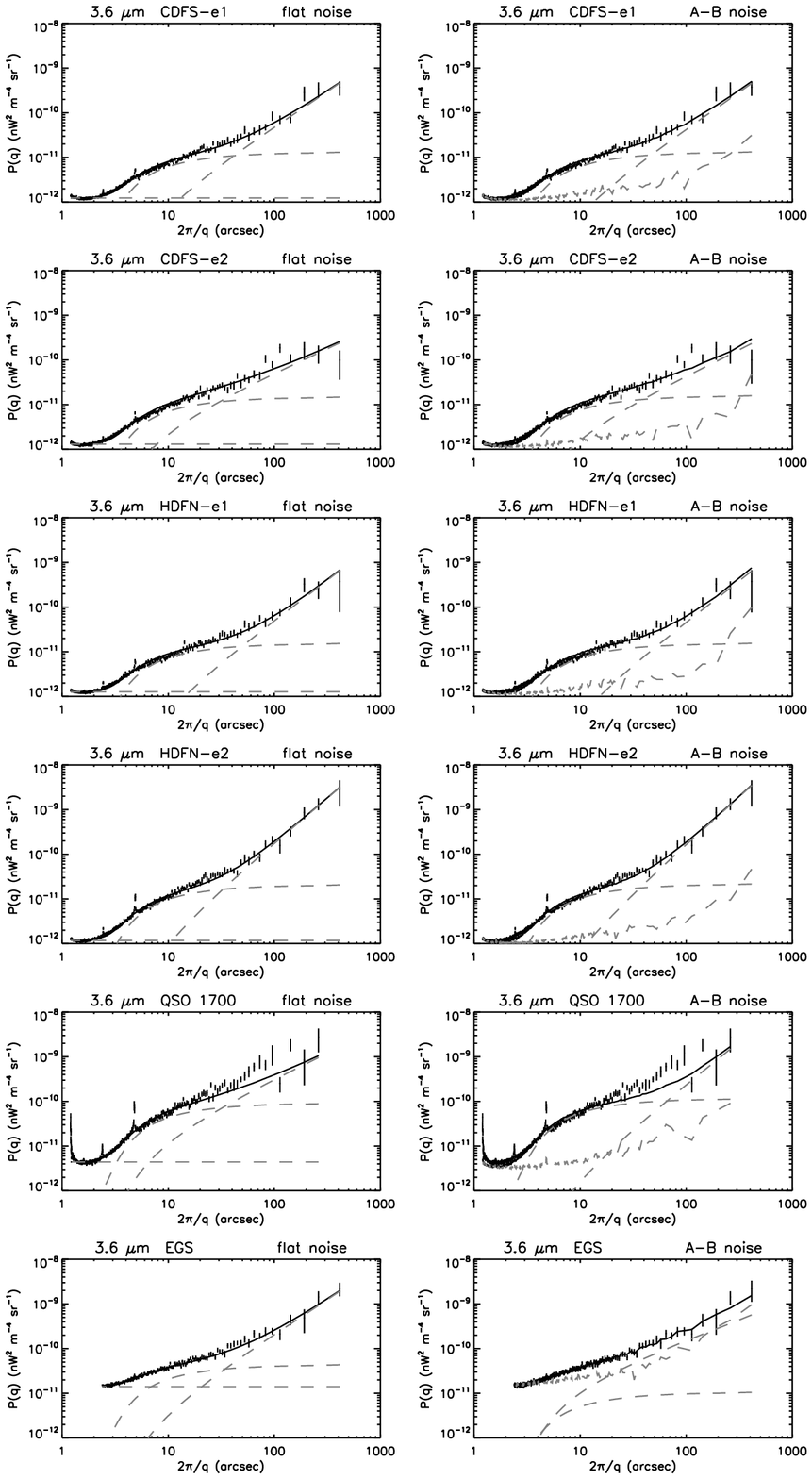}
   \includegraphics[height=1.5in]{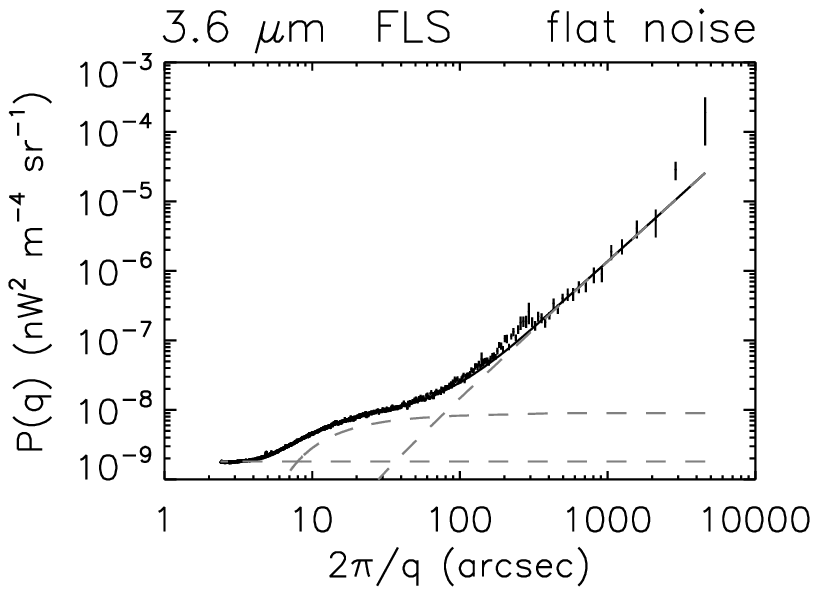}
   \caption{Fits to the 3.6 $\micron$ power spectra for our six main fields, and the FLS.
   The left column shows fits as characterized by Equation (4): (a) flat instrument noise
   components, (b) a flat shot noise component convolved with the PRF, and (c) a power
   law component, also convolved with the PRF. The solid line indicates the sum of
   these three components (dashed lines). The middle column shows the fits as characterized
   by Equation (5), where the measured (A-B)/2 noise takes the place of the flat instrument
   noise component. The FLS result for the flat noise fit is shown separately, as
   it spans a different range of angular scale and power than the other fields.
   }
   \label{fig:fit_plots1}
\end{figure}

\begin{figure}[ht] 
   \includegraphics[height=6in]{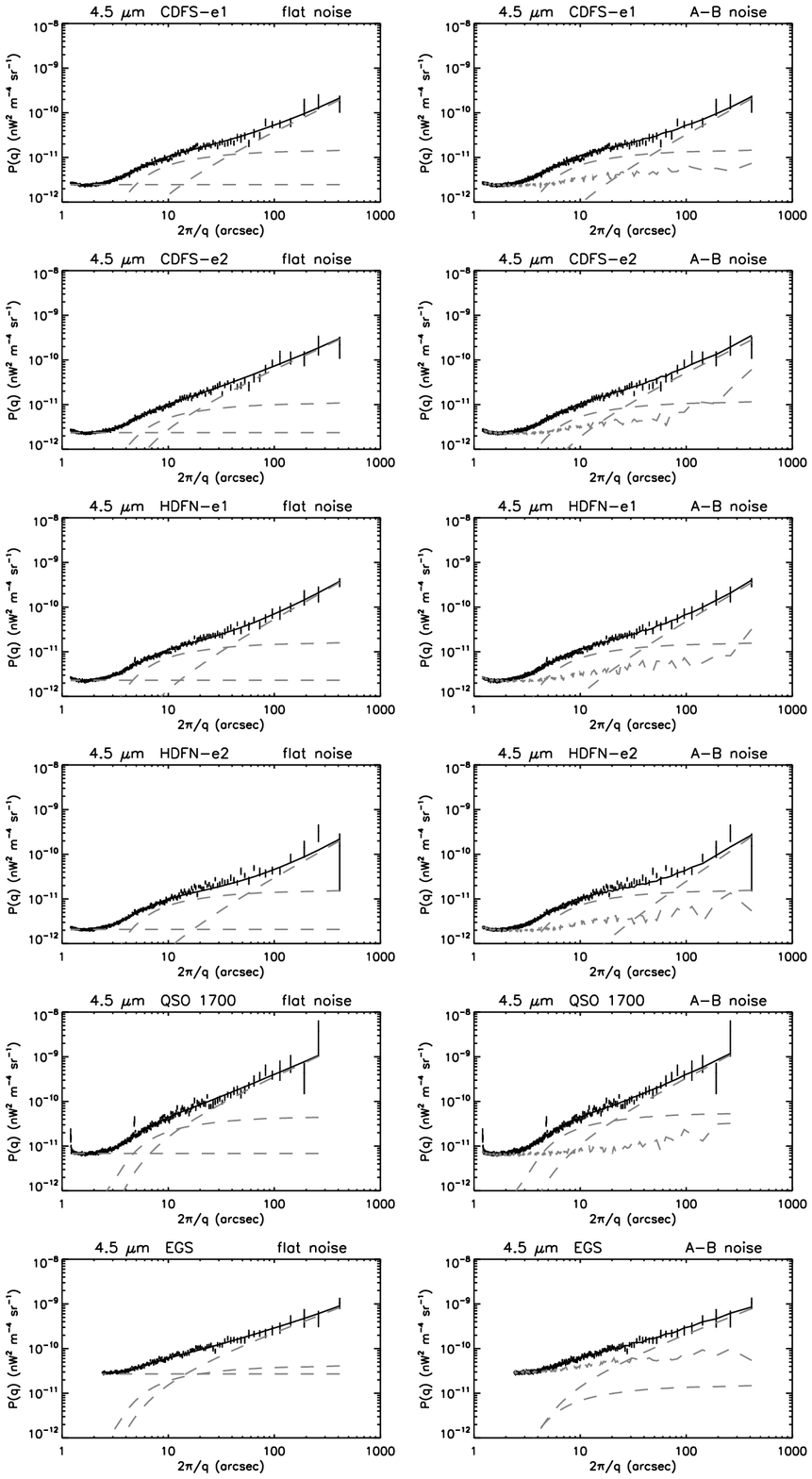}
   \includegraphics[height=1.5in]{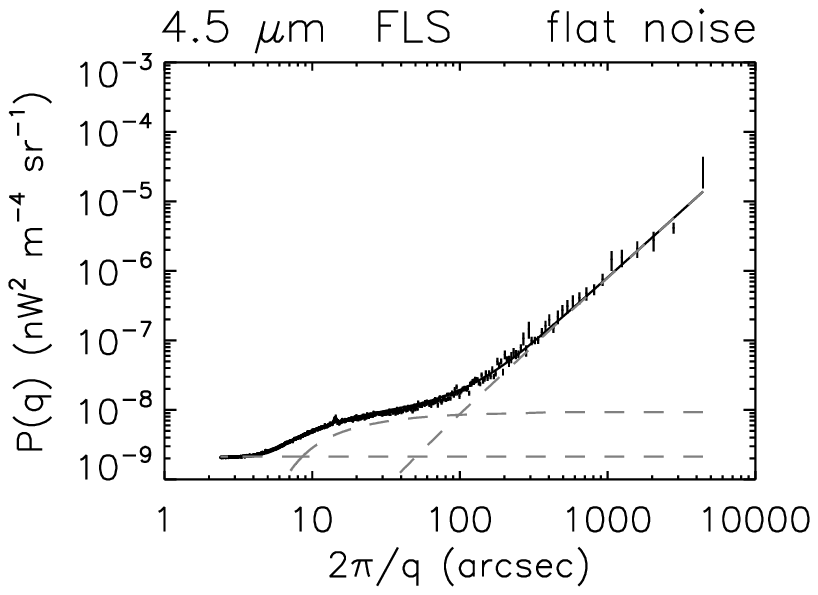}
   \caption{Same as Fig. \ref{fig:fit_plots1}, except for 4.5 $\micron$.}
   \label{fig:fit_plots2}
\end{figure}

\begin{figure}[ht] 
   \includegraphics[height=6in]{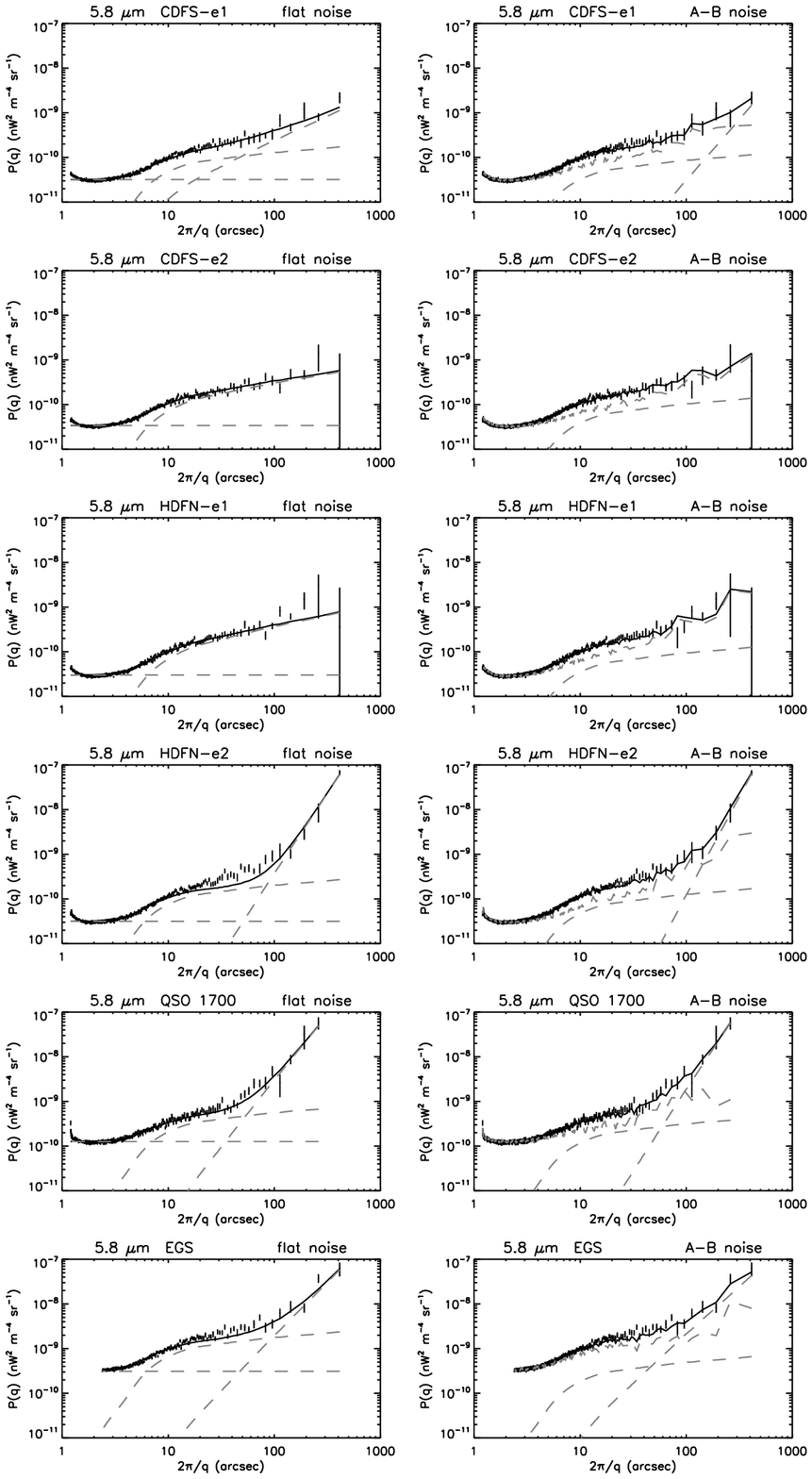}
   \caption{Same as Fig. \ref{fig:fit_plots1}, except for 5.8 $\micron$, and without FLS results.}
   \label{fig:fit_plots3}
\end{figure}

\begin{figure}[ht] 
   \includegraphics[height=6in]{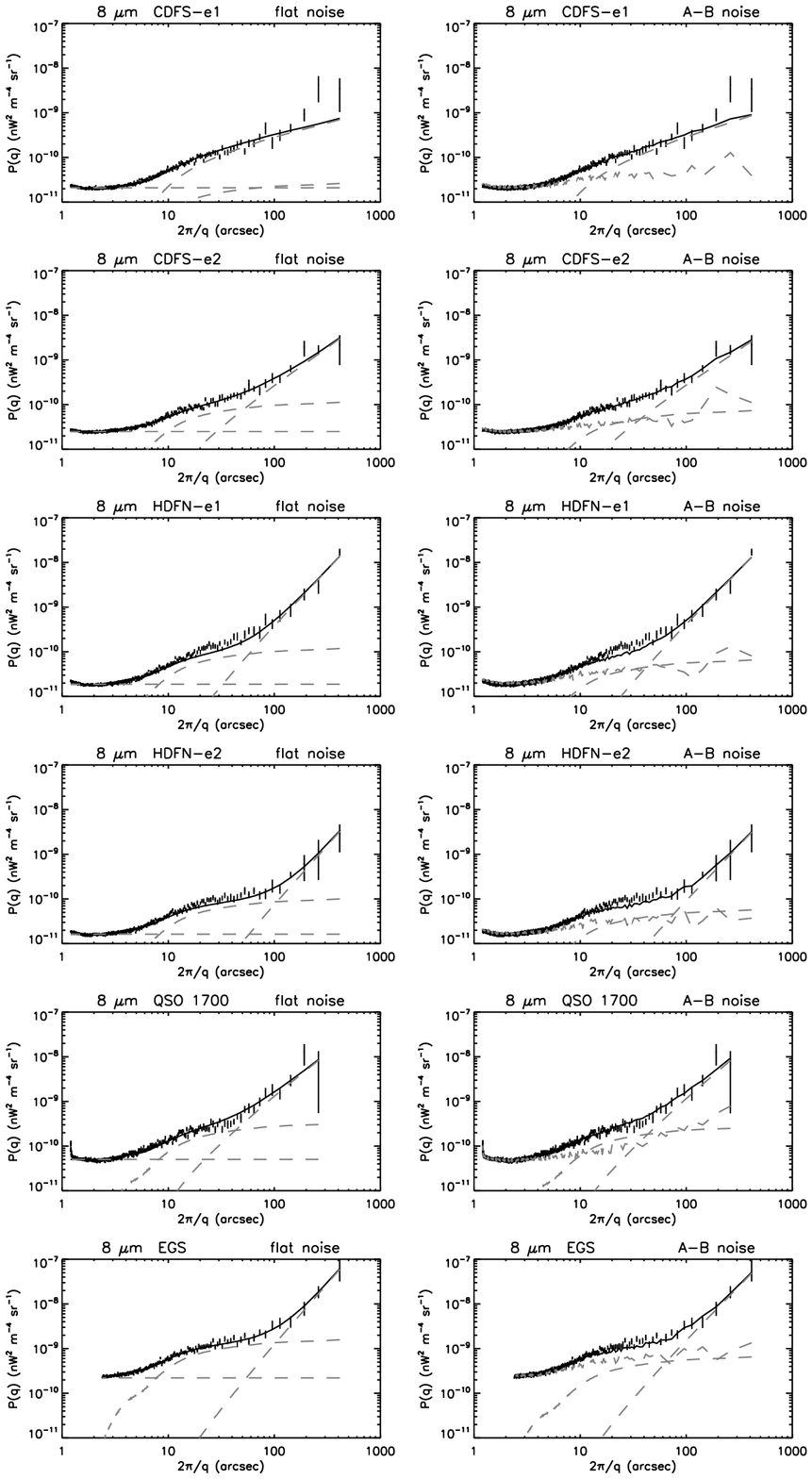}
   \caption{Same as Fig. \ref{fig:fit_plots1}, except for 8 $\micron$, and without FLS results.}
   \label{fig:fit_plots4}
\end{figure}

The parameters (and formal uncertainties) derived via these fits are presented
in Tables \ref{tab:fit_params} and \ref{tab:fit_covars}.
Figures \ref{fig:fit_plots1} - \ref{fig:fit_plots4} show the results graphically.
The uncertainties listed in Table \ref{tab:fit_params} are the formal uncertainties for
the given parameters. However, Table \ref{tab:fit_covars} shows that in many cases there
is significant covariance between parameters, and thus assessing agreement or
disagreement of results in different fields requires caution.

At 3.6 and 4.5 $\micron$, the limiting factor in the results is
the instrument noise, as characterized by $a_3$. As the instrument
noise decreases from field to field, the shot noise (after
subtracting the source model to reach zero skewness) decreases
correspondingly. The EGS and FLS fields
differ from the overall trend because they have been mapped with
$1.2''$ pixels, rather than $0.6''$ pixels. Therefore, the noise
at the pixel scale is reduced by a factor of 2. This allows the pixel-based
source modeling to run to fainter levels, leading to a
corresponding decrease in the shot noise in the residual image.
The amplitude of the power law component decreases as the shot noise
decreases. This indicates that a large fraction of the power seen in the shallower
fields (QSO 1700, and EGS) arises from faint sources, which are
modeled and removed in the deeper GOODS fields. If the GOODS fields are
only model-subtracted to the same shot noise level as the QSO 1700
or EGS fields, then they have a power law component with a
comparable amplitude to the shot noise in the QSO 1700 and EGS fields.

Results at 5.8 and 8 $\micron$, are more erratic because in some cases the best-fit
parameters include a negligible amplitude for one of the components. In such cases there
are also extremely large covariances between the parameters. Aside from the anomalous cases,
it is apparent that the power law component at these wavelength is steeper than that at
3.6 and 4.5 $\micron$.

\begin{figure}[ht] 
   \plotone{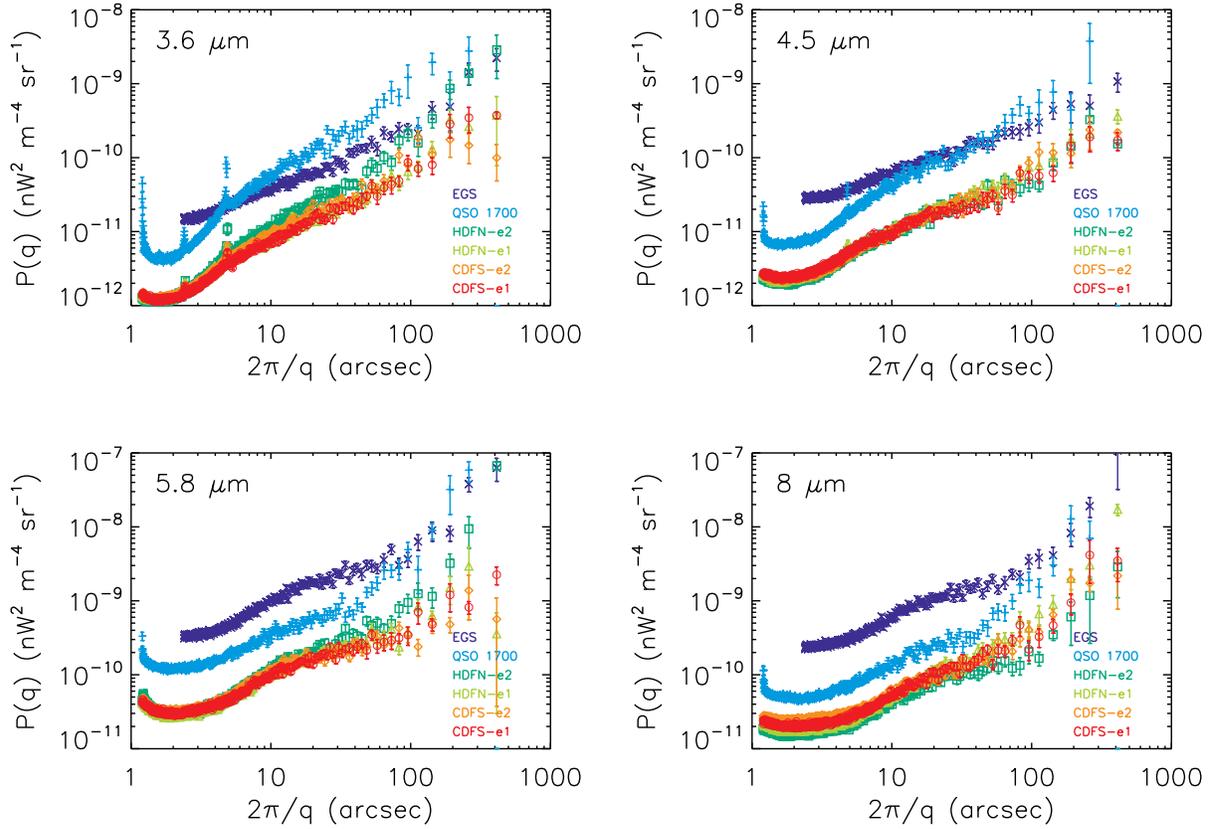}
   \caption{Final power spectra for the different fields. This figure provides a more
   visual comparison of similarities and difference between the power spectra than the
   numerical details of Tables 3 and 4. The FLS results are not shown due to their different
   ranges of spatial scale and power.
   }
   \label{fig:final_power}
\end{figure}

\begin{figure}[ht] 
   \plotone{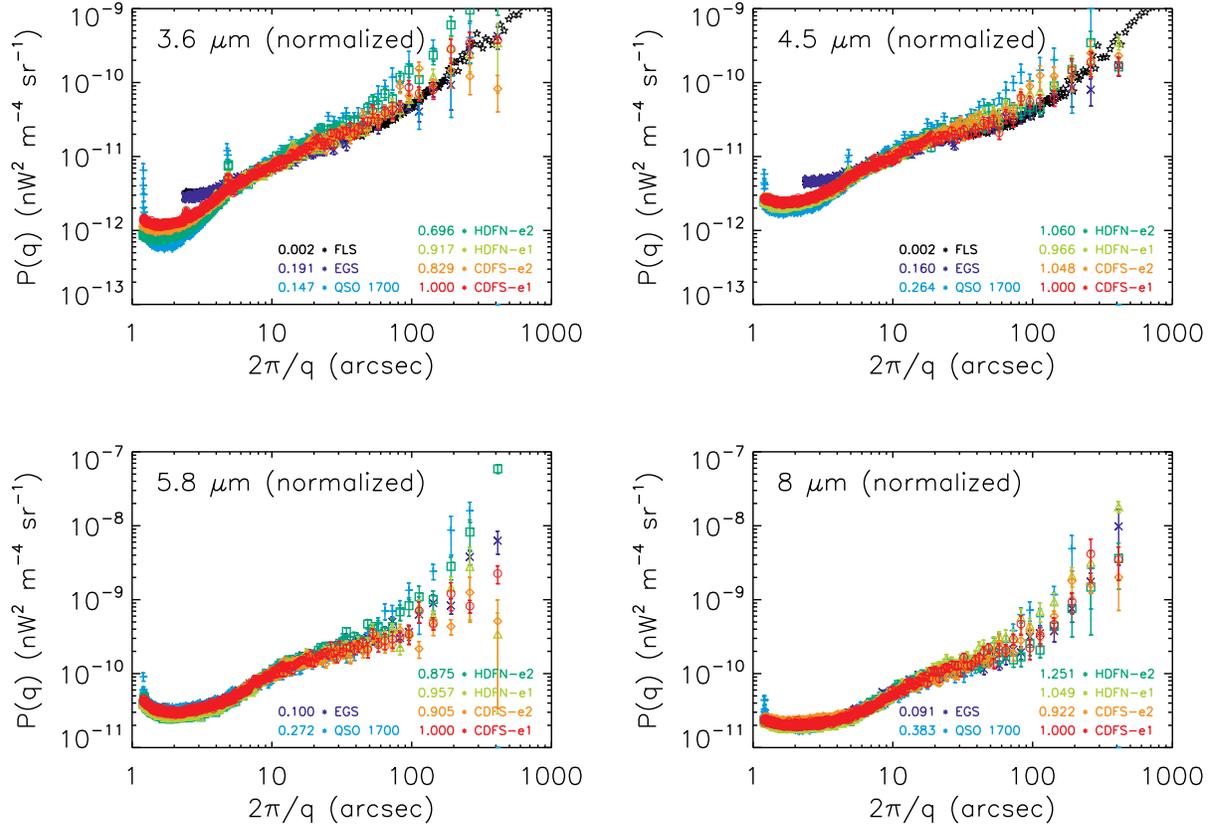}
   \caption{Final power spectra for the different fields, after normalizing all spectra
   to match that of the CDFS-e1 at $2\pi/q > 5''$. The figure legends cite the
   normalization factors required for each field. }
   \label{fig:final_power_norm}
\end{figure}

Final power spectra are compared in Figure \ref{fig:final_power},
and in Figure \ref{fig:final_power_norm} the spectra are all
normalized to match that of CDFS-e1 at $2\pi/q > 5''$. These
figures provide a more visual comparison, as an alternative to the
quantitative details of Tables \ref{tab:fit_params} and
\ref{tab:fit_covars}.
The reduced $\chi^2_{\nu}$ values for the comparison of these
normalized power spectra are given in Table \ref{tab:chi_norm}.
The largest discrepancy amongst the GOODS fields
is seen to be the HDFN-e2 field at 3.6 and 5.8
$\micron$. This is a result of residual detector artifacts induced
by the $K_S = 10.2$ magnitude star 2MASS 12373797+6216308, which
is present in this field. It is $\sim2$ magnitudes brighter than
any other star in any of the GOODS fields. When the 3.6 $\micron$
HDFN-e2 field is cropped to a smaller size to exclude this star's
artifacts, the large scale power spectrum becomes more similar to
those of the other GOODS fields.
As discussed in \S2.2, spikes in the power spectra caused by the 
detector muxbleed artifact are visible at $2\pi/q = 4.8''$ and $2.4''$.
These are only present at 3.6 and 4.5 $\micron$, and are strongest 
in the QSO 1700 data. At these specific frequencies the power is 
factors of 2-4 higher than adjacent frequencies, but there is no 
expected or apparent effect on the power spectrum at low frequencies.

%%%%%%%%%%%%%%%%%%%%%%%%%%%%%%%%
\section{Discussion}

The preceding discussion detailed the analysis of CIB fluctuations
in the source-subtracted deep {\it Spitzer} data. We have shown that
after removing foreground sources there remains a significant CIB
fluctuations component. This component exceeds the instrument
noise and is approximately isotropic on the sky consistent with
its cosmological origin. This section is devoted to discussing the
implications of the KAMM results. We start with summing up the
requirements that any qualified source-subtracted CIB data analysis
must meet, following which we discuss the cosmological
implications of our results. We conclude this section with a
comprehensive comparison between our results and the various other
measurements/constraints of the CIB.

\subsection{Requirements for CIB fluctuations studies: a summary}

By design, CIB fluctuations studies are necessary in order to
uncover populations which cannot be resolved because they are {\it
fainter than either the sensitivity limit or the confusion limit 
of the present-day instruments}.
Recent years have seen increased interest and activity in
measuring CIB fluctuations. We feel it is important to summarize a
{\it minimal} set of requirements any quality study of CIB
fluctuations should meet. These requirements cover three major
aspects of the problem: I) preparing the maps that accurately
isolate the source-subtracted CIB fluctuations down to the (faint)
levels expected from first stars; II) tools required to
correctly analyze the processed (and clipped) data; and III)
details required for robust cosmological interpretation of the
results.

$\bullet$ I. {\it Map assembly}:

1. Maps of diffuse emission should be constructed carefully
removing artifacts down to levels well {\it below} those of
the expected cosmological signal. In practice, this means that the
maps should not have any structure at levels above $\delta F \sim
0.01$ nW m$^{-2}$ sr$^{-1}$ at $\sim$ arcminute scales in IRAC
channels.

2. No correlations should be introduced when constructing the
maps.

3. Observations must be carried out in a way that enables spatial structure 
on the sky to be distinguished from structure of the telescope and 
instrument on the scales of interest. The images produced should not 
contain any direct or indirect spatial filtering that modifies 
the spatial structure of the sky in unknown ways.

4. Because of temporal variations of the zodiacal light, data
should be collected in as short time intervals as possible. For
the GOODS fields combining data separated by 6 months (E1 and E2)
is not reliable when trying to measure fluctuations as faint as
$\lesssim 0.1$ nW m${^-2}$ sr$^{-1}$.

$\bullet$ II. {\it Analysis tools}:

5. Instrument noise, both its amplitude and the power spectrum,
must be estimated from $A-B$ maps. It is particularly necessary
for shot noise estimates etc.

6. When removing foreground sources, one should be careful with
the effects from the remaining mask. If the fraction of removed
pixels is small (in the IRAC images we find that it is typically
$\lesssim 30\%$) one can compute the power spectrum using FFTs;
otherwise the correlation function must be evaluated instead. In
any case, one must demonstrate that the power spectra recovered
are consistent with the computed correlation function, which is
immune to mask effects.

7. The instrument beam (PRF) must be reconstructed and its 
large- and small-scale properties must be understood.

8. Thorough checks must be done to verify that no artifacts mimic the 
signal found.

$\bullet$ III. {\it Interpretation}:

9. A cosmological signal must be isotropic on the sky; this must be
demonstrated with data whenever possible.

10. Foreground contributions must be evaluated: cirrus emission
via estimates and zodiacal emission via measurements at different
epochs (e.g. E1--E2 in GOODS measurements).

\subsection{Source-subtracted CIB fluctuations from {\it Spitzer}}

\begin{figure}[ht] 
   \plotone{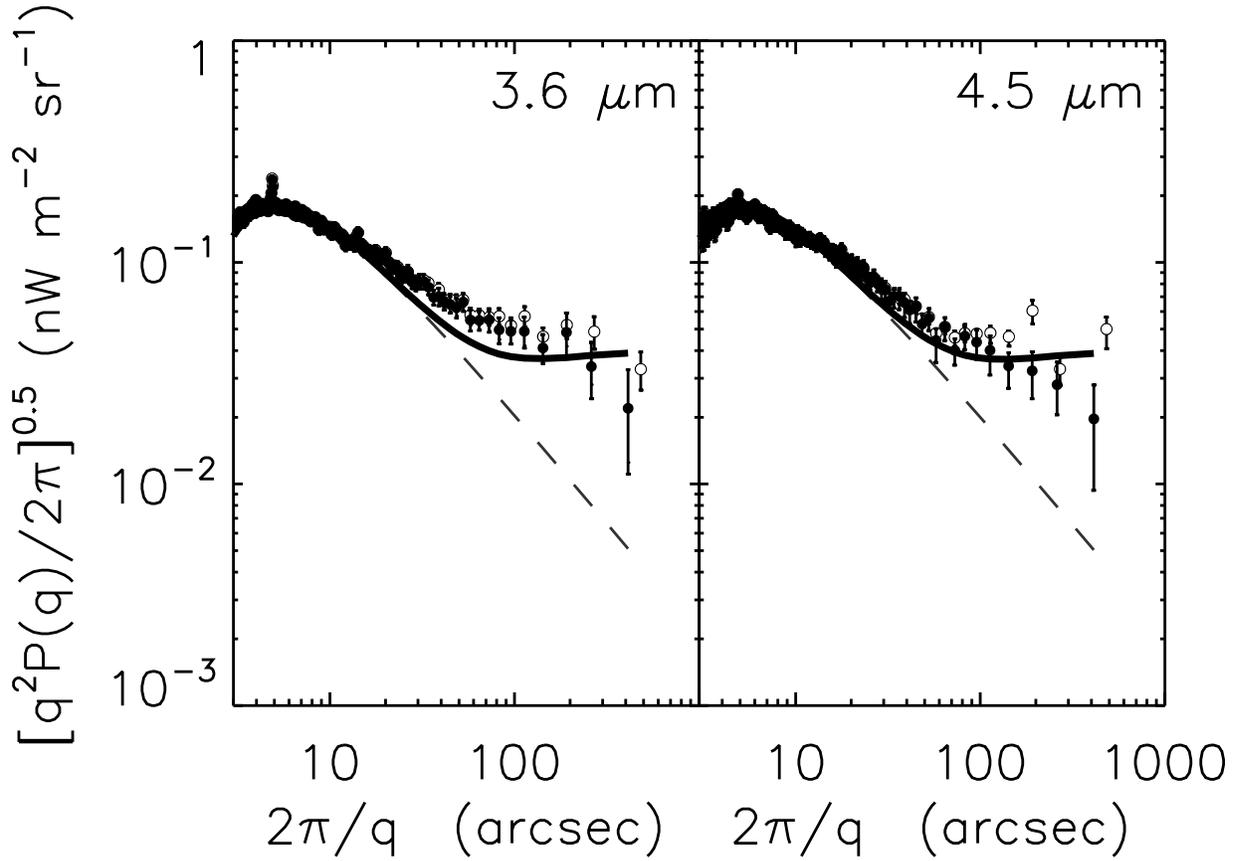}
   \caption{Average power spectra for the four source-clipped and model-subtracted
   GOODS fields at 3.6 and 4.5 $\micron$.
   The filled black circles indicate results when power along the Fourier
   transform axes is excluded. The open circles indicate results when power along
   the Fourier transform axes is retained. Error bars correspond to 1-$\sigma$ uncertainties.
   The dashed line indicates a fit to the shot noise at small angular scales. The
   thick solid line represents at simple model fit to the data assuming emissions originate
   at high $z$ with the concordance $\Lambda$CDM model. The amplitudes of the large-scale
   component are identical at 3.6 and 4.5 $\mu$m indicating that the color of the arcminute-scale
   fluctuations is approximately flat. (See text for details.)}
   \label{fig:cib_ak}
\end{figure}

Fig. \ref{fig:cib_ak} shows the final source-subtracted CIB
fluctuations obtained by averaging over all four GOODS areas at
the optimal Model iteration as discussed above. KAMM3 show that
the signal is made of two components: 1) small scales are
dominated by the shot noise component produced by the variance of
the remaining sources with the beam, and 2) the large scale CIB
fluctuations are produced by the clustering of the sources
producing the CIB. At the two longest wavelength IRAC Channels
(5.8 and 8 $\micron$) the instrument noise does not allow us to
eliminate foreground galaxies to a sufficiently interesting
shot noise level, so only results at 3.6 and 4.5 $\mu$m are shown
in the figure and discussed in this section.

The dashed lines show the shot noise fits obtained by linear
regression to the data. The remaining shot noise level at 3.6 and
4.5 $\mu$m from the populations with counts $dN/dm$ of AB
magnitude $m$ is:
\begin{equation}
P_{\rm SN}= \int S_0^2 10^{-0.8m} \frac{dN}{dm} dm \simeq
(1.4-1.7) \times 10^{-11} {\rm nW^2 m^{-4} sr^{-1}}
 \label{eq:shotnoise}
 \end{equation}
Here $S_0 = 3631$ Jy and the RHS gives the numerical values of
the remaining shot noise at both Channels of Fig.
\ref{fig:cib_ak}. More generally Equation (\ref{eq:shotnoise}) can be
rewritten as $P_{\rm SN} \simeq S_0 10^{-0.4m} F_{\rm CIB}(m)$,
where $F_{\rm CIB}(m)$ is the mean CIB produced by the sources
with typical magnitude $m$. 
For $dN/dm$ with power law slopes as observed at 3.6, and 4.5 $\micron$,
and at near-IR wavelengths (e.g. Fazio et al. 2004, Thompson et al. 2007a), 
the shot noise of Equation (\ref{eq:shotnoise}) will be dominated by the brightest
of the sources that are not excluded, i.e. the extrapolation of the normal
galaxy counts. Thus, the observed shot noise level is a strong upper
limit on the shot noise that is associated with the sources that 
produce the large scale CIB fluctuations.

Fig. 1 of KAMM3 shows the shot noise expected from the observed
source counts at 3.6 and 4.5 $\mu$m; the residual shot noise
levels imply that we have removed galaxy populations to at least
AB $m\sim 26.5-27$. The large-scale CIB fluctuations must thus
arise in fainter sources. KAMM4 show that the correlations between
the ACS galaxies and the source-subtracted CIB maps are very small
and, on arcminute scales, are within the statistical noise. Thus,
at most, the remaining ACS sources contribute to the shot noise
levels in the residual KAMM maps, but not to the large scale
correlations. In other words, the sources that produce the
large-scale CIB fluctuations detected at {\it Spitzer} IRAC
wavelengths are not present in the ACS source catalog. At the same
time, there are excellent correlations between the ACS source maps
and the sources {\it removed} by KAMM prior to computing the
remaining CIB fluctuations, which testifies to the high accuracy
of our Model subtraction procedure. Since the ACS galaxies do not
contribute to the source-subtracted CIB fluctuations at 3.6 and
4.5 $\mu$m, the latter must arise at $z>7$ as is required by the
Lyman break (at rest, $\lambda\sim 0.1 \mu$m) getting redshifted
past the ACS $z$-band with a central wavelength $\simeq 0.9 \mu$m.
This would place the sources producing the KAMM signal within the
first 0.7 Gyr. If the KAMM signal were to originate in lower $z$
galaxies which escaped the ACS GOODS source catalog because they
are below the catalog flux threshold, they would have to be
extremely low-luminosity systems ($< 2\times 10^7 h^{-2}L_\odot$
at $z$=1) and these galaxies would also have to cluster very
differently from their ACS counterparts (KAMM4).

We now turn to estimating the levels of the CIB required by our
results; the methodology of this follows KAMM3. We believe the
absence of correlations between the source-subtracted IRAC maps
and ACS maps places the sources responsible for the CIB
fluctuations at $z>7$. At $z=5-10$ one arcminute subtends comoving
scales of 2-3 $h^{-1}$ Mpc. Such scales are in the linear regime
at these epochs. So at $1+z > \Omega^{-1/3}$ in the flat Universe
with cosmological constant, the amplitude of the density
fluctuation is related to that at present via $\delta(z)\simeq
\delta(z=0) 1.3 (1+z)^{-1}$. (The numerical factor of 1.3 for
comes from the fact that density fluctuations grow very little at
$(1+z) < \Omega^{-1/3}$). The present-day density field is
normalized to the rms density contrast over a sphere with radius
$r_8=8h^{-1}$Mpc of $\sigma_8\sim 1$ at the present epoch.

One can now estimate the order of magnitude of the relative CIB
fluctuations. The cosmological parameters relevant to
such an estimate are well approximated as: the comoving angular
diameter distance, $d_A \simeq 5.4 [(1+z)/{6}]^{0.3}
h^{-1}$Gpc and the cosmic time $t\simeq 1.2[(1+z)/6]^{1.5}$
Gyr. The normalization scale $r_8$ would thus subtend an angle of
$\theta_8 \simeq 4^\prime [(1+z)/6]^{-0.3}$ and fall in the
middle of the angular scales where we detect the clustering
component of the CIB fluctuations. The relative fluctuation in the
projected 2-dimensional power spectrum, $\Delta$, on that angular
scale $\theta_8$, produced from sources located at mean value of
$\bar{z}$ and spanning the cosmic time $\Delta t$, would be
\begin{equation}
 \frac{\delta F_{\rm CIB}}{F_{\rm CIB}}(\theta_8) \sim \sigma_8 (1+\bar{z})^{-1}(r_8/c\Delta t)^{1/2}\simeq
0.02 \sigma_8 (\frac{\bar{z}}{10})(\frac{\Delta t}{{\rm
Gyr}})^{-1/2}
 \label{eq:delta8}
 \end{equation}
neglecting the amplification due to biasing. For the first star
systems forming in the concordance $\Lambda$CDM Universe, biasing
can amplify the fluctuations by a factor of $\sim (3-5)$ (Cooray
et al. 2004, Kashlinsky et al. 2004), so the relative fluctuations
are likely to be at most $\la 10\%$. {\it Thus in order to produce
a CIB fluctuation of amplitude $\delta F\sim 0.05-0.1$ nW m$^{-2}$ sr$^{-1}$
at arcminute scales these populations had to produce at least
$F_{\rm CIB} \ga 0.5-1$ nW m$^{-2}$sr$^{-1}$ in CIB flux.}

More generally, the fluctuations in the CIB generated by sources
clustered with the 3-dimensional power spectrum $P_3$ is given by
the Limber equation which can be written as follows (Kashlinsky
2005a):
\begin{equation}
\frac{q^2P(q)}{2\pi} = \Delta t \int_{\Delta t}
\left(\frac{dF}{dt}\right)^2 \Delta^2(qd_A^{-1}) dt \;\; ; \;
\Delta^2(k) \equiv \frac{1}{2\pi} \frac{k^2P_3(k)}{c\Delta t}
\label{eq:limber}
\end{equation}
where $d_A$ is the comoving angular diameter distance to $z$,
$\Delta t$ is the cosmic time spanned by the emitters and $\Delta$
is the rms fluctuation in the emitters counts over a cylinder of
length $c\Delta t$ and radius $\sim 2\pi/k$. Equation (\ref{eq:limber})
shows that any given shape in $\Delta(k)$ does not necessarily
translate into a similar shape in the angular spectrum of CIB
fluctuations. It is important to bear in mind that the relation
between the measured $P(q)$ and the underlying $P_3(k)$ can be
quite convoluted and, in general, depends on the evolution of the
CIB rate production, $dF/dt$, and other parameters.

It is of interest to consider how well populations described by a
pure $\Lambda$CDM model at high $z$, with a power spectrum $P_{\rm
\Lambda CDM}$ deduced from WMAP observations, fit the KAMM
measurements. In the case of the first stars, such populations are
likely to be biased, i.e. the luminous sources form at the high
peaks of the underlying density field. The relation between the
underlying $P_{\rm \Lambda CDM}$ and $P_3(k)$ is likely to be
non-linear even on linear scales depending on the height of the
peaks (Jensen \& Szalay 1986, Kashlinsky 1998), but as a toy model we approximate
here that $P_3(k) \propto P_{\rm \Lambda CDM}$. 
We further assume that $dF/dt$ does not vary significantly over the corresponding 
epochs, which can happen if the emitters span a narrow range of cosmic times. 
In this case the CIB fluctuations are reduced to $\sqrt{q^2P/2\pi} = 
\Delta t\ dF/dt\ \bar{\Delta}$, where $\bar{\Delta}^2$ being the suitably 
averaged value of $(\Delta t)^{-1}\int \Delta^2 dt$ which we assumed 
dominated by a narrow range of epochs. The values of
$\Delta$ for the concordance $\Lambda$CDM at high $z$ collapse
onto a line of universal shape whose amplitude is proportional to 
$\sigma_8 (1+z)^{-1} (\Delta t)^{-1/2}$; this model spectrum, with
an amplitude obtained by regression to the data, is shown with
a thick solid line in Fig. \ref{fig:cib_ak}. The shape of the
$\Lambda$CDM model power spectrum dictates that the amplitude of
that line does not change appreciably between $\sim 0.5^\prime$
and 10$^\prime$ in approximate agreement with the data. Although
this simple model (with the addition of a shot noise component) 
provides a decent fit to the data, one can
notice, however, that it deviates at places from the measurements.
This is not unexpected given the simplicity (and likely
inaccuracy) of this toy model, and may be indicative of the
particular form of $dF/dt$ and/or biasing over the epochs of
emissions contributing to these CIB fluctuations. 

As we remove populations to progressively lower levels of $P_{\rm
SN}$, we should eventually remove also the populations producing
the clustering CIB component. This shot noise level, at which the
clustering component goes away (or is substantially reduced) will
provide one with information on the typical fluxes of the sources.
One can already set an {\it upper} limit (see KAMM3) from the
current analysis where we reach the shot noise levels given in
Table 3, whereas the levels of the residual fluctuations from
clustering require CIB fluxes of $F_{\rm CIB} \ga 1$ nW m$^{-2}$ sr$^{-1}$:
 \begin{equation}
 f_\nu(m) \leq 10 \; \left(\frac{P_{\rm SN}}{10^{-11}{\rm nW^2 m^{-4} sr^{-1}}}\right)
 \left(\frac{F_{\rm CIB}}{1\; {\rm nW m{^-2} sr^{-1}}}\right)^{-1} \;{\rm nJy}
 \label{eq:popflux}
 \end{equation}
The magnitude of $f_\nu(m)$ is already comparable to that expected for
Population III systems (see Fig. \ref{fig:colors_ak} below - the
first stars are expected to form in mini-haloes of a few million
solar mass and convert a substantial fraction of that mass into
stars.). At the same time, Table 1 and Fig. 1 of KAMM2 possibly
already show (modest) decrease in the levels of the clustering
component of the CIB fluctuations.

Thus if we remove sources to the shot noise levels significantly
lower than in the earlier analyses, we should reach into the shot
noise produced by the first systems and characterize their
individual flux levels. It is impossible to achieve that with the
existing {\it Spitzer} deep fields because in any such analysis the
instrument noise must be at least several times below the required
shot noise (in order to have reliable source modeling for removal
while keeping intact the underlying instrument noise structure).
However, since the instrument noise amplitude scales $P \propto
t_{\rm integration}^{-1}$, in order to reach the shot noise levels
required to identify what kind of populations produce the CIB
anisotropies down to the nJy flux level, we would need to reduce
the instrument noise by another factor of $\sim 4-5$ compared to
the GOODS data. This can be achieved in approximately 100 hours
per pixel integrations reaching the shot noise levels
corresponding to an appreciable drop in the clustering component
over a field of at least $5^\prime\times 10^\prime$ in size. The
small Ultra-Deep field of the GOODS data located within the HDF-N
region has, in principle, sufficient integration time for reaching
the desired shot noise levels, but was obtained over several
Epochs and only covers a region of $5'$ on the side. Our analysis
of these data showed that we cannot extract the required large
scale ($>2^\prime$) information reliably from these data. It also
is in a region with cirrus emission estimated to be several times
brighter than in the Lockman Hole.

Additional information on the populations responsible for these
CIB fluctuations, can be obtained from the fact that the
significant flux ($>$1 nW m$^{-2}$sr$^{-1}$) required to explain
the amplitude of the fluctuations must be produced within the
short time available at these high $z$ (cosmic times $<$0.5-1
Gyr). This can be translated into the comoving luminosity density
associated with these populations, which in turn translates into
the fraction of baryons locked in these objects with the
additional assumption of their $\Gamma\equiv M/L$ (KAMM3). The
smaller the value of $\Gamma$, the fewer baryons are required to
explain the CIB fluctuations detected in the KAMM studies. It
turns out that in order not to exceed the baryon fraction observed
in stars, the populations producing these CIB fluctuations had to
have $\Gamma$ much less than the solar value, typical of the
present-day populations (KAMM3). This is consistent with the
general expectations of the first stars being very massive.

Assuming the first stars were massive and radiating close to the
Eddington limit, the level of the near-IR CIB is directly related
to the fraction of baryons processed by these stars and for the
integrated NIRBE levels of $\sim 30$ nW m$^{-2}$ sr$^{-1}$ claimed by the
various {\it IRTS} and DIRBE-based analyses this fraction is $f_* \sim
2-3\%$ (Kashlinsky 2005b). The minimal CIB fluxes required to
explain the clustering component of the KAMM measurements are,
however, much smaller and {\it would require the baryon fraction
to be as small as} $f_*\ga 0.1\%$.

\subsection{Source subtracted CIB fluctuations and comparison to other
measurements}

\subsubsection{Comparison to earlier CIB fluctuation measurements}

Previous detections of near-IR CIB fluctuations involved analysis
based on DIRBE (Kashlinsky \& Odenwald 2000), {\it IRTS} (Matsumoto et
al. 2005) and deep 2MASS (Kashlinsky et al. 2002, Odenwald et al. 
2003) data. None of these are directly comparable to the KAMM
measurements because of the different wavelengths involved,
different angular scales and most importantly different levels of
source subtraction.

The DIRBE analysis of Kashlinsky \& Odenwald (2000) measured the
CIB fluctuations at $\sim 0.5^\circ$ at wavelengths overlapping
with {\it Spitzer} (DIRBE bands L and M), while {\it IRTS} measurements were
effectively done at $\lambda \sim 2\ \micron$. Neither the DIRBE nor
{\it IRTS} analysis allowed for any significant source subtraction
because of the poor angular resolution and large confusion
noise, and thus both measured the total CIB fluctuations, making it
impossible to isolate the high $z$ contributions.

Analysis of deep 2MASS data (Kashlinsky et al. 2002, Odenwald et al. 
2003) enabled more removal of foreground galaxy populations and
measured CIB fluctuations out to $\sim 1^\prime$ from remaining
sources in J, H, K photometric bands (1.2, 1.6 and 2.2 $\micron$).
This was because of the much better angular resolution, but
atmospheric airglow and thermal fluctuations limited source removal to $m_{\rm
Vega} \sim 19-19.5$. As
discussed explicitly by Kashlinsky et al. (2002), that analysis
detected CIB fluctuations from ordinary galaxies at $z\ga 1$; Fig.
2d,e there makes clear that their signal comes from sources at
these redshifts and is not dominated by Population III sources (see
also Fig. 5 of Kashlinsky et al. 2004).

To conclude, none of these earlier measurements allow for a
model-independent analysis and robust
comparison vis-\`a-vis the KAMM {\it Spitzer} measurements.

\subsubsection{Comparison to $\gamma$-ray limits}

Aside from direct determination of the CIB, indirect limits on it
can be set by studying absorption of gamma-ray sources due to
two-photon absorption. This reaction, $\gamma \gamma_{\rm
CIB}\rightarrow e^+ e^-$, happens above a threshold $E_\gamma
E_{\gamma_{\rm CIB}} \ge (m_{\rm e}c^2)^2$ and being
electrodynamic in nature has cross section $\sim\sigma_{\rm
Thomson}$. Hence, for photons of the right energies it can
provide efficient absorption over cosmological distances (Nikishev
1962). However, difficulties here are 1) interpretation usually
requires assumptions about the original unabsorbed gamma-ray
source spectrum and 2) the amount of absorption at each gamma-ray
energy is not caused by a single energy IR photon, but is a
complex integral over the entire range of CIB photons energies
above the reaction threshold. Nevertheless, two recent studies
(Dwek et al. 2005, Aharonian et al. 2006) suggested NIRBE levels
significantly smaller than those indicated by the DIRBE and
{\it IRTS} analyses.

The HESS team results (Aharonian et al. 2006) have received
particular attention and we address them below in light of the
KAMM results. Their analysis involved modeling CIB with a
scaled spectral template representing the CIB from normal galaxies,
with or without the addition of a NIRBE component represented by the
{\it IRTS} residual emission. Based on this model and
assuming that that the intrinsic hardness of the blazar spectra,
$dN/dE \propto E^{-\Gamma}$, is $\Gamma \geq 1.5$, Aharonian et al. 
(2006) conclude that the full NIRBE suggested by {\it IRTS} would lead to more
attenuation at $\sim 1-2$ TeV than the known blazar physics allows.

\begin{figure}[ht] 
   \plotone{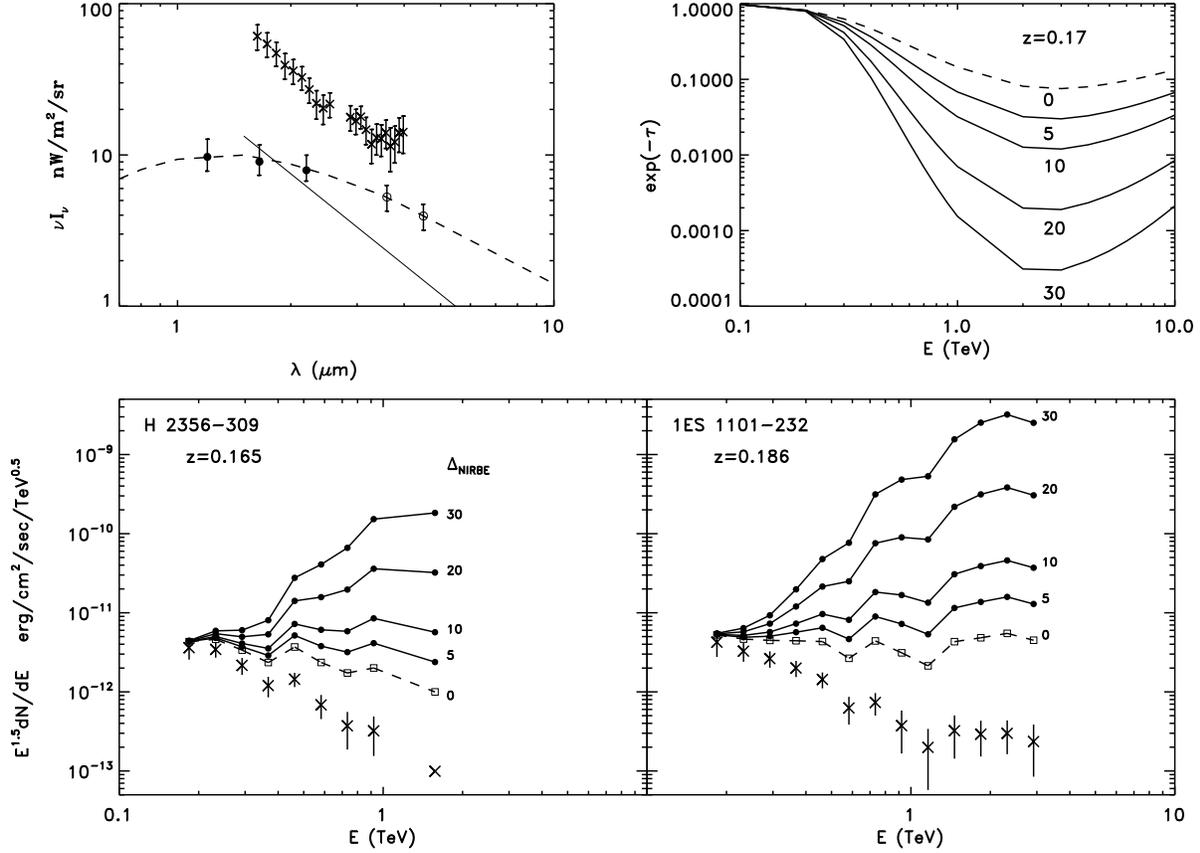}
   \caption{{\it Top}: (Left) Circles show the net observed fluxes from deep galaxy counts from Fig. 9 of
Kashlinsky (2005a and references therein). Dashed line is the
interpolated CIB from these galaxies used in the calculation.
Crosses correspond to the {\it IRTS} measurements from Matsumoto
et al. (2005). Solid line shows the modeled NIRBE from Pop III
with $\nu I_\nu \propto \lambda^{-\alpha}$, $\alpha=2$, and a
Lyman cutoff, normalized to the integrated flux of $\Delta_{\rm
NIRBE} = 10$ nW m$^{-2}$ sr$^{-1}$. (Right) Attenuation factor for
a source at $z$=0.17 over the range of energies of HESS for CIB
with the shown value of NIRBE from Pop III, $\Delta_{\rm NIRBE}$
in nW m$^{-2}$ sr$^{-1}$. Note that at $z\simeq 0.2$ the most
sensitive range for probing NIRBE is around 2 TeV. {\it Bottom}:
HESS measured spectra for the two blazars (Aharonian et al. 2005)
are shown with crosses. Open squares correspond to the intrinsic
spectra in the absence of any NIRBE (dashed line in top left
panel). Circles show the spectra corrected for additional
absorption due to CIB photons produced at $z>$10 and the NIRBE
values shown near each line. }
   \label{fig:gray_ak}
\end{figure}

It is important to emphasize that even assuming the blazar physics
limits adopted by Aharonian et al. (2006) the HESS data still permit
significant CIB fluxes from the epochs identified with the Pop~III
era (Kashlinsky \& Band 2007). A property of any such emission
would be a part of the CIB with a Lyman break corresponding to
(e.g.) $z\ga 10$. As an example of such CIB we computed the
intrinsic (corrected for absorption) blazar spectra assuming that
the NIRBE contribution from Pop~III scales as $\nu I_\nu \propto
\lambda^{-\alpha}$ with $\alpha=2$ and a Lyman limit cutoff corresponding
to the Pop~III era ending at $z_3$=10 and normalized to the shown
levels of the integrated NIRBE flux, $\Delta_{\rm NIRBE}$ in
nW m$^{-2}$ sr$^{-1}$ (Kashlinsky \& Band 2007). 
The results are not
sensitive to the assumed slope of $\nu I_\nu$, which was adopted
because it is in approximate agreement with the
{\it IRTS} data as shown in Fig. \ref{fig:gray_ak} (see also e.g. Dwek
et al. 2005). In addition we assumed the CIB from the observed
galaxies populations to be given by that from the measured galaxy
counts, as shown in Fig. \ref{fig:gray_ak}. The figure shows that
the attenuation due to CIB levels claimed by the {\it IRTS} and DIRBE
measurements is probably too strong assuming $\Gamma \ge 1.5$, but
{\it smaller levels of NIRBE are still allowed by the data}
in that they lead to $\Gamma \leq 1.5$. In particular the HESS
data require the levels of NIRBE due to Pop~III (i.e. with Lyman
cutoff in the CIB at 1 $\micron$) to be $\la$5 nW m$^{-2}$ sr$^{-1}$
leaving $\la 1\%$ of the baryons to have gone through Pop~III.
The KAMM results, $\Delta_{\rm NIRBE} \sim 1$ nW m$^{-2}$ sr$^{-1}$, 
are fully consistent with the HESS blazar data.

The HESS use of the blazar hardness index limit,
$\Gamma \ge 1.5$, has been questioned by several authors who
developed specific models that can reproduce much harder blazar
spectra (Katarzynski et al. 2006, Stecker \& Scully 2008, Krennrich
et al. 2008). This could then allow much higher values of
$\Delta_{\rm NIRBE}$. This situation is expected to be resolved with the data from the
recently launched {\it Fermi} mission which should measure spectra of high-$z$
gamma-ray bursts (GRBs) and blazars out to 300 GeV (Kashlinsky
2005b). Regardless of the precise amount of the near-IR CIB from them,
Pop~III objects likely left enough photons to provide a large
optical depth for high-energy photons from high-$z$ GRBs. Even if
the NIRBE levels from Pop~III were significantly smaller than
claimed by the {\it IRTS} and DIRBE analysis, $\Delta_{\rm NIRBE} \sim
30$ nW m$^{-2}$ sr$^{-1}$, there should still be almost complete damping in
the spectra of high-$z$ gamma ray sources at energies $\ga 260
(1+z)^{-2}$ GeV. Such damping should provide an unambiguous
feature of the Pop~III era and {\it Fermi} observations could provide
important information on the emissions from the Pop~III era
(Kashlinsky 2005b).

\subsubsection{Comparison to HST/NICMOS measurements and colors}

Thompson et al. (2007a,b) have used very deep NICMOS data for a
small ($~\sim144''\times 144''$) field to study the contribution
from resolved galaxies and conclude that the CIB at 1.1 and 1.6
$\micron$ is much smaller than the {\it IRTS} (Matsumoto et al.
2005) and the DIRBE (e.g. Cambresy et al. 2001) results suggest.
From analysis of the NICMOS background fluctuations Thompson et al. 
(2007a,b) measure source-subtracted arcminute-scale fluctuations 
of $\delta F \la 1 $ nW m$^{-2}$ sr$^{-1}$ at 1.1 and 1.6 $\micron$. 
This is broadly consistent with our detected fluctuations of $\delta F(\sim 1')
\sim 0.05-0.07$ nW m$^{-2}$ sr$^{-1}$ at 3.6 and 4.5 $\micron$ (KAMM1-4 and above). 
Neither the IRAC nor NICMOS analyses provide direct measurements of the mean
level of any unresolved background (including a NIRBE). The direct measurement of the
mean NIRBE intensity with NICMOS is prevented by the fact that the zodiacal
background subtraction is derived using a median image constructed from all
individual exposures in the data set. The subtraction of the median image from
each single image may not alter the fluctuations of any NIRBE, but it will
remove the median intensities of any and all unresolved backgrounds. This
includes the zodiacal light (as intended), but also includes any instrumental,
geocentric, Galactic, or extragalactic components.

\begin{figure}[ht] 
   \plotone{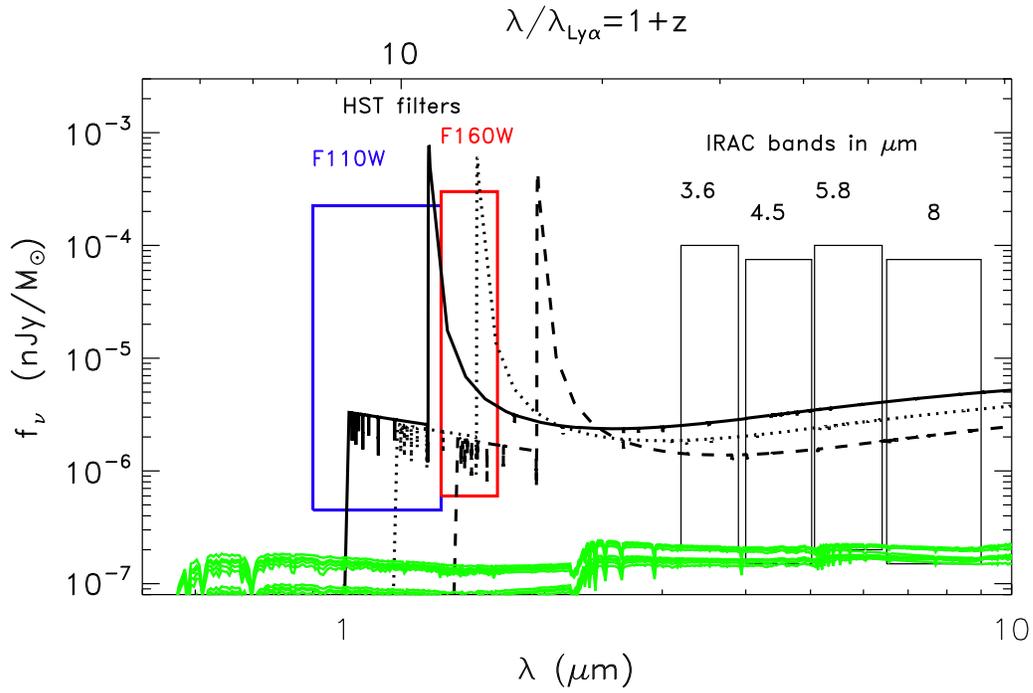}
   \caption{The spectra from Pop~III systems are shown for $z$=10 (thick
solid line), 12 (dotted) and 15 (dashed). The lines are drawn from
Santos et al. (2002) for the case when processing of the radiation
takes place in the gas inside the nebula. The {\it HST} and IRAC filters
are shown. Green lines show the flux spectra for star-bursts at
$z$=5 with the Salpeter-Scalo IMF and ages of 0.5 and 1 Gyr; the
lines span metallicities from 0 to $5\times 10^{-3}Z_\odot$.}
   \label{fig:colors_ak}
\end{figure}

In principle, the colors of the source-subtracted CIB fluctuations provide additional
information on the nature of the sources producing them. Thompson
et al. (2007b) proposed that the NICMOS and NICMOS-to-IRAC colors are
consistent with {\it normal} stellar populations originating at
$z\la 8$. At higher redshifts, the Lyman break begins to move through the F110W
NICMOS filter, and the expected colors of galaxies begin to redden until 
$z\gtrsim 14.4$ when the sources completely drop out of the F110W filter.
For sources with Pop III SEDs dominated by Ly $\alpha$ emission, such as 
the model shown in Fig. \ref{fig:colors_ak} (Santos et al. 2002), the expected 
colors would change more slowly with redshift, until a very abrupt drop off 
when Ly $\alpha$ shifts out of the F110W filter at $z\gtrsim 10.5$.
In KAMM4, the lack of correlation between the IRAC fluctuations and faint
sources in the ACS $z$ band is presented as evidence that any high redshift 
contributors to the fluctuations are $z$-band dropouts, thus lying at $z\gtrsim6.5$.
However, beyond these robust constraints imposed by the Lyman break (and perhaps the 
Lyman $\alpha$ line), a detailed interpretation of the colors of the fluctuations 
requires assumptions about the intrinsic SEDs of contributing systems, 
{\it and} their abundance and evolution with $z$. 
Full analysis of the colors of the fluctuations should also 
include demonstration of a correlation of the fluctuations at different 
wavelengths, to ensure that the different wavelengths are not 
dominated by different populations of sources.

In the context of Pop III emission, such 
as discussed by Santos et al. (2002), the J and H band fluxes are 
dominated by Lyman $\alpha$ emission from the first stars that lie at $5.4 < z < 13.8$, 
whereas the IRAC filters would probe emission reprocessed by IGM and halo gas. 
Lyman $\alpha$ photons diffuse out of their original sources by scattering 
off neutral hydrogen before reionization (Loeb \& Rybicki 1999). 
The density and structure of the halo gas and IGM are only weakly constrained 
at present, so the ratio of the J and H band emission to the mid-IR emission 
is very model- and epoch-dependent.
Figure \ref{fig:cib_ak} shows that
the clustering of the populations producing the KAMM signal at 3.6
and 4.5 $\mu$m is reasonably described by the concordance $\Lambda$CDM
model with sources at high $z$ with the 3.6/4.5 $\mu$m color approximately
expected for populations described by a model SED from Santos et al. (2002)
as shown in Fig. \ref{fig:colors_ak}. Because these
particular Pop III models are dominated by Lyman $\alpha$ emission  
they could be made to fit a wide range of colors by placing the 
Population III systems at
suitable redshift ranges and/or varying their abundances with $z$
in a suitable fashion. For instance, such SEDs would lead to smaller 
3.6/4.5 $\mu$m colors if there is strong evolution in the number density
of the sources between $z\sim 15$ and $z\sim 10$. Mathematically, one 
can construct such evolution models as eq. \ref{eq:limber} and Fig. 
\ref{fig:colors_ak} show. 

\section{Prospects for future measurements}

This section discusses future prospects for isolating and
identifying the nature of the populations responsible for the KAMM
signal and their epochs. Progress in this can be achieved with the
following three experiments which involve 1) large angular scales
range, 2) larger wavelengths range, and 3) deeper integrations.

\subsection{Larger Angular Scales}
If the populations producing the KAMM signal lie at epochs of the
first stars at high $z$, and assuming that the
structures are seeded via the concordance $\Lambda$CDM model, the
angular spectrum of source-subtracted CIB fluctuations should
exhibit a peak at angular scales corresponding to the horizon
scale at the matter-radiation equality projected to that redshift
(Cooray et al. 2004, Kashlinsky et al. 2004, Kashlinsky 2005a).
This peak should then subtend angular scales $\simeq
0.2^\circ-0.3^\circ$ and can be identified with suitable mappings
of regions covering sufficiently large areas.

\begin{figure}[ht] 
   \plotone{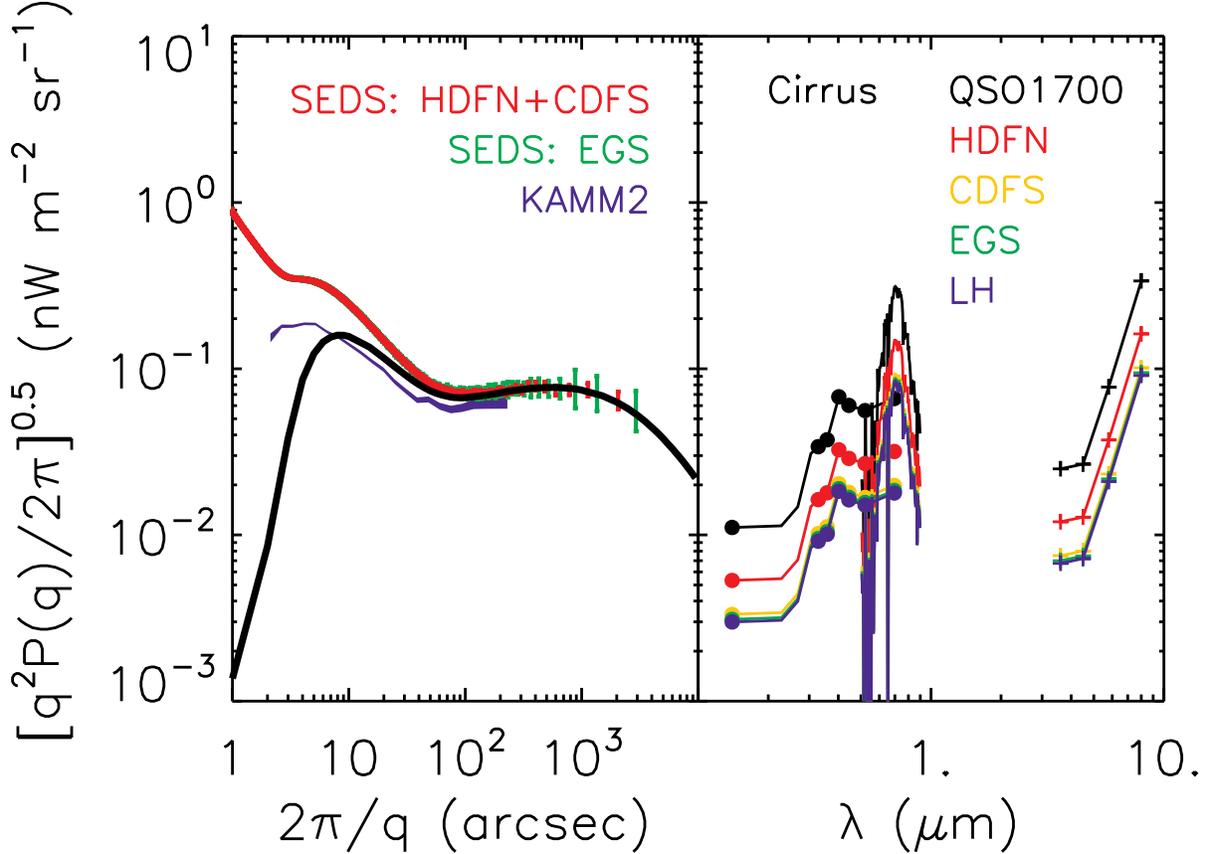}
   \caption{Expectations for measurements of CIB fluctuations at
   larger angular scales, and limitations due to interstellar dust.
   The left panel compares present CIB fluctuation measurements in the GOODS fields
   (blue line; KAMM2), with the expected results (red and green points with
   error bars) that may be obtained from the {\it Spitzer} SEDS project
   which will cover wider regions nearly as deeply. The larger fields should begin
   to reveal the $\sim1000''$ peak in the fluctuation spectrum that is expected if
   the fluctuations are dominated by the first luminous stars in the Universe (black line).
   For several different fields characterized by different \ion{H}{1} column density, the
   right panel shows the estimated noise level (expected to be nearly flat,
   or independent of $q$) due to thermal emission at IRAC wavelengths (adapted from
   Arendt et al. 1998),
   and due to extended red emission (ERE; Gordon et al. 1998) at $\sim0.7$ $\micron$.
   At shorter UV and visible wavelengths, scattered starlight
   becomes important (Leinert et al. 1998; Haikala et al. 1995; Mattila 1990; 
   Guhathakurta \& Tyson 1989; Toller 1981).
   The wavelength range from 1 - 5 $\micron$ is a window where minimal contamination
   by the ISM is expected.}
   \label{fig:future_ak}
\end{figure}

Fig. \ref{fig:future_ak} (left panel) shows the {\it expected}
results that may be obtained by The {\it Spitzer} Extended Deep
Survey
(SEDS)\footnote{http://ssc.spitzer.caltech.edu/geninfo/es/}. This
project will provide data to angular scales as large as 1 degree,
with sufficient depth to detect foreground galaxies to $\sim 0.15\
\mu$Jy (5 $\sigma$). The CIBER rocket experiment (Bock et al.
2005) is designed to detect spatial structure on scales up to 2
degrees, though with a much more limited capability for excluding
faint foreground sources because of very shallow exposures.

\subsection{Wider Range of Wavelengths}
If these populations originate at high $z$ their emissions below
the Lyman break, at rest wavelengths $\la 0.1 \mu$m, should
effectively have been absorbed by the local IGM (Santos et al
2002, Schaerer 2002). Thus one should probe the level of residual
diffuse light at wavelengths $\la 1 \mu$m and see if they
correlate with the source-subtracted CIB maps at IRAC bands.
Progress here has been made by in KAMM4 where it was shown that
the maps of deep ACS sources in GOODS observations exhibit
completely negligible correlations with the source-subtracted maps
at IRAC wavelengths. This likely places the sources producing the
KAMM signal at $z\ga 6.5$ unless they originate in extremely
low-luminosity local galaxies with $m_{\rm AB} \ga 28-29$ that
somehow escaped the ACS source catalog detection. Still, it would
be desirable to compare directly the residual diffuse light maps
in visible bands with those observed by IRAC. (The levels of
residual maps artifacts in the ACS maps have prevented KAMM4 from
doing such direct comparison).

Fig. \ref{fig:future_ak} (right panel) shows the estimated ISM (cirrus)
spectra in our deep fields and the Lockman Hole (a region of minimum
\ion{H}{1} column density). Emission from small grains and PAHs in the ISM rises
sharply at wavelengths $>5$ $\micron$. At wavelengths of $\sim0.7$
$\micron$ the extended red emission (ERE; Gordon et al. 1998),
is the likely limiting factor, with the directly scattered starlight
being an underlying continuum at all wavelengths. Though {\it Spitzer}
is incapable of observing at shorter wavelengths, CIBER will
perform its fluctuation measurements at 0.8 and 1.6 $\micron$,
and also includes a low resolution spectrometer to search for a
Lyman break redshifted into the CIB.
The Wide Field Camera 3 on the {\it Hubble
Space Telescope} and the NIRCAM instrument on the {\it James Webb
Space Telescope (JWST)} should also provide good opportunities to
extend CIB fluctuation studies to short wavelengths.

\subsection{Deeper Integration}
Finally, KAMM have already reached very low residual shot noise
levels of $P_{\rm SN} \simeq (1-2)\times 10^{-11}$ nW$^2$ m$^{-4}$ sr$^{-1}$
at 3.6 and 4.5 $\micron$ and there is tentative evidence that the
clustering signal already starts diminishing as lower levels of
the shot noise are reached. This may imply that in our modeling
we are already beginning to remove the very populations producing
the large-scale fluctuations. This is not surprising since as
shown in KAMM3 and discussed above, the low levels of the
shot noise coupled with the relatively significant levels of the
arcminute scales source-subtracted CIB fluctuations, imply that
the sources producing the latter must have individual fluxes $\la
20-30$ nJy. This is already in the range of fluxes expected even
for Population III systems as Fig. \ref{fig:colors_ak} shows.
Hence, it is reasonable to expect that mapping a suitable region
of up to $\la 10^\prime$ on the side with sufficiently long
integrations ($\ga$ 100 hr/pixel) should reach the shot noise
levels, $P_{\rm SN} \sim (\rm a\; few) \times 10^{-12} \propto
1/t$ nW m$^{-2}$ sr$^{-1}$ where the clustering component of the CIB
disappears completely, or is significantly diminished. This would
then probe the flux levels of the individual sources producing the
KAMM signal at arcminute scales. At this depth {\it Spitzer}
becomes limited by confusion. However, with its much larger
aperture, {\it JWST} will not be limited by confusion until much fainter
levels, and should provide a much clearer picture of the clustering of
very faint sources on angular scale $>100''$.

%%%%%%%%%%%%%%%%%%%%%%%%%%%%%%%%
\section{Summary}

This paper provides the details behind our prior
analysis (KAMM1 -- KAMM4) of the spatial fluctuations or power
spectrum of the CIB. We show the extent to which the final results
do or do not depend on the details of the data reduction and
analysis.

For various deep {\it Spitzer} IRAC data sets, we show that the self-calibration
that we apply to the data effectively removes spatial and temporal artifacts
well enough to probe fluctuations in the source-subtracted CIB down to
levels well below $\sim 0.1$ nW/m$^2$/sr on arcminute angular scales.
Some of the relatively strong and large--scale artifacts that we remove
are seen to be present in the current v0.30 release of the independently
processed GOODS data. The self-calibration procedure does not add 
artificial spatial correlations to the data.

We describe in detail the masking and modeling of the resolved sources.
Various checks on these procedures involving either modest changes
in their parameters, or the construction of test images demonstrate that
the CIB fluctuations are not directly related to sources that could
be identified above the sensitivity (or confusion) limits of the
given observations. Table \ref{tab:checks} itemizes the tests performed.

At 3.6 and 4.5 $\micron$ the residual CIB fluctuations can be
reasonably characterized using 2 components: shot noise, produced
by the variance of sources too faint to be individually detected
(dominant at small scales); and a clustered component which
dominates on scales $\ga 30'$. As a simplified representation of
the power spectra of each field, we provide the parameters of
fitting each power spectrum with these components, plus either 
a flat noise spectrum or an empirical estimate of the instrument noise.

We summarized the requirements that must be met by studies of such
faint cosmological signals. We reiterate that the sources
producing the large scale signal must have a very small shot noise
component, while contributing significant fluctuations on
arcminute scales. The latter component can be reasonably fit by a
high-$z$ population within a $\Lambda$CDM concordance model, with
net CIB fluxes at 3.5 and 4.5 $\mu$m of $\ga 1$ nW/m$^2$sr. The
low levels of the shot noise imply that individual sources
producing the large scale CIB fluctuations must be individually
very faint, $\la 20$ nJy. We then demonstrate that this population
and its CIB level is consistent with the available data on
high-energy $\gamma$-ray absorption, and HST NICMOS data on CIB
and its colors.

Finally, we discuss future prospects for testing
the nature of the CIB fluctuations: the currently approved SEDS
survey in warm Spitzer mission would enable us to extend the
measurements to sub-degree scales and probe the peak in the
spatial spectrum of the fluctuations at $\sim 0.2^\circ$ expected
from a high-$z$ population in the $\Lambda$CDM concordance
cosmology. Deeper IRAC integrations over a smaller region are
recommended in that they could detect the shot noise levels where
the clustering component disappears or is appreciably diminished;
this would identify the flux range of the individual sources
contributing to the latter. We point out that, while diffuse light
measurements below 1 $\micron$ could in principle probe the Lyman
break of these populations, such measurements may be limited by
increased levels of scattered Galactic starlight light (and
extended red emission) in the ISM.

\acknowledgements Support was provided by the National Science
Foundation through grant NSF AST 04-06587. This work is based on
archival data obtained with the Spitzer Space Telescope, which is
operated by the Jet Propulsion Laboratory, California Institute of
Technology under a contract with NASA. Additional support for the
First Look Survey (FLS) portion of this work was provided by an award issued by JPL/Caltech 
(NASA Spitzer NM0710076). 

%%%%%%%%%%%%%%%%%%%%%%%%%%%%%%%%

\clearpage

\clearpage

\clearpage

\clearpage 

\clearpage

%%%%%%%%%%%%%%%%%%%%%%%%%%%%%%%%
\begin{deluxetable}{ccccccccc}
\tabletypesize{\scriptsize}
\tablewidth{0pt}
\tablecaption{Background Fluctuation Data Sets}
\tablehead{
\colhead{} &
\colhead{Analyzed Field} &
\colhead{Image Pixel}&
\colhead{Integration} &
\colhead{} &
\multicolumn{4}{c}{Expected 3$\sigma$ Sensitivity (AB mag)}\\
\cline{6-9}
\colhead{Field} &
\colhead{Size (arcmin)} &
\colhead{Scale (arcsec)}&
\colhead{Time (hr)} &
\colhead{}&
\colhead{3.6$\micron$}&
\colhead{4.5$\micron$}&
\colhead{5.8$\micron$}&
\colhead{8$\micron$}
}
\startdata
GOODS\tablenotemark{1}    &   9.7$\times$9.7  & 0.6 & $\sim$21  & & 26.9 & 26.1 & 24.0 & 23.8\\
QSO 1700                  &  11.5$\times$5.1  & 0.6 & $\sim7.8$ & & 26.3 & 25.6 & 23.5 & 23.3\\
EGS\tablenotemark{2}      &  12.8$\times$7.7  & 1.2 & 1.4       & & 25.4 & 24.7 & 22.6 & 22.4\\
FLS                       &   108$\times$108  & 1.2 & 0.017     & & 22.3 & 19.4 & 20.0 & 19.9\\
\enddata
\tablenotetext{1}{HDFN-e1, HDFN-e2, CDFSe-1, CDFS-e2}
\tablenotetext{2}{``patch 4'', epoch 2}
\label{tab:data}
\end{deluxetable}

%%%%%%%%%%%%%%%%%%%%%%%%%%%%%%%%
\begin{deluxetable}{llll}
\tabletypesize{\scriptsize}
\tablewidth{0pt}
\tablecaption{Background Analysis Checks}
\tablehead{
\colhead{Issue} &
\colhead{Test} &
\colhead{Result} &
\colhead{Reference}
}
\startdata
1. Excess power on           & Compare power spectra before & Ignoring power on the FFT      & Fig. \ref{fig:zero_axes}, \S3.2\\
preferred axes?              & and after blanking axes      & axes can mitigate artifacts    & \smallskip \\
2. Self-calibration          & Compare different channels   & Self-calibration does remove   & Figs. \ref{fig:offsets}-\ref{fig:compare34}, \S4.1\\
effectiveness?               & Compare to GOODS processing  & some artificial patterns       & \smallskip \\
3. Optimal depth for         & Check residual skewness and  & Zero skewness is most robust   & Figs. \ref{fig:skewness}-\ref{fig:figure5_EGS}, \S4.2\\
resolved source models?      & correlation with model       & and simplest criterion         & \smallskip \\
4. Correct PRF for           & Test with modified PRFs      & PRF at 4.5 $\micron$ may be slightly & Fig. \ref{fig:figure6_PRF}, \S4.2\\
resolved source models?      &                              & too sharp, but not a problem   & \smallskip \\
5. Sensitivity to            & Alter clipping masks and     & Little sensitivity to          & Figs. \ref{fig:figure_mask1}-\ref{fig:figure_mask4}, \S4.2\\
clipping fraction?           & add random clipping          & variation of masked area       & \smallskip \\
6. Results related to        & Calculate power spectrum     & Very unlike power spectrum     & Fig. \ref{fig:dither}, \S4.3\\
dither pattern?              & of dither pattern            & of residual intensity          & \smallskip \\
7. Results related to dither & Calculate power spectrum     & Calibration errors would yield & Fig. \ref{fig:dither_offsets}, \S4.3\\
pattern + calibration?       & of dithered detector offset  & distinct large scale power     & \smallskip \\
8. Results related to        & Calculate power spectrum     & Test power is too flat, or too & Figs. \ref{fig:power_mask_CDFSe1}-\ref{fig:power_mask_EGS}, \S4.3\\
foreground sources and mask? & of foreground sources, mask, & weak to produce observed       & \\
                             & ``halo'' image               & large scale power              & \smallskip \\
9. Similar results in        & Compare parameterized fits   & Large scale power has similar  & Figs. \ref{fig:fit_plots1}-\ref{fig:final_power_norm}, \S5\\
different fields?            & to different power spectra   & shape in different fields, but & Tables \ref{tab:fit_params}-\ref{tab:chi_norm} \\
                             &                              & scales with shot noise (depth) & \smallskip \\
10. Similar structure at     & Calculate cross-correlation  & Significant correlation and    & KAMM1\\
different wavelengths?       & coefficients and colors      & constant color indicate celestial & \\
                             &                              & origin at 3.6 and 4.5 $\micron$ & \smallskip \\
11. Possible zodiacal light  & Constrain by re-observation  & Indicates that fluctuations in & KAMM1\\
structures?                  & at different epochs          & zodiacal light are smaller than & \\
                             &                              & those in the observed background & \smallskip \\
12. Possible ISM (cirrus)    & Constrain by observations of & ISM could dominate large-scale  & KAMM1\\
structures?                  & regions at various H~I column & fluctuations at 8 $\micron$, but & \\
                             & density                      & should be unimportant at shorter $\lambda$ & \\
\enddata
\label{tab:checks}
\end{deluxetable}

%%%%%%%%%%%%%%%%%%%%%%%%%%%%%%%%%%%%%%%%%%%%
\begin{deluxetable}{llccccccccc}
\tabletypesize{\scriptsize}
\tablewidth{0pt}
\tablecaption{Power Spectrum Parameters}
\tablehead{
\colhead{$\lambda$ ($\micron$)} &
\colhead{Field} &
\colhead{$10^{11}a_0$} &
\colhead{$a_1$} &
\colhead{$10^{11}a_2$} &
\colhead{$10^{11}a_3$} &
\colhead{$\chi^2_{\nu}$} &
\colhead{$10^{11}b_0$} &
\colhead{$b_1$} &
\colhead{$10^{11}b_2$} &
\colhead{$\chi^2_{\nu}$}
}
\startdata
3.6 & CDFS-e1    &  5.10$\pm$ 0.11 &  1.56$\pm$ 0.03 &  1.31$\pm$ 0.01 &  0.12$\pm$ 0.00 &  2.73 &  4.58$\pm$ 0.11 &  1.62$\pm$ 0.03 &  1.33$\pm$ 0.01 &  4.01 \\
3.6 & CDFS-e2    &  5.30$\pm$ 0.13 &  1.07$\pm$ 0.02 &  1.48$\pm$ 0.03 &  0.13$\pm$ 0.00 &  3.25 &  4.62$\pm$ 0.13 &  1.14$\pm$ 0.03 &  1.58$\pm$ 0.02 &  6.26 \\
3.6 & HDFN-e1    &  5.32$\pm$ 0.12 &  1.77$\pm$ 0.04 &  1.53$\pm$ 0.01 &  0.13$\pm$ 0.00 &  3.01 &  4.93$\pm$ 0.12 &  1.81$\pm$ 0.04 &  1.55$\pm$ 0.01 &  5.67 \\
3.6 & HDFN-e2    & 18.89$\pm$ 0.35 &  1.98$\pm$ 0.03 &  2.06$\pm$ 0.02 &  0.12$\pm$ 0.00 &  2.98 & 17.80$\pm$ 0.36 &  2.09$\pm$ 0.03 &  2.16$\pm$ 0.01 &  6.98 \\
3.6 & QSO 1700   & 33.02$\pm$ 1.19 &  1.15$\pm$ 0.05 &  9.15$\pm$ 0.19 &  0.44$\pm$ 0.00 &  4.48 & 25.36$\pm$ 1.28 &  1.88$\pm$ 0.12 & 11.59$\pm$ 0.09 & 15.06 \\
3.6 & EGS        & 22.14$\pm$ 0.48 &  1.52$\pm$ 0.04 &  4.35$\pm$ 0.09 &  1.41$\pm$ 0.01 &  0.99 & 16.17$\pm$ 0.56 &  0.89$\pm$ 0.06 &  1.05$\pm$ 0.26 &  3.25 \\
3.6 & FLS        &  1600$\pm$ 9.12 &  1.93$\pm$ 0.01 &   896$\pm$ 1.44 &   181$\pm$ 0.06 &  3.76 &  \nodata        &  \nodata        &  \nodata        &  \nodata \\
4.5 & CDFS-e1    &  4.07$\pm$ 0.10 &  1.11$\pm$ 0.03 &  1.44$\pm$ 0.03 &  0.25$\pm$ 0.00 &  2.55 &  3.52$\pm$ 0.10 &  1.25$\pm$ 0.04 &  1.46$\pm$ 0.02 &  2.29 \\
4.5 & CDFS-e2    &  6.61$\pm$ 0.14 &  1.04$\pm$ 0.03 &  1.09$\pm$ 0.03 &  0.24$\pm$ 0.00 &  2.47 &  5.59$\pm$ 0.15 &  1.14$\pm$ 0.03 &  1.16$\pm$ 0.03 &  2.07 \\
4.5 & HDFN-e1    &  5.82$\pm$ 0.14 &  1.28$\pm$ 0.03 &  1.60$\pm$ 0.02 &  0.23$\pm$ 0.00 &  2.83 &  5.19$\pm$ 0.15 &  1.36$\pm$ 0.03 &  1.56$\pm$ 0.02 &  2.24 \\
4.5 & HDFN-e2    &  3.23$\pm$ 0.08 &  1.29$\pm$ 0.06 &  1.53$\pm$ 0.02 &  0.21$\pm$ 0.00 &  2.45 &  2.69$\pm$ 0.08 &  1.57$\pm$ 0.06 &  1.56$\pm$ 0.02 &  2.12 \\
4.5 & QSO 1700   & 38.53$\pm$ 1.05 &  1.05$\pm$ 0.03 &  4.56$\pm$ 0.19 &  0.68$\pm$ 0.00 &  2.25 & 36.04$\pm$ 1.05 &  1.20$\pm$ 0.04 &  5.49$\pm$ 0.16 &  2.41 \\
4.5 & EGS        & 24.42$\pm$ 0.71 &  0.87$\pm$ 0.03 &  4.09$\pm$ 0.35 &  2.70$\pm$ 0.02 &  1.08 & 22.59$\pm$ 0.71 &  0.88$\pm$ 0.04 &  1.48$\pm$ 0.31 &  2.31 \\
4.5 & FLS        &   957$\pm$ 4.58 &  1.92$\pm$ 0.00 &   926$\pm$ 1.51 &   212$\pm$ 0.07 &  3.00 &  \nodata        &  \nodata        &  \nodata        &  \nodata \\
5.8 & CDFS-e1    &  32.3$\pm$  1.2 &  0.88$\pm$ 0.04 &  17.3$\pm$  0.6 &   3.2$\pm$  0.0 &  7.30 &   4.2$\pm$  1.0 &  2.52$\pm$ 0.20 &  11.6$\pm$  0.3 &  3.12 \\
5.8 & CDFS-e2    &  41.0$\pm$ 12.9 &  0.20$\pm$ 0.08 &   0.0$\pm$ 12.9 &   3.4$\pm$  0.0 &  6.78 &   0.0$\pm$  0.8 &  2.38$\pm$ \nodata &  13.9$\pm$  0.3 &  3.34 \\
5.8 & HDFN-e1    &  49.5$\pm$  6.4 &  0.30$\pm$ 0.06 &   0.0$\pm$  6.6 &   3.0$\pm$  0.0 &  8.88 &   0.0$\pm$  1.3 &  2.37$\pm$ \nodata &  12.6$\pm$  0.3 &  3.51 \\
5.8 & HDFN-e2    &  50.1$\pm$  2.1 &  3.40$\pm$ 0.04 &  27.1$\pm$  0.3 &   3.2$\pm$  0.0 &  9.41 &  16.3$\pm$  1.6 &  4.21$\pm$ 0.07 &  17.0$\pm$  0.4 &  2.94 \\
5.8 & QSO 1700   & 406.2$\pm$ 13.1 &  2.72$\pm$ 0.07 &  74.9$\pm$  1.3 &  12.7$\pm$  0.0 &  4.08 & 249.1$\pm$ 13.3 &  3.38$\pm$ 0.08 &  42.3$\pm$  1.5 &  3.08 \\
5.8 & EGS        & 261.8$\pm$ 11.2 &  2.19$\pm$ 0.05 & 237.7$\pm$  3.1 &  31.0$\pm$  0.2 &  1.54 & 284.1$\pm$ 13.0 &  1.92$\pm$ 0.06 &  65.9$\pm$  3.5 &  2.58 \\
5.8 & FLS        &  8710$\pm$   35 &  2.23$\pm$ 0.00 & 18400$\pm$27.84 &  2730$\pm$ 0.92 & 19.88 &  \nodata        &  \nodata        &  \nodata        &  \nodata \\
8   & CDFS-e1    &  33.0$\pm$  1.4 &  0.53$\pm$ 0.05 &   2.6$\pm$  1.3 &   2.1$\pm$  0.0 &  2.96 &  33.4$\pm$  0.7 &  0.67$\pm$ 0.02 &  -0.0$\pm$  0.0 &  4.13 \\
8   & CDFS-e2    &  30.9$\pm$  0.8 &  1.60$\pm$ 0.04 &  11.3$\pm$  0.2 &   2.5$\pm$  0.0 &  2.04 &  31.7$\pm$  0.9 &  1.49$\pm$ 0.05 &   7.3$\pm$  0.3 &  3.06 \\
8   & HDFN-e1    &  44.3$\pm$  1.3 &  2.42$\pm$ 0.03 &  11.8$\pm$  0.2 &   1.9$\pm$  0.0 &  4.47 &  48.4$\pm$  1.4 &  2.32$\pm$ 0.03 &   6.5$\pm$  0.2 &  9.29 \\
8   & HDFN-e2    &   8.7$\pm$  0.4 &  2.55$\pm$ 0.07 &  10.0$\pm$  0.1 &   1.6$\pm$  0.0 &  3.64 &  10.7$\pm$  0.4 &  2.38$\pm$ 0.07 &   5.7$\pm$  0.2 &  7.33 \\
8   & QSO 1700   & 151.9$\pm$  4.3 &  1.83$\pm$ 0.07 &  32.2$\pm$  0.7 &   5.0$\pm$  0.0 &  3.68 & 134.6$\pm$  4.7 &  1.95$\pm$ 0.08 &  26.4$\pm$  0.8 &  2.10 \\
8   & EGS        & 144.6$\pm$  7.2 &  2.62$\pm$ 0.07 & 157.8$\pm$  2.4 &  22.1$\pm$  0.1 &  1.28 & 177.8$\pm$  7.8 &  2.34$\pm$ 0.07 &  65.2$\pm$  2.4 &  3.18 \\
8   & FLS        &  1730$\pm$  8.3 &  2.49$\pm$ 0.01 &  5860$\pm$10.32 &  1130$\pm$ 0.38 & 22.83 &  \nodata        &  \nodata        &  \nodata        &  \nodata \\
\enddata
\tablecomments{Units for $a_0$, $a_2$, $a_3$, $b_0$, and $b_2$ are nW$^2$ m$^{-4}$ sr$^{-1}$.}
\label{tab:fit_params}
\end{deluxetable}

%%%%%%%%%%%%%%%%%%%%%%%%%%%%%%%%%%%%%%%%%%%%
\begin{deluxetable}{llccccccccc}
\tabletypesize{\scriptsize}
\tablewidth{0pt}
\tablecaption{Power Spectrum Parameter Covariances}
\tablehead{
\colhead{$\lambda$ ($\micron$)} &
\colhead{Field} &
\colhead{$\frac{C(a_0,a_1)}{\sigma_{a_0}\sigma_{a_1}}$} &
\colhead{$\frac{C(a_0,a_2)}{\sigma_{a_0}\sigma_{a_2}}$} &
\colhead{$\frac{C(a_0,a_3)}{\sigma_{a_0}\sigma_{a_3}}$} &
\colhead{$\frac{C(a_1,a_2)}{\sigma_{a_1}\sigma_{a_2}}$} &
\colhead{$\frac{C(a_1,a_3)}{\sigma_{a_1}\sigma_{a_3}}$} &
\colhead{$\frac{C(a_2,a_3)}{\sigma_{a_2}\sigma_{a_3}}$} &
\colhead{$\frac{C(b_0,b_1)}{\sigma_{b_0}\sigma_{b_1}}$} &
\colhead{$\frac{C(b_0,b_2)}{\sigma_{b_0}\sigma_{b_2}}$} &
\colhead{$\frac{C(b_1,b_2)}{\sigma_{b_1}\sigma_{b_2}}$}
}
\startdata
3.6 & CDFS-e1    & -0.422 & -0.438 &  0.081 &  0.608 & -0.100 & -0.256 & -0.478 & -0.427 &  0.549 \\
3.6 & CDFS-e2    & -0.389 & -0.597 &  0.136 &  0.847 & -0.164 & -0.268 & -0.442 & -0.609 &  0.824 \\
3.6 & HDFN-e1    & -0.403 & -0.352 &  0.064 &  0.495 & -0.080 & -0.245 & -0.430 & -0.346 &  0.463 \\
3.6 & HDFN-e2    & -0.304 & -0.292 &  0.055 &  0.521 & -0.091 & -0.253 & -0.414 & -0.288 &  0.437 \\
3.6 & QSO 1700   &  0.117 & -0.156 &  0.069 &  0.883 & -0.229 & -0.331 & -0.489 & -0.363 &  0.536 \\
3.6 & EGS        & -0.420 & -0.480 &  0.219 &  0.652 & -0.257 & -0.613 & -0.125 & -0.383 &  0.924 \\
3.6 & FLS        &  0.692 &  0.258 & -0.046 &  0.505 & -0.094 & -0.297 &  \nodata &  \nodata &  \nodata \\
4.5 & CDFS-e1    & -0.475 & -0.611 &  0.172 &  0.822 & -0.195 & -0.338 & -0.558 & -0.602 &  0.720 \\
4.5 & CDFS-e2    & -0.149 & -0.431 &  0.157 &  0.865 & -0.238 & -0.367 & -0.278 & -0.505 &  0.831 \\
4.5 & HDFN-e1    & -0.565 & -0.614 &  0.157 &  0.741 & -0.173 & -0.338 & -0.627 & -0.599 &  0.684 \\
4.5 & HDFN-e2    & -0.379 & -0.495 &  0.140 &  0.768 & -0.178 & -0.337 & -0.472 & -0.429 &  0.508 \\
4.5 & QSO 1700   &  0.304 & -0.033 &  0.058 &  0.885 & -0.280 & -0.405 &  0.177 & -0.130 &  0.856 \\
4.5 & EGS        & -0.609 & -0.765 &  0.445 &  0.904 & -0.437 & -0.637 & -0.591 & -0.747 &  0.892 \\
4.5 & FLS        &  0.306 &  0.008 &  0.001 &  0.372 & -0.067 & -0.289 &  \nodata &  \nodata &  \nodata \\
5.8 & CDFS-e1    & -0.765 & -0.839 &  0.217 &  0.863 & -0.187 & -0.309 & -0.964 & -0.244 &  0.225 \\
5.8 & CDFS-e2    & -0.990 & -0.999 &  0.367 &  0.995 & -0.341 & -0.364 & -0.911 & -0.300 &  0.249 \\
5.8 & HDFN-e1    & -0.973 & -0.993 &  0.386 &  0.992 & -0.350 & -0.381 & -0.783 & -0.255 &  0.193 \\
5.8 & HDFN-e2    & -0.903 & -0.097 &  0.027 &  0.085 & -0.023 & -0.304 & -0.973 & -0.075 &  0.071 \\
5.8 & QSO 1700   & -0.638 & -0.226 &  0.053 &  0.222 & -0.050 & -0.284 & -0.802 & -0.140 &  0.116 \\
5.8 & EGS        & -0.820 & -0.337 &  0.164 &  0.274 & -0.129 & -0.563 & -0.794 & -0.391 &  0.323 \\
5.8 & FLS        & -0.317 & -0.100 &  0.017 &  0.151 & -0.024 & -0.235 &  \nodata &  \nodata &  \nodata \\
8   & CDFS-e1    & -0.746 & -0.869 &  0.312 &  0.967 & -0.263 & -0.328 &  0.567 &  0.001 &  0.003 \\
8   & CDFS-e2    & -0.636 & -0.539 &  0.129 &  0.521 & -0.110 & -0.304 & -0.625 & -0.602 &  0.588 \\
8   & HDFN-e1    & -0.788 & -0.227 &  0.059 &  0.205 & -0.052 & -0.309 & -0.764 & -0.269 &  0.241 \\
8   & HDFN-e2    & -0.692 & -0.342 &  0.093 &  0.215 & -0.057 & -0.299 & -0.681 & -0.404 &  0.253 \\
8   & QSO 1700   & -0.264 & -0.366 &  0.103 &  0.599 & -0.147 & -0.327 & -0.359 & -0.390 &  0.540 \\
8   & EGS        & -0.790 & -0.389 &  0.220 &  0.306 & -0.171 & -0.610 & -0.744 & -0.409 &  0.317 \\
8   & FLS        &  0.228 & -0.020 &  0.004 &  0.149 & -0.023 & -0.217 &  \nodata &  \nodata &  \nodata \\
\enddata
\tablecomments{Units for $a_0$, $a_2$, $a_3$, $b_0$, and $b_2$ are nW$^2$ m$^{-4}$ sr$^{-1}$.}
\label{tab:fit_covars}
\end{deluxetable}

\begin{deluxetable}{lccccc}
\tabletypesize{\scriptsize}
\tablewidth{0pt}
\tablecaption{$\chi^2_{\nu}$ for Comparison of Normalized Power Spectra ($2\pi/q > 5''$)}
\tablehead{
\colhead{Field} &
\colhead{CDFS-e2} &
\colhead{HDFN-e1} &
\colhead{HDFN-e2} &
\colhead{QSO 1700} &
\colhead{EGS}
}
\startdata
CDFS-e1 & 1.20/1.16/1.18/0.97 & 1.05/1.08/1.04/1.45 & 1.84/1.22/1.75/1.14 & 2.41/2.44/1.54/1.19 & 1.64/1.07/1.12/1.10\\
CDFS-e2 & \nodata             & 1.35/0.74/1.67/2.02 & 1.70/1.13/2.38/1.27 & 2.01/1.79/1.82/1.28 & 2.25/1.55/1.51/1.01\\
HDFN-e1 & \nodata             & \nodata             & 2.33/0.95/1.61/1.41 & 2.73/1.83/1.85/1.40 & 1.67/1.67/1.34/1.57\\
HDFN-e2 & \nodata             & \nodata             & \nodata             & 1.18/2.06/2.05/1.35 & 4.13/1.70/1.54/0.90\\
QSO 1700& \nodata             & \nodata             & \nodata             & \nodata             & 4.34/3.17/1.77/1.26\\
\enddata
\tablecomments{Results at 3.6/4.5/5.8/8 $\micron$. Number of degrees of freedom, $\nu = 115$.}
\label{tab:chi_norm}
\end{deluxetable}

\end{document}